\documentclass{article}
\usepackage[utf8]{inputenc}
\usepackage{newtxmath} 
\usepackage{newtxtext} 
\usepackage{fullpage}
\usepackage{graphicx}

\usepackage{color}
\usepackage{hyperref}

\usepackage{amssymb}
\usepackage{latexsym}

\usepackage{hyperref}
\usepackage{amsmath, bm}
\usepackage{tabularx}
\usepackage[table]{xcolor}
\usepackage{subcaption}
\usepackage{authblk}

\pdfoutput=1

\usepackage{nomencl}

\makenomenclature

\usepackage{soul}

\newcommand{\gray}[1]{\color{gray}#1}
\newcommand{\implementation}[1]{#1} 

\newcommand\SOAR{\mathit{SOAR}}

\usepackage{tikz}
\usetikzlibrary{calc,trees,positioning,arrows,chains,shapes.geometric,%
    decorations.pathreplacing,decorations.pathmorphing,shapes,%
    matrix,shapes.symbols,fit,decorations.text,%
    graphs,graphs.standard,quotes}

\tikzset{
  >=stealth,
  auto,
  x=4em,
  y=4em,
}
\pgfdeclarelayer{underbackestground}
\pgfdeclarelayer{backestground}
\pgfdeclarelayer{background}
\pgfdeclarelayer{foreground}
\pgfsetlayers{underbackestground,backestground,background,main,foreground}

\newcommand{\state}[0]{\bm{\psi}}
\newcommand{\modelerror}[0]{\bm{\beta}}
\newcommand{\observation}[0]{\bm{y}}
\newcommand{\singleobservation}[0]{y}
\newcommand{\observationerror}[0]{\bm{\epsilon}}
\newcommand{\interpolationoperator}[0]{I_{\Omega}}
\newcommand{\coarseningoperator}[0]{I_{\Omega}^T}
\newcommand{\innovation}[0]{\bm{d}}
\newcommand{\mean}[1]{\overline{#1}}

\newcommand{\dx}[0]{\Delta x}
\newcommand{\dy}[0]{\Delta y}
\newcommand{\dt}[0]{\Delta t}
\newcommand{\coarsedx}[0]{\tilde{\dx}}
\newcommand{\coarsedy}[0]{\tilde{\dy}}

\newcommand{\eref}[1]{(\ref{eq:#1})}

\newcommand{\reffig}[1]{Figure~\ref{fig:#1}}
\newcommand{\refsec}[1]{Section~\ref{sec:#1}}

\newcommand{\appref}[1]{\ref{app:#1}}

\begin{document}

\title{Massively Parallel Implicit Equal-Weights Particle Filter \\ for Ocean Drift Trajectory Forecasting}


\author[1,2]{H\aa{}vard~Heitlo~Holm\footnote{Corresponding author: havard.heitlo.holm@sintef.no}}
\author[3,4]{Martin~Lilleeng~S\ae{}tra}
\author[5,6]{Peter~Jan~van~Leeuwen}


\affil[1]{
SINTEF~Digital,
Mathematics~and~Cybernetics,
P.O.~Box 124~Blindern,
NO-0314~Oslo,
Norway.}
\affil[2]{
Norwegian~University~of~Science~and~Technology,
Department~of~Mathematic,
NO-7491~Trondheim,
Norway.}
\affil[3]{
Norwegian Meteorological Institute, 
P.O.~Box 43~Blindern,
NO-0313~Oslo,
Norway.}
\affil[4]{
Oslo Metropolitan University,
Department of Computer Science,
P.O.~Box 4~St. Olavs plass,
NO-0130~Oslo,
Norway.
}
\affil[5]{
Department of Atmospheric Science,
Colorado State University,
3915 W. Laporte Ave.
Fort Collins, CO 80521,
USA}
\affil[6]{
Department of Meteorology,
University of Reading,
Earley Gate, 
Reading RB6 6BB,
UK.}

\date{}

\maketitle


\begin{abstract}
Forecasting ocean drift trajectories are important for many applications, including search and rescue operations, oil spill cleanup and iceberg risk mitigation. 
In an operational setting, forecasts of drift trajectories are produced based on computationally demanding forecasts of three-dimensional ocean currents. 
Herein, we investigate a complementary approach for shorter time scales by using a recent state-of-the-art implicit equal-weights particle filter applied to a simplified ocean model.
To achieve this, we present a new algorithmic design for a data-assimilation system in which all components -- including the model, model errors, and particle filter -- take advantage of massively parallel compute architectures, such as graphical processing units.
%
%
Faster computations can enable in-situ and ad-hoc model runs for emergency management, and larger ensembles for better uncertainty quantification. 
Using a challenging test case with near-realistic chaotic instabilities, we run data-assimilation experiments based on synthetic observations from drifting and moored buoys, and analyse the trajectory forecasts for the drifters.
Our results show that even sparse drifter observations are sufficient to significantly improve short-term drift forecasts up to twelve hours.
With equidistant moored buoys observing only 0.1\% of the state space, the ensemble gives an accurate description of the true state after data assimilation followed by a high-quality probabilistic forecast. 

\end{abstract}






\section{Introduction}
\label{sec:introduction}

Prediction of drift trajectories in the ocean has many applications that are important to society and the environment.
Examples include search and rescue operations, recovering objects lost at sea, planning of boom placements for oil spill cleanup, and preventing collisions between icebergs and offshore installations.
To produce high-quality drift trajectory forecasts, it is important to have a good representation of ocean currents. This is not an easy task, as ocean currents have large natural variability and there are typically few available observations.
Furthermore, the size of ocean low- and high-pressure systems, so-called \emph{eddies}, is much smaller than their atmospheric counterparts, and it is challenging to place them correctly in typical grid resolutions used by operational ocean models today. 

The operational approach for drift trajectory prediction is to use the currents from the most recent ocean forecasts directly~\cite{dagestad2018}.
These are imported from computationally expensive ocean circulation models, which solve the dynamic state of the ocean in three dimensions. 
Typically, a large portion of the simulation run-time is spent on the data assimilation, which uses available real-world observations to correct the modeled ocean states.
At the Norwegian Meteorological Institute (MET Norway), the 4D-Var data-assimilation window is 48 hours and the forecast length 66 hours. Both figures are limited by available computational resources. Common forecast ranges for ocean circulation models are three to five days.
Operational drift trajectory forecasts at MET Norway are produced by OpenDrift~\cite{dagestad2018}, which is an offline trajectory model. 
It reads the ocean current forecasts produced by the ocean circulation models, and uses these to predict drift trajectories.
Although OpenDrift is computationally efficient, the ocean circulation models still require access to supercomputers.

This paper explores the option of using a state-of-the-art particle filter method applied to a simplified ocean model for efficient drift trajectory forecasting.
The aim is to build a data-assimilation system that can run efficiently on commodity-level desktop computers, and also be extendable to supercomputers.
We achieve this by using a simplified ocean model and a data-assimilation method that both are able to take advantage of massively parallel accelerator hardware, such as the graphical processing unit (GPU).
This work is not intended as a substitute of current operational systems, but as a complementary approach, in which the predicted currents may even be updated with in-situ observations, e.g., during ongoing search and rescue operations.
Furthermore, by enabling research models to run on individual desktop and laptop computers, researchers are able to do more rapid prototyping.
At the same time, this work will contribute to more efficient simulations also on supercomputers, since all algorithms may be extended to run on multiple GPUs and compute nodes.

The paper is organized as follows:
We start by reviewing related work relevant for Lagrangian data assimilation with accelerated particle filters.
In \refsec{dataassimilation}, we describe the data-assimilation problem and summarize the key concepts of so-called proposal-distribution particle filters.
We present the simplified ocean model and model errors in \refsec{modelSection}, whereas \refsec{iewpf_applied} offers a detailed description of an algorithm for running the chosen particle filter on this model.
The latter two sections also discuss how the GPU is used for efficient implementation of the computationally intensive components.
In \refsec{results}, we present experimental results of drift trajectory ensemble forecasts on an identical-twin experiment with near-realistic chaotic behavior.
Furthermore, we show and discuss the statistical validity of the forecasts and examine the computational performance of the simulations.
Finally, \refsec{conclusion} contains a summary and concluding remarks.



\paragraph{Related work}
Particle filters, and more generally Sequential Monte Carlo (SMC) methods, constitute a large class of numerical methods for statistical inference. 
It is well-known that the standard particle filter is prone to degeneracy in high-dimensional systems~\cite{snyder2008_obstacles_highdimPF, pjvl_2009_pf_review, snyder2015_perfboundsPF_optprop}, and there have been several attempts at designing particle filters without this limitation.
A few such particle filters have been used on high-dimensional, near-realistic applications in the geosciences.
Ades and van Leeuwen~\cite{pjvl_2010_ewpf} use the equivalent-weights particle filter on a high-dimensional, simplified, ocean model based on the barotropic equations, showing that it is possible to avoid the degeneracy problem in high-dimensional systems, at the cost of a biased estimate. Although the scheme performed well, the bias grows with ensemble size. 
Poterjoy, Sobash and Anderson~\cite{poterjoyLocalNWP} use a local particle filter on a weather research and forecasting model, in which the bootstrap particle filter is applied locally to observations, and particle states are merged in state space between the locations of the observations. However, it remains problematic to glue particles from these local updates together to full particles that span the whole model domain. The smoothing needed can easily destroy delicate balances in the flow. Furthermore, the minimum size of the local areas is set by physical length-scale constraints, typically meaning that too many observations are within a local domain to avoid degeneracy. In practise, a minimum weight value is set, meaning that not all information is extracted from the observations. Hence, also localisation is not solving the problem. A recent review by van Leeuwen et al.\cite{pjvl_2019_ReviewPF} discusses most recent developments on particle filters for high-dimensional geophysical systems.



Several implementations of standard particle filters for parallel architectures such as GPUs exist, but mainly within other scientific disciplines than geosciences.
Lopez et al.~\cite{LOPEZ2015116} present GPU-implementations of a particle filter (with sequential importance resampling) and auxiliary particle filter to detect anomalies in manufacturing processes, and show sufficient performance for real-time application. Gelencs{\'e}r-Horv{\'a}th et al.~\cite{Gelencser-Horvath2013} introduce a modified cellular particle filter with Metropolis resampling on the GPU for real-time applications. LibBi~\cite{2013LibBi} is a software package for state-space modelling and Bayesian inference capable of utilizing GPUs. Several particle filters are implemented in LibBi, e.g., particle Markov Chain Monte Carlo (pMCMC) and SMC$^2$. Other methods in LibBi include the Extended Kalman Filter (EKF) and parameter optimisation routines. Bai and Hu~\cite{Bai:2013:CMP:2499968.2499979} demonstrate particle filter-based data assimilation for simulation of wildfire spread, with parallel sampling and weight computation based on the MapReduce programming model. In a more recent work, Bai et al.~\cite{Bai2016-7046371} describe more efficient routing of particles between processing units in the resampling step of a distributed particle filter. 


Other data-assimilation methods have also been subject to GPU-accelerations.
Blattner and Yang~\cite{yang2012} give a performance study of a GPU-implementation of the local ensemble transform Kalman filter, Wei and Huang~\cite{wei2011} explore a GPU-based implementation of the EKF, and  Quinn and Abarbanel~\cite{QUINN20118168} present a general path integral Monte Carlo approach applied to a neuron model.
They all report massive speed-ups on the order 100-1000 over CPU implementations. 
Theoretical speed-up based on hardware specifications for FLOPS and memory bandwidth is on the order 10~\cite{lee_etal_10}.

Assimilation of Lagrangian data is challenging due to the potential complexity of the trajectories and the need for transforming the data into Eularian velocity data (for fixed-grid or spectral numerical models). Apte, Jones and Stuart~\cite{apte2008} use particle smoothing for assimilating Lagrangian data from drifters and present three methods for sampling from the exact posterior probability density function based on the Langevin equation and the Metropolis-Hastings algorithm. Their methods are shown to produce better results than the ensemble Kalman filter using perturbed observations. Spiller, Apte and Jones~\cite{spiller2013} use both particle filtering and smoothing (exact posterior sampling) for assimilating Lagrangian data from gliders and drifters. They propose a new observation operator to deal with the high uncertainty in the locations of the observations. Spiller et al.~\cite{SPILLER20081498} investigate the divergence of a particle filter for the point-vortex model. They introduce backtracking particle filters and show that the filters outperform EKF for the two-point vortex system. Other methods than particle filters and smoothers have also been successfully implemented~\cite{kuznetsov2003,apte2013,carrier2014,slivinski2015,SLIVINSKI2017131}. 


\paragraph{Paper contribution}
%
We present an efficient GPU-implementation of the recent implicit equal-weights particle filter applied to a simplified ocean model.
The data-assimilation algorithm and the numerical scheme for evolving the ocean model are both designed to take advantage of massively parallel architectures.
The data-assimilation system is tailored for observations of the ocean current obtained from either free drifting buoys or moored buoys.
We show numerical experiments for assimilating a challenging test case with near-realistic chaotic behavior, along with drift trajectory forecasts.
To the best of our knowledge, there exists no previous massively parallel versions of a state-of-the-art particle filter applied to a near-realistic geophysical application for data assimilation. 





\section{The data-assimilation problem}
\label{sec:dataassimilation}

There are many potential sources for errors in the simulation of atmospheric and oceanographic processes.
These errors may arise from physical processes missing in the mathematical model, discretization errors in the numerical method, sub-grid effects that can not be resolved in the discretized model, and uncertainties in model parameters, initial conditions, forcing and boundary conditions.
Hence, we do not only wish to simulate the behavior of the unknown physical state, denoted by $\state$, but rather its probability density function (pdf), $p(\state)$. 
As geophysical applications tend to be very high-dimensional and driven by nonlinear processes, an analytic description of $p(\state)$ is generally unobtainable, and an ensemble-based Monte-Carlo simulation is one way to measure the uncertainties in the system.
In its simplest form, ensemble-based statistical simulation consists of a set of $N_{e}$ independent state vectors $\{\state_i\}_{i=1,...,N_{e}}$, which are initialized according to uncertainties in the model parameters and initial conditions.
The state of each ensemble member is then simulated independently according to the model equation,
\begin{equation}
	\state^{n}_i = M\left(\state_i^{n-1}\right) + \modelerror^{n-1}_i, \quad \mathrm{for} \; n = 1, 2, ...,
	\label{eq:model_equation}
\end{equation}
in which the model $M$ evolves the solution deterministically from time $t^{n-1}$ to $t^n$, and $\modelerror_i^{n-1}$ is an optional stochastic variable that represents realizations of the errors in the model.
The pdf of the system can then be represented through the statistical properties of the resulting ensemble, e.g., as
\begin{equation}
	p(\state^n) = \frac{1}{N_{e}} \sum_{i=1}^{N_{e}} \delta \left(\state^n - \state_i^n \right),
    \label{eq:montecarlo}
\end{equation}
in which $\delta$ is the Dirac delta function.
\nomenclature{$\state$}{The state vector consisting of the model variables}
\nomenclature{$p(\cdot)$}{Probability density function (pdf)}
\nomenclature{$\state_i$}{Particle number $i$ / ensemble member $i$}
\nomenclature{$N_{e}$}{Number of particles / ensemble size}
\nomenclature{$\{\state_i\}_{i=1,...,N_e}$}{Ensemble consisting consisting of $N_e$ particles / ensemble members}
\nomenclature{$M(\state)$}{Deterministic model operator, evolving the state one time step forward}
\nomenclature{$\modelerror$}{Stochastic model error}
\nomenclature{$\delta(\cdot)$}{Dirac's delta. Evaluates to one if its argument is zero, and zero otherwise.}

If an observation $\observation^n$ of the system is available at time $t^n$, this information can be used to improve the obtained probability density.
Typically, the observation is also influenced by uncertainty, as
\begin{equation}
	\observation^n = H\left(\state_{true}^n\right) + \observationerror^n,
    \label{eq:observation}
\end{equation}
in which $H$ is the observation operator that maps the true state $\state_{true}^n$ to observation space and $\observationerror^n$ is a stochastic observation error.
The observations typically only cover parts of the system, so that the size of the observation vector (denoted $N_y$) is smaller than the size of state vector (denoted $N_{\state})$.
This is particularly true for geophysical systems, for which it is normal that  $N_y \ll N_{\state}$ (e.g., $\observation$ can be the value and direction of the ocean current at a single point in space and time).
Because of this, we can not simply replace the observed parts of $\state^n$ with the values in $\observation^n$ directly, and we have to consider the conditional pdf $p\left(\state^n | \observation^n \right)$.
\nomenclature{$\observation^n$}{Observation at time $t^n$}
\nomenclature{$\state_{true}^n$}{True state at time $t^n$.}
\nomenclature{$H(\state)$}{Observation operator, mapping a state $\state$ to observation space.}
\nomenclature{$\epsilon^n$}{Observation error, typically $\epsilon \sim N(0, R)$.}
\nomenclature{$N_y$}{Size of the observation space}
The data-assimilation problem consists of finding this conditional density, and its fundamental building block is Bayes theorem:
\begin{equation}
	p(\state^n | \observation^n) = \frac{p(\observation^n | \state^n) p(\state^n)}{p(\observation^n)}.
	\label{eq:bayesTheorem}
\end{equation}
The original pdf $p(\state^n)$ is here termed the \emph{prior probability}, as it represents our understanding of the system prior to assimilating the information in the observation.
The \emph{likelihood} $p(\observation^n | \state^n)$ expresses the probability of observing $\observation^n$ under the assumption that $\state^n$ is the true state of the system.
The \emph{marginal probability} $p(\observation^n)$, i.e., the probability of observing $\observation^n$, acts mainly as a normalization constant and ensures that the resulting \emph{posterior probability density} is a pdf.

\subsection{Standard particle filter}
\label{sec:standardParticleFilter}

The \emph{standard particle filter} is an ensemble-based data-assimilation technique that uses a direct evaluation of Bayes theorem.
Each particle (equivalent to an ensemble member), $\state_i$, is assigned a weight $w_i$ that gives the relative importance of that particle in the ensemble.
Typically, all $N_e$ particles are initialized with weight $w_i^0 = 1/N_{e}$, as they are sampled independently from the pdf of the initial conditions, $p(\state^0)$.
Each particle is then simulated independently according to \eref{model_equation} until observation time $t^n$. 
By applying \eref{bayesTheorem} directly with \eref{montecarlo} as the prior density, and by considering the marginal probability as a normalization constant, the posterior distribution is expressed as
\begin{equation}
	\begin{split}
	p(\state^n | \observation^n) &\propto \sum_{i=1}^{N_{e}}  \frac{ p(\observation^n | \state_i^n) }{\sum_{j=1}^{N_e} p(\observation^n | \state_j^n)}\delta(\state^n - \state_i^n) \\
    						   & = \sum_{i=1}^{N_{e}} w_i^n \delta(\state^n - \state_i^n).
	\end{split}
	\label{eq:standardParticleFilter}
\end{equation}
Here, the likelihood is used to update the weights $w_i^n$ for each particle, so that the posterior is represented by a weighted discrete distribution.
We can evaluate the likelihood if we know the pdf for the observation. 
For instance, if the observation error is Gaussian, $\epsilon^n \sim N(0, R)$, the weight for particle $\state_i$  becomes
\begin{equation}
	w_i \propto  \exp \left[ - \frac{1}{2} \left(\observation^n - H(\state_i^n)\right)^T R^{-1} \left( \observation^n - H(\state_i^n) \right) \right].
    \label{eq:standardParticleFilterWeight}
\end{equation}
\nomenclature{$w_i^n$}{Weight of particle $i$ at time $t^n$.}
\nomenclature{$R$}{Covariance matrix for the observation error $\observationerror$.}


As some particles inevitably end up with very low weights, they no longer carry significant statistical value.
To improve the statistical coverage in the high-probability regions, the ensemble is \emph{resampled} according to the weight distribution in \eref{standardParticleFilterWeight}, so that $\{\state_i^n\}_{i=1,...,N_e} \sim p(\state^n| \observation^n)$.
All weights for the resampled particles are then reset to $1/N_e$.
This is known as sequential importance resampling.
Several schemes can be used for this resampling~\cite{pjvl_2009_pf_review}, and in this work we consider the residual resampling scheme~\cite{Liu98sequentialmonte}.
Note that if the model \eref{model_equation} has $\modelerror = 0$, it is important that duplicated particles are given a perturbation to avoid ensemble collapse and completely overlapping particle trajectories.
With a stochastic model, however, exact duplications will evolve differently through independent realizations of $\modelerror_i$.

One of the main advantages of the standard particle filter is that it preserves all physical properties throughout the simulation, as the final particles are generated from successful simulation runs and not through manipulation of the state vectors.
A drawback, however, is that the ensemble is prone to collapse when the dimension of the observation space increases~\cite{snyder2008_obstacles_highdimPF, pjvl_2009_pf_review, snyder2015_perfboundsPF_optprop}.
In high-dimensional systems, all particles end up in the tail of the likelihood, with the consequence that only very few particles (perhaps even just one) gain a much higher weight than all others.
The distribution then collapses as all $N_e$ particles are resampled from few (or a single) particles that have non-zero weights.
This problem is often referred to as the \emph{curse of dimensionality}.

\subsection{The implicit equal-weights particle filter}
\label{sec:iewpf}
One technique used for overcoming the curse of dimensionality is to sample the states $\state_i^n$ from a \emph{proposal density}, $q$, with an appropriate compensation in the weights.
First, \eref{model_equation} shows that the pdf of the state at time $t^n$ is related to that of the previous time by the Markovian property
\begin{equation}
	\begin{split}
	p(\state^n) &= \int p(\state^{n} | \state^{n-1}) p(\state^{n-1})\; d\state^{n-1} \approx \frac{1}{N_e} \sum_{i=1}^{N_e} p(\state^{n} | \state_i^{n-1}),
    \end{split}
    \label{eq:markovianProperty}
\end{equation}
where we assumed that all particles have the same weight at time $t^{n-1}$.
In the standard particle filter, we draw the evolution of the particle from $p(\state^{n} | \state_i^{n-1})$, which is equivalent to solving the model equation for one time step.
We can choose it differently, by first multiplying and dividing the argument of the integral by a proposal density $q$ and then draw the particle evolution from that density,
\begin{equation}
	\begin{split}
	p(\state^n) &= \frac{1}{N_e} \sum_{i=1}^{N_e} \frac{p(\state^{n} | \state_i^{n-1})} {q_i(\state^n|\state_{1:N_e}^{n-1},y^n)}  q_i(\state^n|\state_{1:N_e}^{n-1},y^n).
    \end{split}
    \label{eq:applyingProposalDistribution}
\end{equation}
We have large freedom in how to choose $q$, but the support of $q$ is required to be equal to or larger than the support of $p(\state^n | \state_i^{n-1})$, and it should preferably be easy to sample from.
Here, the proposal is chosen to be conditioned on the observation $\observation^n$ and all particle states at the previous time step, $\state_{1:N_e}^{n-1}$, and it depends on the parent state $\state_i^{n-1}$ via index $i$.
Using the proposal density in Bayes theorem \eref{bayesTheorem} gives us
\begin{equation}
	p(\state^n | \observation^n) =  \frac{1}{N_e} \sum_{i=1}^{N_e}
	\frac{p(\observation^n | \state^n) p(\state^n|\state_i^{n-1})}{p(\observation^n) q_i(\state^n | \state_{1:N_e}^{n-1}, \observation^n)}  q_i(\state^n | \state_{1:N_e}^{n-1}, \observation^n).
	\label{eq:proposalDensity}
\end{equation}
By now sampling $\state^n_i \sim q_i(\state^n | \state_{1:N_e}^{n-1}, \observation^n)$, the posterior becomes
\begin{equation}
	p(\state^n | \observation^n) = \sum_{i=1}^{N_e} w_i^{n} \delta(\state^n - \state_i^n), \quad \mathrm{with} \quad w_i^{n} = \frac{ p(\observation^n | \state_i^n) p(\state_i^n | \state_i^{n-1})}{N_e p(\observation^n) q_i(\state_i^n | \state_{1:N_e}^{n-1}, \observation^n)}.
	\label{eq:sampledProposalDensity}
\end{equation}
\nomenclature{$\state_{1:N_e}^{n-1}$}{All particle states $\state_i, ..., \state_{N_{\state}}$}
\nomenclature{$q(\cdot)$}{Proposal density, typically used to ease the sampling from a complex distribution.}
\nomenclature{$\state_i^{n,a}$}{The analysis state for particle $i$ at time $t^n$, and mean of the optimal proposal density.}
\nomenclature{$P$}{Covariance matrix of the optimal proposal density.}

One choice of $q$ is the \emph{optimal proposal density}~\cite{Doucet2000}, in which $q_i(\state^n | \state_{1:N_e}^{n-1}, \observation^n) = p(\state_i^n | \state_i^{n-1}, \observation^n)$.
By considering a linear observation operator $H$ and Gaussian model and observation errors, $\modelerror \sim N(0, Q)$ and $\observationerror \sim N(0, R)$, the optimal proposal density is equivalent to $N(\state_i^{n,a}, P)$, with
\begin{equation}
	\state_i^{n,a} = M(\state_i^{n-1}) + QH^T\left(HQH^T + R \right)^{-1} \innovation_i^n
	\label{eq:opd_update}
\end{equation}
and
\begin{equation}
	P = \left( Q^{-1} + H^T R^{-1} H \right)^{-1},
	\label{eq:opd_covariance}
\end{equation}
in which 
\begin{equation}
    \innovation_i^n := \observation^n - H M(\state_i^{n-1})
    \label{eq:innovation}
\end{equation}
is called the \emph{innovation} for particle $i$.
The proposal is optimal in the sense that it gives optimal variance in the weights for proposals of the form $q(\state^n | \state_i^{n-1}, \observation^n)$, but as it turns out, it is not sufficient to avoid ensemble degeneracy~\cite{snyder2008_obstacles_highdimPF, snyder2015_perfboundsPF_optprop, pjvl_2013_exploration_ewpf}.
\nomenclature{$\innovation_i^n$}{The innovation for particle $\state_i$ at time $t^n$.}

The main particle filter we will use in this work is an extension of the implicit equal-weights particle filter (IEWPF). 
In the IEWPF~\cite{pjvl_2016_iepfw}, $q$ is chosen similar but not identical to the implicit particle filter~\cite{chorin_2013_ipf} by choosing the new particles as
\begin{equation}
	\state_i^n = \state_i^{n,a} + \alpha_i^{1/2} P^{1/2} \xi_i,
	\label{eq:iewpfUpdate}
\end{equation}
in which $\xi_i$ is a draw from
the standard multivariate Gaussian distribution $\xi_i \sim N(0, I)$ and $\alpha_i$ is a function of both $\xi$ and $\psi_i^{n-1}$.
Furthermore, we choose $\alpha_i$ such that the weights of all particles become equal to a target weight, which is equal to the lowest optimal proposal weight of all the particles. 
This choice is needed to ensure that we keep all particles in the ensemble, but comes with two drawbacks. 
Firstly, when the number of particles increases, the worst particle will be located further and further away from the observations, so the scheme enforces all particles to move further away from the observations. 
Secondly, numerical experiments show that the spread of the particles becomes underestimated in low-dimensional systems (its behaviour in high-dimensional systems is harder to assess as we do not know the true answer). 
Not withstanding these negatives, the IEWPF is the first particle filter that has uniform weights in high-dimensional systems.

To alleviate these two issues, Skauvold et al.~\cite{skauvold2019_2sIEWPF} extended the scheme by proposing an update equation for each particle of the form:
\begin{equation}
	\state_i^n = \state_i^{n,a} + \alpha_i^{1/2}P^{1/2}\xi_i + \beta^{1/2} P^{1/2} \nu_i,
	\label{eq:updateEquationTwoStage}
\end{equation}
in which $\nu_i$ is a second random vector $\nu_i \sim N(0,I)$ and $\beta$ is a covariance scaling parameter common to all particles.
The introduction of the new term enables us to remove the underestimation of the particle spread by tuning $\beta$. Furthermore, we can choose $\alpha_i$ and $\beta$ such that the target weight is equal to the mean of the optimal proposal weights. The consequence of this choice is that the particles are not forced away from the observations when the ensemble size increases. With this, both problems are solved, and this new scheme is the basis for our numerical experiments. Details of the scheme are given in \appref{iewpfDetails}.

\section{Simplified ocean model for massively parallel architectures}
\label{sec:modelSection}

Traditional ocean circulation models~\cite{roms,nemo} are generally written to resolve as many of the physical processes in the ocean as possible, and typically consider conservation of mass, momentum, energy, and tracers (salt and temperature) in three dimensions.
This makes them very computationally demanding and limits the feasible number of ensemble members. 
The number of members in an operational ensemble prediction system today is usually between 10 and 100.
Instead of a full three-dimensional ocean circulation model, we assume that the vertical velocities are negligible compared to the horizontal movement, and let the nonlinear shallow-water equations in a rotational domain serve as a simplified model. 
Thus, we vastly reduce the state space of the problem. 
In operational settings, the simplified model may be initialized based on the most recent ocean state from a traditional ocean circulation model, and be used for to forecast short-term ocean currents.
Furthermore, drift of Lagrangian objects in the ocean are typically driven by the ocean currents, wind, and wave-induced forces (Stokes drift)~\cite{christensen2018}, whereas in this work we only consider the contribution from the ocean currents.

The shallow-water equations are in the class of hyperbolic conservation laws, which are often solved using explicit finite-volume methods~\cite{LevequeFVM2004}. 
This class of problems is well-suited for efficient implementation on massively parallel hardware, such as GPUs~\cite{HagenHenriksenHjelmervikLie2007, sw12, AMC10:JS}. By also carefully tailoring the data-assimilation algorithms to use local operations, we are able to run the most computationally demanding parts of the code on the GPU. 
Control flow and intrinsic serial operations, however, are still carried out on the CPU. 
This way, we use each processor type for the task which it is best suited for. 
Through this approach, we can efficiently run an ensemble of a simplified ocean model on commodity-level desktop computers, reducing the requirements for access to supercomputers.

The GPU is an extreme case of a many-core processor, with hundreds or thousands of simple cores. 
Measured in floating-point operations per second (FLOPS), a standard desktop GPU surpasses the performance of the top supercomputer in the world ten years ago~\cite{top500}, and is today roughly ten times as fast as the CPU.
GPUs were initially designed for efficient graphics operations, but have become increasingly popular for general-purpose computing over the last 15 years.
Due to their design for optimized throughput of data-parallel operations and low prices driven by the gaming market, they became attractive accelerators when the steadily increasing CPU clock frequency came to an end~\cite{FreeLunchIsOver}.
Programming languages such as CUDA and OpenCL, and easy access to highly specialized third-party libraries\footnote{BLAS, RNG, FFT, image and signal processing, collective communication primitives, graph analytics, etc.}, debuggers and profilers, have further contributed to make them accessible for a wide range of computational problems.

The programming model of the GPU is accessed through \emph{kernels}, which are programs written in specialized languages for running on the GPU in a SIMD/SIMT (Single Instruction, Multiple Data/Threads) fashion.
The threads are organized in \emph{blocks}, which again are organized in a \emph{grid}. 
The grid (and blocks) can be one-, two- or three-dimensional, and the ideal choice of block-size configuration, denoted by $(b_x, b_y)$, will vary for different kernels and for different GPUs.
Each thread can communicate with other threads in the same block through the \emph{shared memory}, which can be described as a programmable cache or scratchpad memory. 
Communication between threads in different blocks, however, requires costly global synchronization.
The GPU does not share the main CPU memory, and all required data therefore needs to be explicitly transferred between the GPU and CPU.
This operation is relatively expensive and should be minimized for optimal performance.
For a more thorough introduction to GPU computing; see, e.g., Sanders and Kandrot~\cite{cudabyexample}.

\nomenclature{$(b_x, b_y)$}{Block size configuration for a GPU kernel}

To achieve both computational performance and code development efficiency, we treat the computational intensive part of the code and the program flow in different ways.
PyCUDA~\cite{kloeckner_pycuda_2012} is a Python package that exposes the complete CUDA run-time API and allows us to call native GPU kernels written in CUDA directly from Python. 
This way, one can write the program flow, as well as pre- and post-processing of the specific applications, in high-level Python, and at the same time ensure that the computationally expensive simulation loop runs as efficient as possible through low-level CUDA C/C++.
By taking advantage of widely available and popular packages -- including NumPy~\cite{numpyBook} and matplotlib~\cite{matplotlib}, and environments such as the Jupyter Notebook~\cite{jupyternotebook} -- the code and experiments can be developed efficiently through rapid prototyping.

In the remainder of this section we give an overview of the model and the model errors, and show how we utilize the GPU to increase computational efficiency.


\subsection{The simplified ocean model}
\label{sec:simplifiedOceanModel}

The shallow-water equations consider three conserved variables; the elevation $\eta$ of the free ocean surface relative to its equilibrium level, and the volume transport $hu$ and $hv$ along the abscissa and ordinate, respectively.
The equilibrium depth is given by $H_{eq}$ and is here assumed to be constant, so that the full height of the water column becomes $h = H_{eq} + \eta$.
With gravitational acceleration $g$ and Coriolis parameter $f$, the shallow-water equations can be written
\begin{equation}
	\begin{split}\textit{}
		(\eta)_t + (hu)_x + (hv)_y &= 0, \\
		(hu)_t + \left(hu^2 + \frac{1}{2} gh^2\right)_x + (huv)_y &= fhv , \\
		(hv)_t + (huv)_x + \left(hv^2 + \frac{1}{2} gh^2 \right)_y &= -fhu.
	\end{split}
	 \label{eq:swe_full}
\end{equation}
The equations represent a hyperbolic conservation law, and can be written in vector form as
\begin{equation}
	\state_t + F(\state)_x + G(\state)_y = S_f (\state),
	\label{eq:hyperbolic_conservation_law}
\end{equation}
for a state vector $\state = [\eta, hu, hv]^T$.
Here, $F$ and $G$ are flux terms along the absicca and ordinate, respectively, and $S_f$ consists of the source terms due to the Coriolis forces.
\nomenclature{$\eta$}{Ocean surface deviation from equilibrium (dependent variable)}
\nomenclature{$hu$}{Volume transport along the absicca (dependent variable)}
\nomenclature{$hv$}{Volume transport along the ordinate (dependent variable)}
\nomenclature{$H_{eq}$}{Equilibrium water depth (parameter)}
\nomenclature{$h$}{Water depth $h = H_{eq} + \eta$ (dependent variable)}
\nomenclature{$g$}{Gravitational constant (parameter)}
\nomenclature{$f$}{Coriolis parameter}
\nomenclature{$x$}{independent position variable along the absicca}
\nomenclature{$y$}{independent position variable along the ordinate}
\nomenclature{$t$}{independent time variable}
\nomenclature{$F(\state)$}{Flux function along the absicca}
\nomenclature{$G(\state)$}{Flux function along the ordinate}
\nomenclature{$S_f(\state)$}{Source term accounting for the Coriolis forces}
\nomenclature{$[\eta, hu, hv]^T$}{Ocean model state}

The model operator $M(\state)$ will be the numerical scheme that solves \eref{swe_full} and evolves the state forward in time.
We use the high-resolution central-upwind scheme proposed by Chertock et al.~\cite{Chertock2017}, but with a reformulation that avoids the expensive recursive formulation of Coriolis potential terms~\cite{cdklmPracticalities}.
The scheme is designed to be well-balanced with respect to the geostrophic balance, 
\begin{equation}
	hu = - \frac{g H_{eq}}{f} \frac{\partial \eta}{\partial y} \quad \mathrm{and} \quad 
	hv =   \frac{g H_{eq}}{f} \frac{\partial \eta}{\partial x},
	\label{eq:geostrophic_balance_cont}
\end{equation}
which permits rotating steady-state solutions by balancing the gravitational and Coriolis forces.
The numerical scheme is solved on a Cartesian grid $\Omega^M$ consisting of $N_M = n_x \times n_y$ cells.
The size of each cell is $\dx \times \dy$, so that the cell with index $(j,k)$, containing the value $\state_{j,k}$, is the cell centered at 
\begin{equation}
	(x_j, y_k) = \left(\left(j+\tfrac{1}{2}\right)\dx, \left(k + \tfrac{1}{2}\right)\dy\right).
	\label{eq:cell_center}
\end{equation}
The total size of the state vector $\state$ then becomes $N_{\state} = 3 N_M$.
The time integration is solved by a second-order strong-stability-preserving Runge-Kutta method, \implementation{and the storage requirement for the scheme is therefore $2 N_{\state}$, as the full state must be stored for two consecutive time steps.}

The step size of the numerical scheme is limited by the CFL condition,
\begin{equation}
    \dt_{scheme} \leq \frac{1}{4} \min \left\{ \frac{\dx}{\max_{\Omega^M} \left| u \pm \sqrt{g(H_{eq} + \eta)} \right|}, 
                                      \frac{\dy}{\max_{\Omega^M} \left| v \pm \sqrt{g(H_{eq} + \eta)} \right|}  \right\},
    \label{eq:swecfl}
\end{equation}
in which the dominating term is the speed of gravitational waves, $\sqrt{g(H_{eq}+\eta)}$.
Even though such waves occur in the ocean, perhaps most notable through tides, their contribution to drifter motion is limited.
Eddies and other rotation-driven dynamics are much more important, but they operate on longer timescales.
Nevertheless, the CFL-condition in \eref{swecfl} must be satisfied to ensure numerical stability. 
To run the data-assimilation model on a relevant time scale, we decouple the model operator $M$ from the time step of the numerical scheme, and let the fixed model time step $\dt$ consist of as many $\dt_{scheme}$ steps as necessary.
We evaluate the condition in \eref{swecfl} continuously to adapt $\dt_{scheme}$ to the most recent model state, using a Courant number of 0.8.

\nomenclature{$\Omega^M$}{Cartesian grid for the discretization of the model}
\nomenclature{$N_M$}{Total number of grid cells in $\Omega^M$}
\nomenclature{$(n_x \times n_y)$}{Number of grid cells in $\Omega^M$ in each direction}
\nomenclature{$(\dx, \dy)$}{Size of a grid cell in $\Omega^M$}
\nomenclature{$\Omega_{j,k}^M$}{Grid cell with index $(j,k)$ in $\Omega^M$}
\nomenclature{$\state_{j,k}$}{Ocean state in cell $\Omega_{j,k}^M$}
\nomenclature{$t^n$}{Simulation time number $n$.}
\nomenclature{$N_{\state}$}{The size of the state vector $\state$.}
\nomenclature{$\dt_{scheme}$}{Time step of the numerical scheme.}
\nomenclature{$\dt$}{Size of model time step $M: \state(t^n) \rightarrow \state(t^n + \dt)$}

\subsection{Small scale model errors}
\label{sec:modelerrors}
To account for errors in our model (e.g., missing physics), we introduce small-scale perturbations through the stochastic variable, $\modelerror = [\delta \eta, \delta hu, \delta hv]^T$, so that $\modelerror \sim N(0,Q)$.
This model error is generated by sampling a random vector $\xi \sim N(0,I)$ and applying a covariance operator,
\begin{equation}
	\modelerror = Q^{1/2} \xi.
	\label{eq:model_error}
\end{equation}
This error is added to the model state after each model time step $\dt$.
We design the covariance operator based on two requirements.
First, since we aim to implement all components in the data-assimilation system to run efficiently on massively parallel architectures, we design the covariance operator $Q^{1/2}$ in terms of local operations. 
Second, it is important that the stochastic model error does not introduce discontinuities or non-physical model states to the solution. 
\nomenclature{$[\delta \eta, \delta hu, \delta hv]$}{Perturbation of a ocean state $[\eta, hu, hv]$}
\nomenclature{$Q$}{Covariance matrix for the stochastic model perturbation}
\nomenclature{$N(0,Q)$}{Multivariate normal distribution with mean zero and covariance $Q$}
\nomenclature{$\xi$}{Sample from the multivariate standard distribution, $\xi \sim N(0,I)$}

To make the perturbation of the ocean surface $\delta \eta$ sufficiently smooth, it is generated according to a second-order auto-regressive (SOAR) function given by 
\begin{equation}
	\delta \eta_{j,k} = \sum_{a=1}^{n_x} \sum_{b=1}^{n_y} Q_\SOAR^{1/2}\left(\Omega_{j,k}, \Omega_{a,b}\right) \xi_{a,b},
	\label{eq:global_soar_on_eta}
\end{equation}
in which
\begin{equation}
	Q_\SOAR^{1/2} (\Omega_{j,k}, \Omega_{a,b}) = q_0 \left(1 + \frac{\mathrm{dist}(\Omega_{j,k}, \Omega_{a,b})}{L_0}\right) \exp \left[-\frac{\mathrm{dist}(\Omega_{j,k}, \Omega_{a,b})}{L_0}\right].
	\label{eq:soar_function}
\end{equation}
Here, $q_0$ is a scaling parameter for the amplitude of $\delta \eta$, $L_0$ is a measure of the correlation length scale, and $\mathrm{dist}(\Omega_{j,k}, \Omega_{a,b})$ is the euclidean distance between the center of the cells with indices $(j, k)$ and $(a,b)$.
Since the covariance between points that are far from each other relative to $L_0$ becomes zero, the computational work can be limited to operate on local data points only, and this satisfies the first design requirement.
Equation \eref{global_soar_on_eta} can then be written as
\begin{equation}
	\delta \eta_{j,k} = \sum_{a=j-c_\SOAR}^{j+c_\SOAR} \sum_{b=k-c_\SOAR}^{k+c_\SOAR} Q_\SOAR^{1/2}\left(\Omega_{j,k}, \Omega_{a,b}\right) \xi_{a,b},
	\label{eq:local_soar_on_eta}
\end{equation}
in which $c_\SOAR$ is our cut-off value, tuned so that there are no contribution to $\delta \eta_{j,k}$ from a distance larger than $c_\SOAR \min(\dx, \dy)$ from cell $\Omega_{j,k}$.
Operations such as \eref{local_soar_on_eta} are very well suited for implementation on the GPU.
\nomenclature{$Q^{1/2}_\SOAR$}{SOAR covariance operator}
\nomenclature{$Q^{1/2}_\SOAR(\Omega_{j,k}, \Omega_{a,b})$}{The second-order auto-regressive function between two grid cells.}
\nomenclature{$q_0$}{Amplitude parameter for the SOAR function}
\nomenclature{$L_0$}{Length scale parameter for the SOAR function}
\nomenclature{$c_\SOAR$}{Cut-off parameter for the SOAR function}

\begin{figure}[t!]
    \begin{center}
	\scalebox{0.6}{
\begin{tikzpicture}
[nonterminal/.style={
        rectangle,
        minimum size=6mm,
        draw=blue!50!black!50,
        top color=white,
        bottom color=blue!50!black!20
    },
    stage_style/.style={
        rectangle,
        very thick,
        draw=blue!50!black!30,
        top color=white,
        bottom color=white,
        minimum width=400,
        rounded corners=0.2cm,
        align = left,
        top color=blue!50!black!5,
        bottom color=blue!50!black!5
    },
    step_style/.style={
        rectangle,
        very thick,
        draw=blue!50!black!30,
        top color=yellow!50!black!3, 
        bottom color=yellow!50!black!3, 
        rounded corners=0.2cm,
        align = left
    },
    da_style/.style={
        stage_style,
        top color=red!50!black!5,
        bottom color=red!50!black!5
    },
    io_style/.style={
        stage_style,
        minimum width=60, 
        minimum height=30
    },
    math_step/.style={
       minimum width=380,
       text width=360,
       align=left,
    },
    sync_style/.style={
        stage_style,
        minimum width=500
    },
    sync_step/.style={
        math_step,
        minimum width=470,
        text width = 450
    },
    fine_grid_line/.style={
        draw=black,
        fill=black!20,
        thick,
        rounded corners=0.2cm
    },
    coarse_grid_line/.style={
        very thick,
        red!80,
        dashed
    },
    fine_cell_center/.style={
    	circle,
    	draw=black,
        fill=black!40,
        minimum size=7pt,
    	inner sep=0pt
    },
    coarse_grid_point/.style={
    	circle, 
    	fill=red!20,
    	draw=red!80, 
    	minimum size=17pt, inner sep=0pt
    },
    point/.style={coordinate},>=stealth',thick,draw=black,
    tip/.style={->,shorten >=1pt},every join/.style={rounded corners},
    hv path/.style={to path={-| (\tikztotarget)}},
    vh path/.style={to path={|- (\tikztotarget)}},
    fancytitle/.style={
        fill = white,
        text = black,
        very thick, 
        draw=blue!50!black!30,
        rounded corners=0.1cm
    }
]

\foreach \coord in {0,...,9} {
	\draw[fine_grid_line] (\coord, 3) -- (\coord, 9);
}
\foreach \coord in {3,...,9} {
	\draw[fine_grid_line] (0, \coord) -- (9, \coord);
}

\begin{pgfonlayer}{foreground}
	\foreach \x in {0.5,...,8.5} {
		\foreach \y in {3.5,...,8.5} {
			\node[fine_cell_center] at (\x, \y) {};
		}
	}
\end{pgfonlayer}

\begin{pgfonlayer}{main}
	\foreach \x in {1.5, 4.5, 7.5} {
		\foreach \y in {4.5, 7.5} {
			\node[coarse_grid_point] at (\x, \y) {};
		}
	}
\end{pgfonlayer}

\begin{pgfonlayer}{background}
	\foreach \coord in {1.5, 4.5, 7.5} {
		\draw[coarse_grid_line] (\coord, 3) -- (\coord, 9);
	}
	\foreach \coord in {4.5, 7.5} {
		\draw[coarse_grid_line] (0, \coord) -- (9, \coord);
	}
\end{pgfonlayer}

\draw[fine_grid_line] (10, 6.5) -- (10, 7.5);
\draw[fine_grid_line] (11, 6.5) -- (11, 7.5);
\draw[fine_grid_line] (10, 6.5) -- (11, 6.5);
\draw[fine_grid_line] (11, 7.5) -- (10, 7.5);
\node[fine_cell_center] at (10.5, 7){};
\node at (11.2, 7)[anchor=west, align=left] {\Large{Cell with cell center in $\Omega^M$}};

\begin{pgfonlayer}{background}
	\draw[coarse_grid_line] (10.5, 4.5) -- (10.5, 5.5);
	\draw[coarse_grid_line] (10, 5) -- (11, 5);
\end{pgfonlayer}
\node[coarse_grid_point] at (10.5, 5){};
\node[anchor=west, align=left] at (11.2, 5) {\Large{Grid point in $\Omega^R$}};

\end{tikzpicture}
}
    	\caption{Alignment of nested grids with $c_{\Omega} = 3$. The grid $\Omega^M$ contains cells and is used for evolving the numerical model, whereas the grid $\Omega^R$ contains point values and is used for applying the SOAR function on sampled random numbers from $N(0, I)$. For best possible assimilation of observations, an offset can be applied to $\Omega^R$ so that one of its grid points is co-located with the cell in $\Omega^M$ in which the observation was made.}
    	\label{fig:nestedGrids}
    \end{center}
\end{figure}
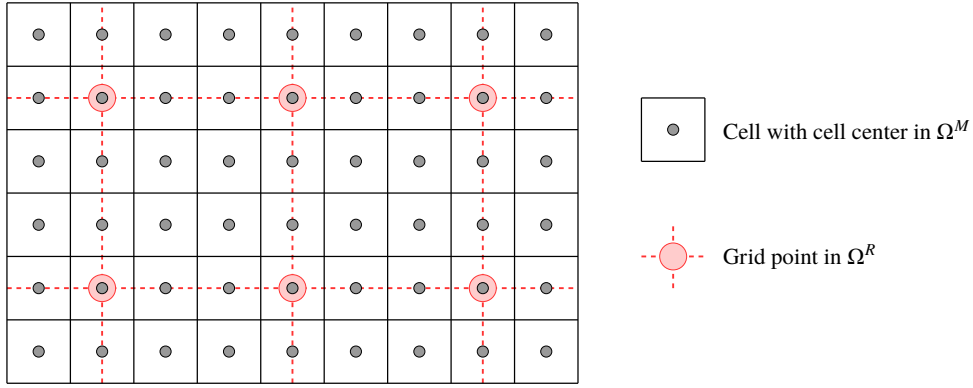

A drawback to the expression in \eref{local_soar_on_eta} is that the computational work and data dependency of the stencil is tightly connected to the ratio between $L_0$ and the cell size. 
To have better control of this workload, we introduce a coarse \emph{random number grid} $\Omega^{R}$, on which the standard normal distributed random numbers $\xi$ are sampled, and apply the SOAR function here.
We choose the discretization of $\Omega^R$ so that we obtain a good trade-off between computational efficiency of \eref{local_soar_on_eta}, while maintaining a good spread of information within the correlated areas.
The coarse grid will have grid cells of size $(\coarsedx, \coarsedy) = c_{\Omega}(\dx, \dy)$, where $c_{\Omega}$ is an odd number representing the coarseness of $\Omega^R$.
Values on $\Omega^R$ are interpreted as point values, and we denote the number of grid points in $\Omega^R$ by $N_R$.
By requiring that $c_{\Omega}$ is odd, we ensure that the point values defined on $\Omega^{R}$ are co-located with cell centers of $\Omega^M$, as show in \reffig{nestedGrids}.
Furthermore, we choose the coarsening factor $c_{\Omega}$ so that the cut-off factor in \eref{local_soar_on_eta} can be chosen as $c_\SOAR = 2$.
After having obtained $\delta \eta$ on $\Omega^{R}$ through \eref{local_soar_on_eta}, we use bicubic interpolation, denoted by the operator $\interpolationoperator$, to obtain cell-averaged values on $\Omega^M$.
\nomenclature{$\Omega^{R}$}{Discrete grid on which random numbers are defined}
\nomenclature{$c_{\Omega}$}{Coarsening factor between the two discretized grids $\Omega^M$ and $\Omega^R$.}
\nomenclature{$(\coarsedx, \coarsedy)$}{Size of a grid cells in $\Omega^R$.}
\nomenclature{$N_R$}{Number of grid cells in $\Omega^R$.}


To avoid that the perturbation $\modelerror$ produces non-physical model states (the second design requirement), we use \eref{geostrophic_balance_cont} to ensure that $\modelerror$ is in geostrophic balance.
By discretizing \eref{geostrophic_balance_cont} with central differences on the $\Omega^M$ grid, $\delta hu$ and $\delta hv$ are found from $\delta\eta$ by
\begin{equation}
	\delta hu_{j,k}= - \frac{g H_{eq}}{f} \frac{\delta \eta_{j,k+1} - \delta \eta_{j,k-1}}{2 \dy} \quad \mathrm{and} \quad 
	\delta hv_{j,k} =   \frac{g H_{eq}}{f} \frac{\delta \eta_{j+1,k} - \delta \eta_{j-1,k}}{2 \dx}.
	\label{eq:geostrophic_balance_disc}
\end{equation}
This operation is denoted by $Q_{GB}^{1/2}$. 
It should be noted that the derivatives of $\delta\eta$ are approximated by \eref{geostrophic_balance_disc}, even though they are analytically available directly from the bicubic interpolation.
The reason is that geostrophic balance is only maintained by the numerical scheme with respect to the grid resolution.
The bicubic surface, however, is continuously defined and will typically contain oscillations on sub-grid scale, meaning that the derivatives of the bicubic surface often will not be represented by the discrete values on the grid.
The central differences in \eref{geostrophic_balance_disc} are therefore better suited for generating an model state that is in balance under the numerical scheme.
\nomenclature{$\interpolationoperator$}{Interpolation operator generating values on $\Omega^M$ based on values on $\Omega^R$.}

\begin{figure*}[t!]
    \centering
    
    \begin{subfigure}[t]{0.3\textwidth}
        \centering
        \includegraphics[width=\textwidth]{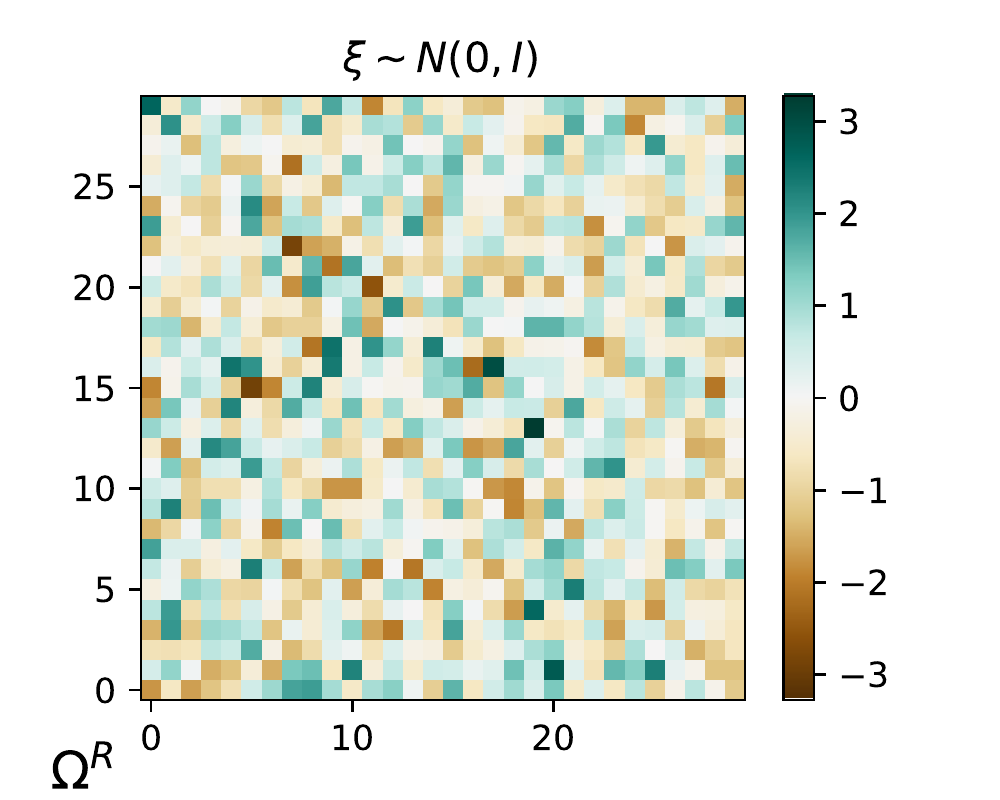}
        \caption{Independent random numbers}
        \label{fig:modelerror:xi}
    \end{subfigure}%
	~
	\begin{subfigure}[t]{0.3\textwidth}
        \centering
        \includegraphics[width=\textwidth]{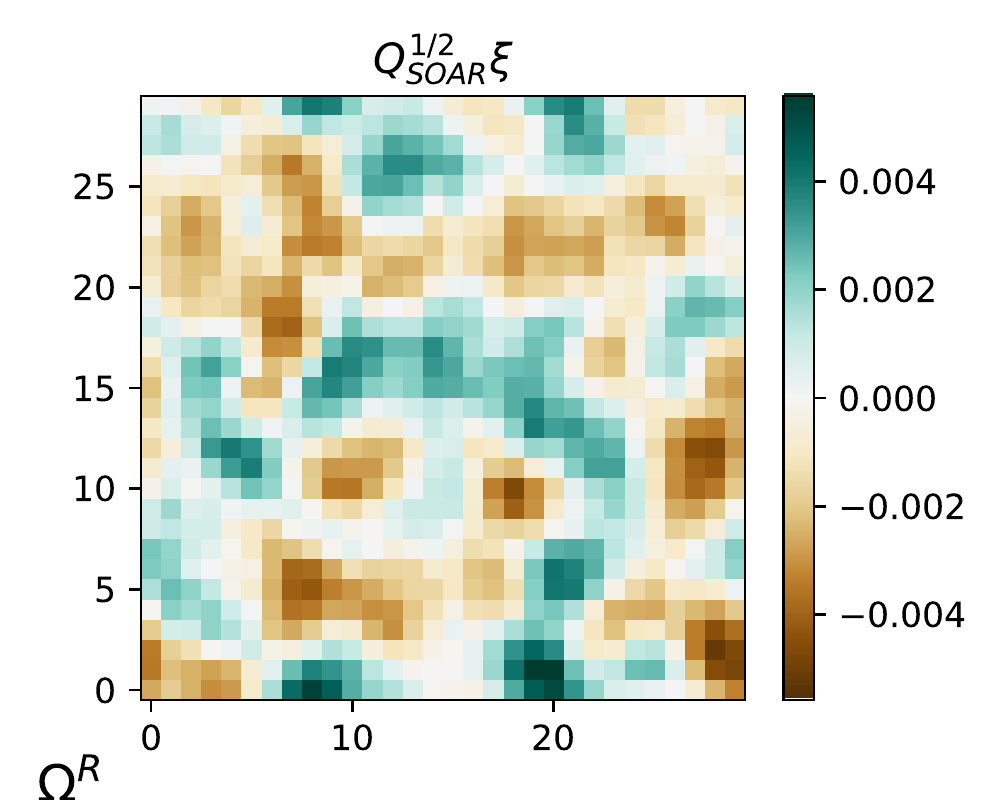}
        \caption{Random coarse field with covariance $Q_\SOAR$}
        \label{fig:modelerror:soar}
    \end{subfigure}

    \begin{subfigure}[t]{0.9\textwidth}
        \centering
        \includegraphics[width=\textwidth]{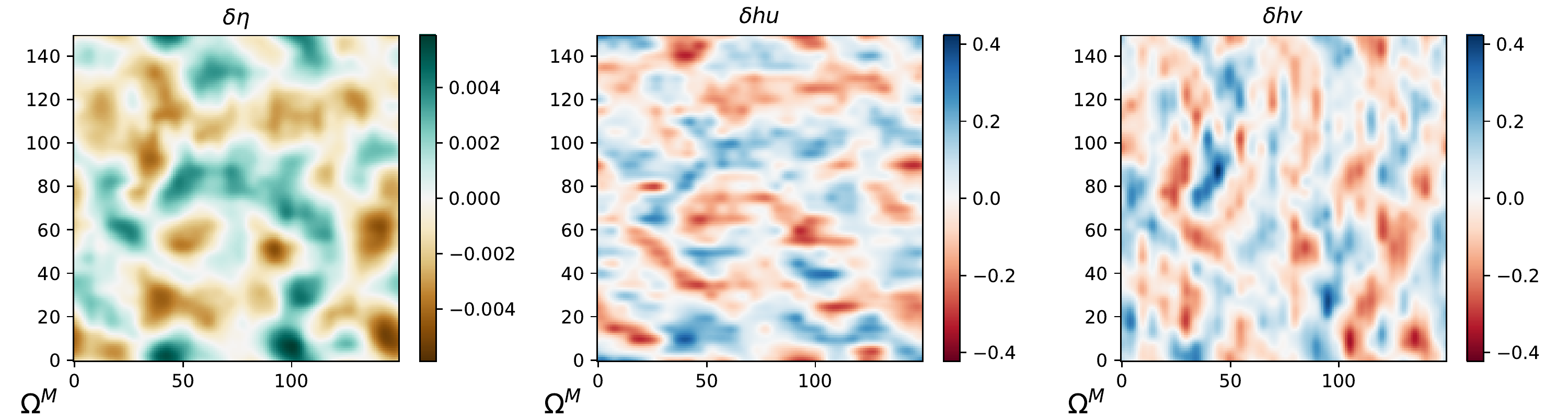}
        \caption{Model perturbation in geostrophic balance.}
        \label{fig:modelerror:modelerror}
    \end{subfigure}
    \caption{The small scale model perturbation $\modelerror = [\delta \eta, \delta hu, \delta hv]^T$ is generated by (a) sampling random numbers from a standard normal distribution $\xi \sim N(0,I)$ on the coarse grid $\Omega^R$; (b) giving the random field the covariance according to the SOAR function; and (c) interpolating the coarse random field onto the fine model grid $\Omega^M$ and calculating the momentum to impose geostrophic balance.}
    \label{fig:modelerror}
\end{figure*}

Evaluating the complete model error now consists of four operations, 
\begin{equation}
	\modelerror = Q^{1/2} \xi =  Q_{GB}^{1/2} \interpolationoperator  Q_\SOAR^{1/2} \xi,
\end{equation}
in which the first step is to sample $\xi \sim N(0, I)$.
Note that $Q^{1/2}_{GB}$ and $Q^{1/2}_\SOAR$ are linear operators, whereas $\interpolationoperator$ is a nonlinear stencil. 
The input and output for each of the operations are
\begin{equation}
	\begin{split}
		Q_\SOAR^{1/2}&: \Omega^{R} \rightarrow \Omega^{R}, \\
		\interpolationoperator&: \Omega^{R} \rightarrow \Omega^M, \\
		Q_{GB}^{1/2} &: \Omega^M \rightarrow 3 \times \Omega^M,
	\end{split}\textit{}
    \label{eq:covarianceOperatorsDefSub}
\end{equation}
making the covariance operator act as 
\begin{equation}
	Q^{1/2} : \Omega^{R} \rightarrow 3 \times \Omega^M.
\label{eq:covarianceOperatorsDef}
\end{equation}
These operations are illustrated in \reffig{modelerror}.
First, the random field $\xi$ is sampled on the coarse grid $\Omega^R$ in \reffig{modelerror:xi}, and the SOAR operator $Q_\SOAR^{1/2}$ is applied to generate a coarse correlated field in \reffig{modelerror:soar}.
Then, the correlated field is interpolated onto the computational grid $\Omega^M$, and $\delta hu$ and $\delta hv$ are computed to be in geostrophic balance to $\delta \eta$ in \reffig{modelerror:modelerror}.


It should be noted that our choice of $Q$ leads to a non-symmetric square root $Q^{1/2}$, and that this implementation-oriented definition of $Q^{1/2}$ makes use of significantly less random numbers than variables in the state vector.
To justify why $Q$ is a covariance matrix, we can imagine that all $3 \times \Omega^M$ variables have a corresponding sampled random number, but all those that are not involved in $Q^{1/2}_\SOAR$ are given very small variance and no correlation to any other variables, so that they become negligible in the above computations.

\subsection{Efficient   of model errors}
The SOAR function, bicubic interpolation, and geostrophic balance are all local stencil operations that are simple to parallelize, as each element of their output can be found independently from all other output elements.
Generation of random numbers $\xi$ can further be done through the cuRAND library available through the CUDA toolkit.
The sampling of $\modelerror$ is therefore well-suited for implementation on the GPU.


The SOAR function in \eref{local_soar_on_eta} with $c_\SOAR = 2$ consists of a stencil operation depending on $5 \times 5$ input values centered on the target cell.
We use one GPU thread per output element.
To minimize the amount of data read from global memory, all threads within the same block cooperate to read the collectively required input data into shared memory.

In the bicubic interpolation $\interpolationoperator$, each value in the fine grid $\Omega^M$ depends on the $4 \times 4$ points in the coarse grid $\Omega^R$ that surrounds its position.
This means that the $c_{\Omega} \times c_{\Omega}$ output values that are located between the same four coarse grid points have overlapping data dependencies.
We still apply one GPU thread per output element, and obtain geostrophically balanced $\delta hu$ and $\delta hv$ within the same kernel.
Each block computes $(b_x + 2) \times (b_y + 2)$ values of $\delta \eta$ and stores them temporarily in shared memory, so that $b_x \times b_y$ values of $\delta hu$ and $\delta hv$ efficiently can be computed using \eref{geostrophic_balance_disc}.

The memory footprint of obtaining $\modelerror$ is two buffers of size $N_R$, holding $\xi$ and the result from $Q_\SOAR^{1/2} \xi$, respectively.
The memory footprint of the random number generator comes in addition to this.
Note that we never store $\modelerror$ itself, but add it directly into the state vector $\state$.

\subsection{Synthetic truth and observations}
\label{sec:truthAndObservations}
The experiments in this paper are so-called identical twin experiments, meaning that the same model equations are used to generate the synthetic true state and to evolve the ensemble.
The true state $\state_{true}$ is generated from a known set of initial conditions by running the numerical scheme with stochastic model errors as described above.
Furthermore, $N_D$ Lagrangian drifters (drifting buoys) are simulated to be advected passively along the ocean current according to a simple forward Euler integration scheme.
\nomenclature{$N_D$}{Number of drifters.}

Our data-assimilation experiments make use of two types of observations.
The first type is based on ``GPS'' tracking of drifters, and we denote the position of drifter $d$ at observation time $t_m$ by $(x_d^m, y_d^m)$\footnote{Note that observations might not be available for each model time step, which is the reason for the use of  subscript $m$ to distinguish observation time step $t^m$ from from model time step $t^n$.}.
Since the location of a drifter at a single point in time gives no information about the underlying ocean currents, we use the difference in two subsequent drifter locations to estimate the drifter velocity, which in our model represents the current.
The observation $\observation_d^n$ then becomes
\begin{equation}
	\observation_d^m = \left[ \frac{x_d^m - x_d^{m-1}}{t_m - t_{m-1}} H_{eq}, 
    							  \frac{y_d^m - y_d^{m-1}}{t_m - t_{m-1}} H_{eq} \right] + \observationerror_d^m,
	\label{eq:drifterObservation}
\end{equation}
in which $\observationerror_d^m \sim N(0, R)$ is the observation error.
Note that the observation is chosen to be an estimate of the state variables $hu$ and $hv$, but where we have ignored the contribution of the unobserved sea-surface level $\eta$. 
This simplifies the observation operator $H$ to be the state values in the cell corresponding to the drifter position. 
If drifter $d$ is observed at location $(x_d^m, y_d^m)$, and this is a point within cell $\Omega^M_{j,k}$, the observation operator applied to a particle state $\state_i^m$ becomes
\begin{equation}
	H\left(\state_i^m, (x_d^m, y_d^m)\right) = \left[(hu_i^m)_{j,k}, (hv_i^m)_{j,k}\right]^T.
	\label{eq:stateObservation}
\end{equation}
The size of the observation vector becomes $N_y = 2N_D$.
\nomenclature{$(x_d^m, y_d^m)$}{Observed location of drifter $d$ at time $t_m$.}
\nomenclature{$t_m$}{Observation time number $m$}

One challenge with the above observation is the unobserved value of the sea-surface level $\eta$, as it in general is not negligible compared to $H_{eq}$, and therefore introduces a bias in \eref{drifterObservation}. 
To compensate for this, we use the best available estimate for $\eta$, namely the simulated $\eta$ for each individual particle, and define the innovation related to drifter $d$ for particle $i$ as
\begin{equation}
    \innovation_{i,d}^m = \observation_d^m \frac{H_{eq} + (\eta_i^m)_{j,k}}{H_{eq}} -  H\left(\state_i^m, (x_d^m, y_d^m)\right).
    \label{eq:etaCompensatedInnovation}
\end{equation}

The second observation type is observations from moored buoys, referred to simply as moorings in the remainder of the paper, that give Eulerian point measures of the current throughout the entire simulation.
To be consistent with \eref{stateObservation} and \eref{etaCompensatedInnovation}, the mooring observations are provided in terms of $hu$ and $hv$, but ignoring the contribution from $\eta$.
The observation from mooring $\mu$, located at $(x_{\mu}, y_{\mu})$ in cell $\Omega^M_{j,k}$, is therefore defined as 
\begin{equation}
    \observation_{\mu}^m = \left[ 
                        (hu_{true}^m)_{j,k} \frac{H_{eq}}{H_{eq} + (\eta_{true}^m)_{j,k}}, 
                        (hv_{true}^m)_{j,k} \frac{H_{eq}}{H_{eq} + (\eta_{true}^m)_{j,k}}
                        \right] + \observationerror_{\mu}^m.
    \label{eq:buoyObservation}
\end{equation}
As for the drifter observations, the size of the mooring observation vector becomes $N_y = 2N_{\mu}$, for $N_{\mu}$ moorings.

\subsection{Adjoint of the model error operators}
\label{sec:adjoint}

Whereas the model error term depends on $Q^{1/2}$ only, the IEWPF algorithm requires that we apply the full $Q$ operator, e.g., in \eref{opd_update}. 
This requires us to express $Q^{1/2, T}$, the adjoint operator for $Q^{1/2} = Q_{GB}^{1/2} \interpolationoperator Q_\SOAR^{1/2}$.
As mentioned in \refsec{modelerrors}, $Q^{1/2}$ is not symmetric for our application. 
The operator $Q^{1/2}_\SOAR$ is linear and symmetric, however, and therefore its own adjoint $Q^{1/2}_\SOAR = Q^{1/2, T}_\SOAR$.
The expression for geostrophic balance is close to linear, and $Q^{1/2, T}_{GB}$ is approximated simply by $H_{eq}+\eta \approx H_{eq}$.
The bicubic interpolation operator, $\interpolationoperator$, however, is nonlinear and its adjoint is therefore challenging to express.
Our solution to this is to approximate $Q^{1/2, T}$ entirely on the coarse grid $\Omega^R$ and define $\coarseningoperator$ to be a coarsening operator.
The approximate adjoint operator for the model errors is then defined as
\begin{equation}
	Q^{1/2, T} \approx Q_\SOAR^{1/2} Q_{GB}^{1/2, T} \coarseningoperator,
	\label{eq:adjointCovarianceOperator}
\end{equation}
with
\begin{equation}
	\begin{split}
	    \coarseningoperator &: 3 \times \Omega^M \rightarrow 3 \times \Omega^R, \\
		Q_{GB}^{1/2, T} &: 3 \times \Omega^R \rightarrow \Omega^R, \\
	\end{split}\textit{}
    \label{eq:adjointCovarianceOperatorsDefSub}
\end{equation}
and $Q_\SOAR^{1/2}$ as in \eref{covarianceOperatorsDefSub}, resulting in
\begin{equation}
		Q^{1/2, T} : 3 \times \Omega^{R} \rightarrow \Omega^R.
    \label{eq:adjointCovarianceOperatorsDef}
\end{equation}
\nomenclature{$Q^{1/2, T}_{GB}$}{Adjoint geostrophic balance operator, operating on the coarse grid.}
\nomenclature{$\coarseningoperator$}{Coarsening/restriction operator mapping values from $\Omega^M$ that can be represented on $\Omega^R$.}

The full $Q$ operator is always applied to the adjoint of the observation operator $H^T$, which maps an observation vector to state space. 
This means that $Q_{GB}^{1/2,T}$ in all practical sense can be considered to operate on $hu$ and $hv$ at a single grid point only.
To preserve the location of these point values, we align the cell containing the observation with a grid point in the coarse grid, by applying an offset on the location of grid points of $\Omega^R$.
The observation values can then be mapped directly from their position in $\Omega^M$ to the corresponding location in $\Omega^R$. 
With an observation $\observation = [\singleobservation_{hu}, \singleobservation_{hv}]^T$ located at grid point $\Omega^{R}_{j,k}$, we apply the adjoint geostrophic balance as
\begin{equation}
	\left(Q^{1/2, T}_{GB} \coarseningoperator H^T \observation \right)_{(l,m)} = 
    \begin{cases}
    	\frac{-g}{H_{eq}f} \frac{1}{2 \coarsedy} \singleobservation_{hu} \quad &\mathrm{if} \; (l,m) = (j,k+1),\\
        \frac{ g}{H_{eq}f} \frac{1}{2 \coarsedy} \singleobservation_{hu} \quad &\mathrm{if} \; (l,m) = (j,k-1),\\
        \frac{ g}{H_{eq}f} \frac{1}{2 \coarsedx} \singleobservation_{hv} \quad &\mathrm{if} \; (l,m) = (j+1,k),\\
        \frac{-g}{H_{eq}f} \frac{1}{2 \coarsedx} \singleobservation_{hv} \quad &\mathrm{if} \; (l,m) = (j-1,k),\\
          0 \qquad \quad \; &\mathrm{otherwise},
    \end{cases}
    \label{eq:adjointGeoBalance}
\end{equation}
for all grid points $\Omega^R_{l,m} \in \Omega^R$. 
\nomenclature{$\singleobservation_{hu}$}{A single element of the observation vector $\observation$. The subscript is an example.}


\section{Efficient implementation of the IEWPF scheme}
\label{sec:iewpf_applied}
The objective of this section is to present how IEWPF can be efficiently implemented for a shallow-water model with additive locally defined model errors, using the GPU as a target architecture. 
The algorithmic nature is not limited to GPUs, and the following approach can be used on other architectures that take advantage of massively parallel operations.
\refsec{iewpf} gave a high-level overview of the method, whereas mathematical details important to the implementation are given in \appref{iewpfDetails}.
This description assumes the use of drifter observations.

In this work, we also rely on using single-precision floating point arithmetic.
This has a potentially huge impact on performance, as some commodity-level GPUs have single- to double precision ratios of up-to 1:32\footnote{Nvidia's GTX series, Maxwell generation GPUs.}.
Hatfield et al.~\cite{hatfield2018} demonstrate how the accuracy of weather forecasts can be improved through reduced-precision data assimilation. They ran a Lorenz '96 ``toy'' atmospheric model and the ensemble square root filter at double-, single-, and half-precision, and measured the performance through mean error statistics and rank histograms. 
By trading reduced precision for increased ensemble size, the authors could reduce the assimilation error and improve forecast accuracy compared to double-precision assimilation.

\begin{figure}
    \begin{center}
    \input{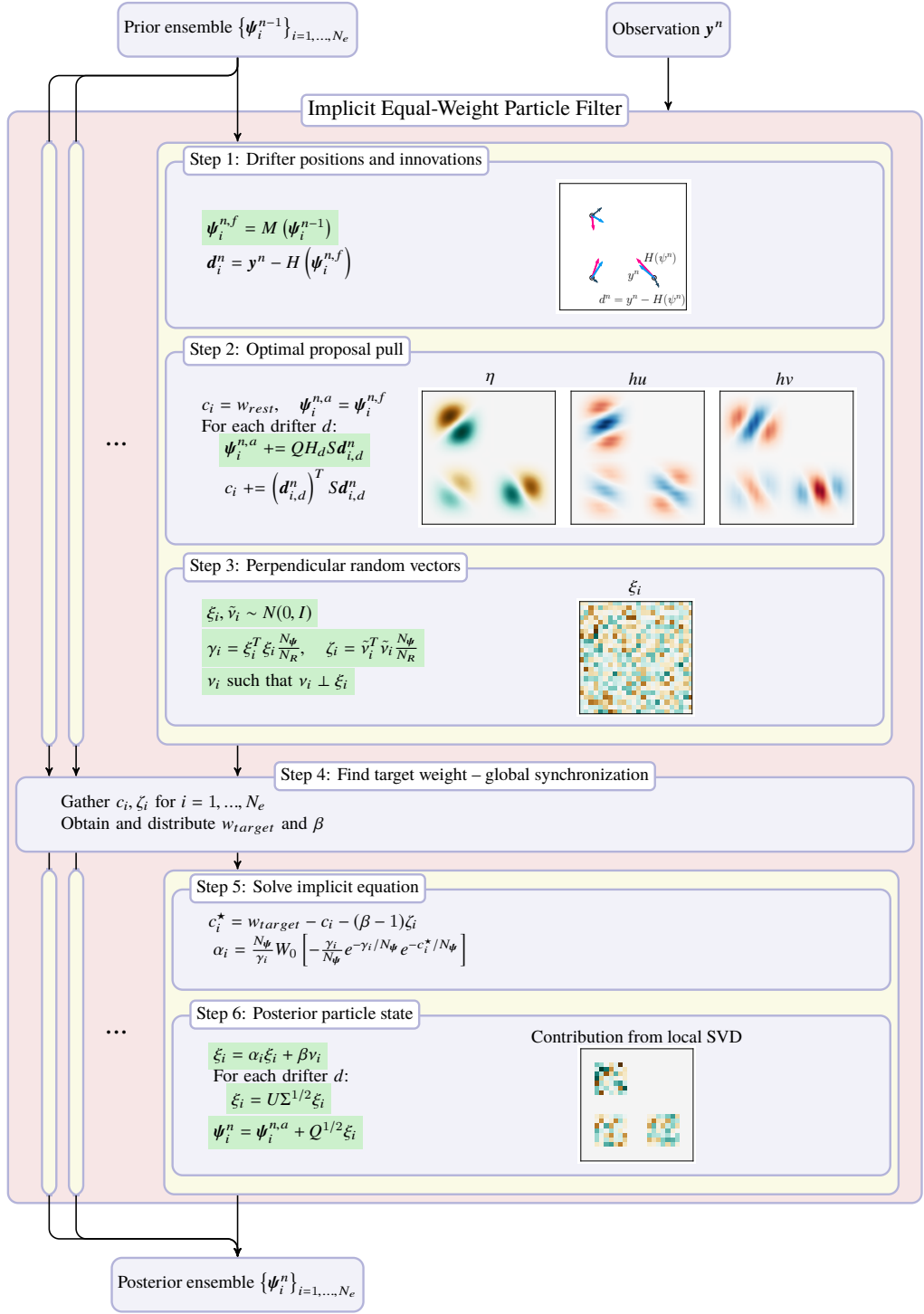}
    \caption{An algorithmic overview of one data-assimilation cycle with the implicit equal-weights particle filter. 
    Operations that consists of massively parallel operations are identified on green background and are implemented on a suitable architecture (such as the GPU).
    During the three first stages, each particle can be handled independently.
    As stage 4 requires the values of $c_i$ and $\zeta_i$ from all particles, this step represents a global synchronization across the entire ensemble. Thereafter, stages 5 and 6 can again be executed independently.}
    \label{fig:iewpfAlgorithm}
    \end{center}
\end{figure}

The starting point of our algorithm is an ensemble of forecast states $\{\state_i^{n-1}\}_{i=1,...,N_e}$ having equal weights at the time step before an observation $\observation^{n}$ is available.
Each particle is then updated through the following pseudo-code:
\begin{enumerate}
    \item Obtain the position of the drifter and find the innovations $\innovation_i^n$.
    \item Pull each particle towards the observation according to the mean of the optimal proposal density \eref{opd_update}. Simultaneously, obtain the value of the $\phi_i$, which is a measure of the innovation and defined in \eref{iewpfPhi}.
    \item Sample $\xi_i, \nu_i \sim N(0,I)$, such that $\xi_i \perp \nu_i$, and find the sizes of the two random vectors. 
    \item Find the parameter $\beta$ and the target weight $w_{target}$.
    \item Solve the implicit equation given by \eref{implicitEquation} for $\alpha_i$ for each particle.
    \item Apply the covariance structures of $P$ to $\xi_i$ and $\nu_i$, and calculate the posterior particle states according to \eref{updateEquationTwoStage}.
\end{enumerate}
\reffig{iewpfAlgorithm} summarizes the algorithm and shows the relevant equations for each step.
The algorithm has only a single synchronization point across all particles at step 4.
Further, equations marked in green identify massively data-parallel operations, for which we can execute efficiently on the GPU. 
The following subsections give details about each of the steps just mentioned. 


\subsection{Observations and innovations}
\label{sec:iewpf_innovations}
The innovation $\innovation_i^n$ is a measure of how well the observed currents $\observation^n$ are represented by each particle state.
To obtain this value for IEWPF, each particle is evolved forward in time to the observation time $t^n$ by the model, $\state_i^{n,f} = M(\state_i^{n-1})$, but without adding the stochastic model error.
The observation also contains the location of each drifter, which is used to look up the relevant parts of the particle state vectors according to \eref{drifterObservation} and \eref{stateObservation}.

\subsection{Optimal proposal particles}
\label{sec:iewpf_optimal_proposal_particles}
Based on the innovations $\innovation_i^n$, each particle state is pulled towards the observation according to the mean of their individual optimal proposal density, given by \eref{opd_update}.
For simplicity, we start by considering a case with a single drifter located in cell $\Omega_{j,k}^M$.
In this case, the matrix $S = (HQH^T + R)^{-1}$ becomes a $2 \times 2$ matrix only and represents a combination of the uncertainty or covariance structure from the model error in observation space and the observation error. 
Since the correlation pattern that makes up the covariance matrix $Q$ for the model error is the same across the entire domain, $HQH^T$ (and thus also $S$) becomes independent of the observed drifter position.
This means that $S$ can be computed and stored once and for all ahead of the assimilation loop. 
For now, we assume $S$ is already available, and look at how the particle states are pulled towards the observation.
Thereafter, we will get back to how $S$ is pre-computed.

\begin{figure*}[t!]
    \centering
    \begin{subfigure}[t]{0.2\textwidth}
        \centering
        \includegraphics[width=\textwidth]{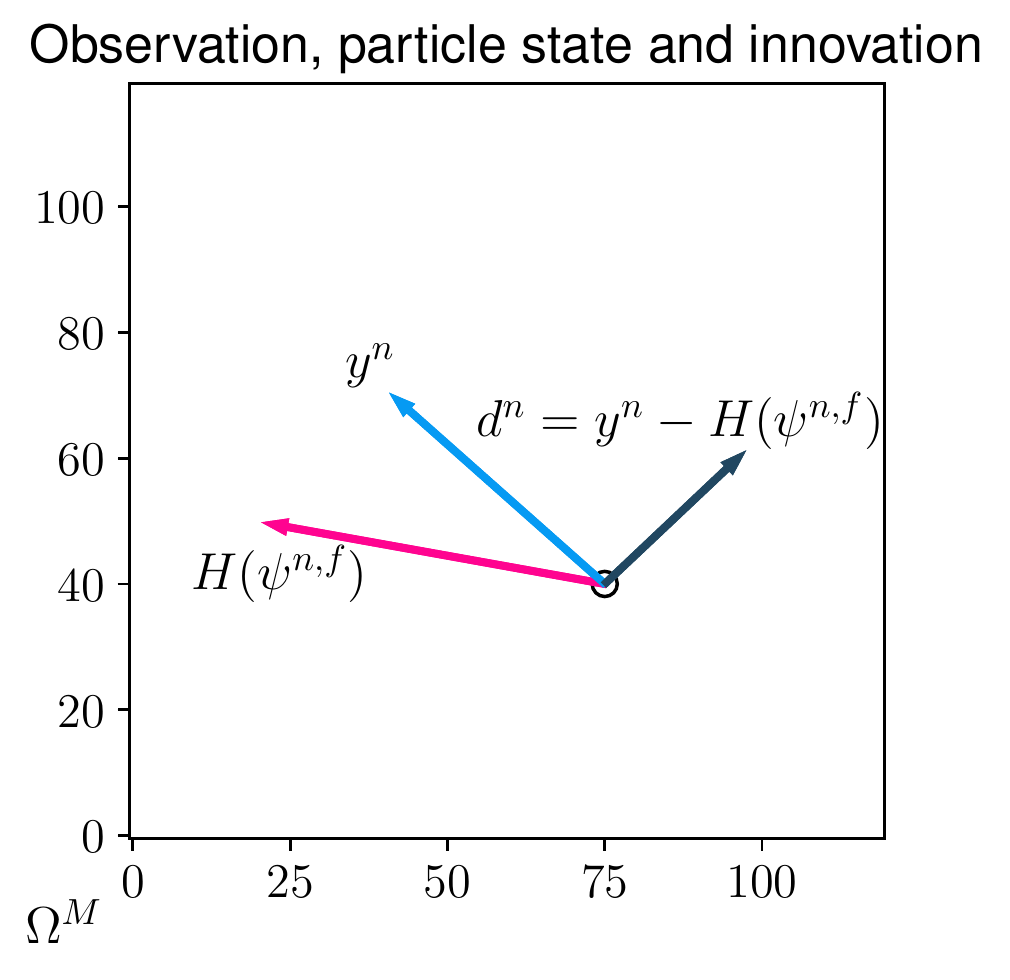}
        \caption{Observation $\observation^n$, observed forecasted particle state $H(\state^{n,f})$, and the innovation $\innovation^n$.}
        \label{fig:optproppull:1}
    \end{subfigure}%
    \quad
	\begin{subfigure}[t]{0.7\textwidth}
        \centering
        \includegraphics[width=\textwidth]{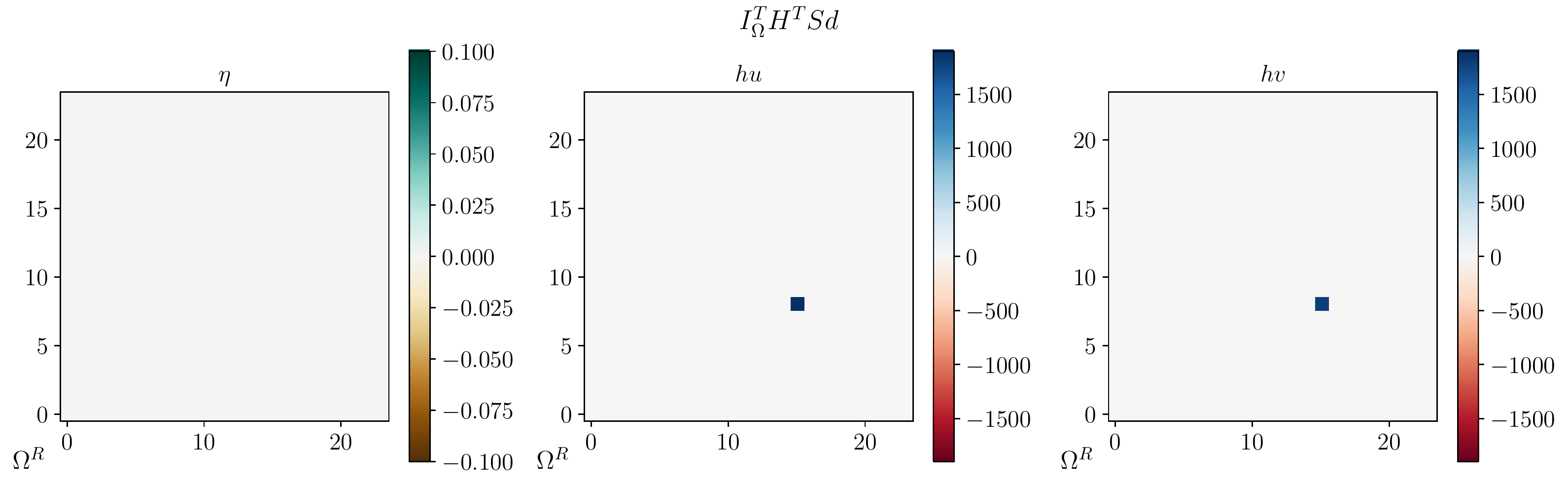}
        \caption{Innovation $\innovation^n$ scaled with the model and observation uncertainty $S$ in the coarse grid state space $\Omega^R$. The coarse grid has been centered onto the observation position.}
        \label{fig:optproppull:2}
    \end{subfigure}    
    
    \begin{subfigure}[t]{0.25\textwidth}
        \centering
        \includegraphics[width=\textwidth]{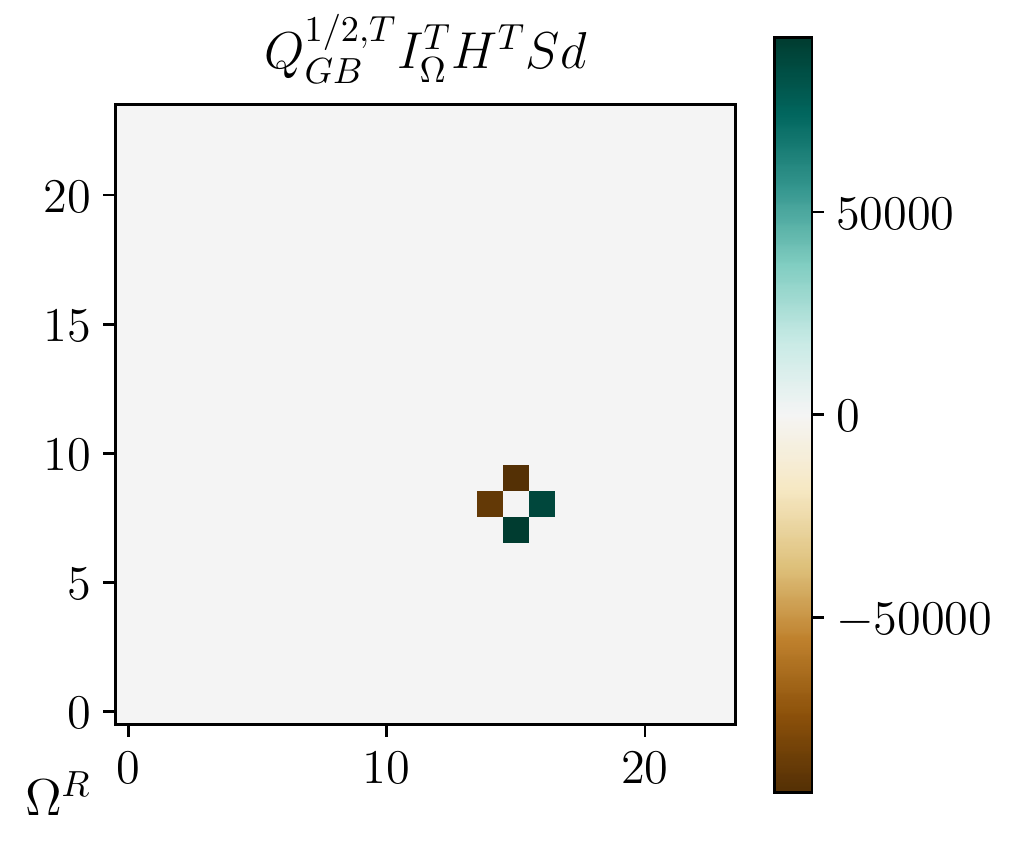}
        \caption{$Q^{1/2,T}_{GB}$ takes the values from $hu$ and $hv$ over to neighbouring grid points in $\eta$.}
        \label{fig:optproppull:3}
    \end{subfigure}
    \quad
    \begin{subfigure}[t]{0.23\textwidth}
        \centering
        \includegraphics[width=\textwidth]{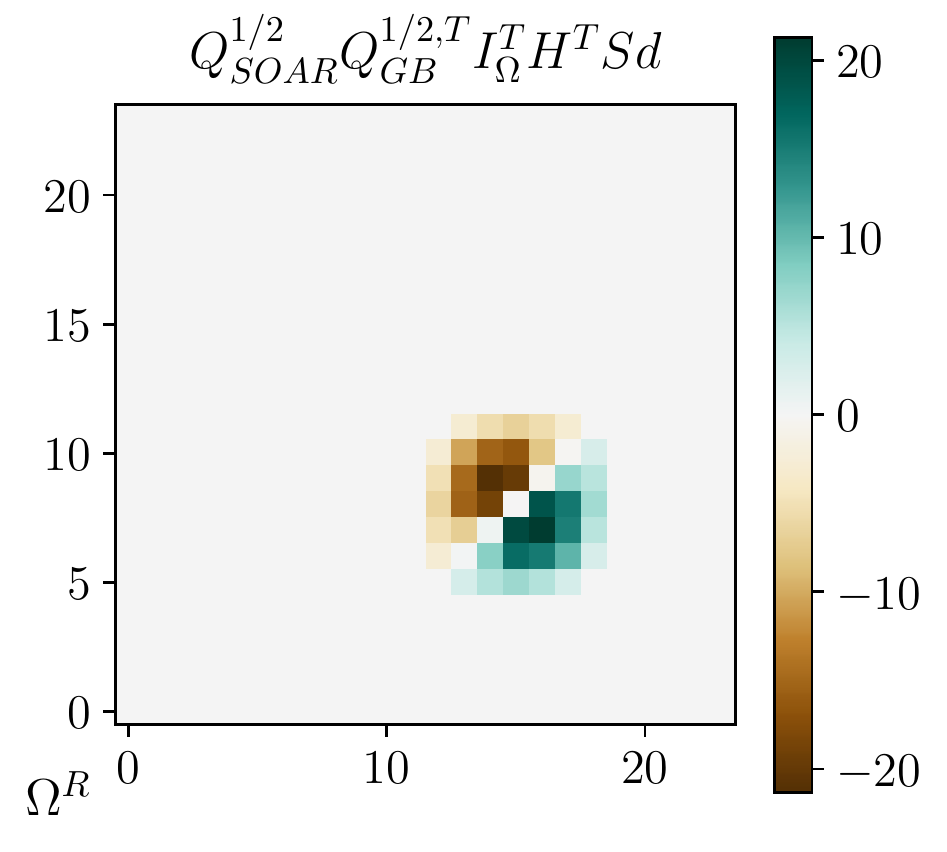}
        \caption{Correlation applied through $Q^{1/2}_\SOAR$.}
        \label{fig:optproppull:4}
    \end{subfigure}
    \quad
    \begin{subfigure}[t]{0.25\textwidth}
        \centering
        \includegraphics[width=\textwidth]{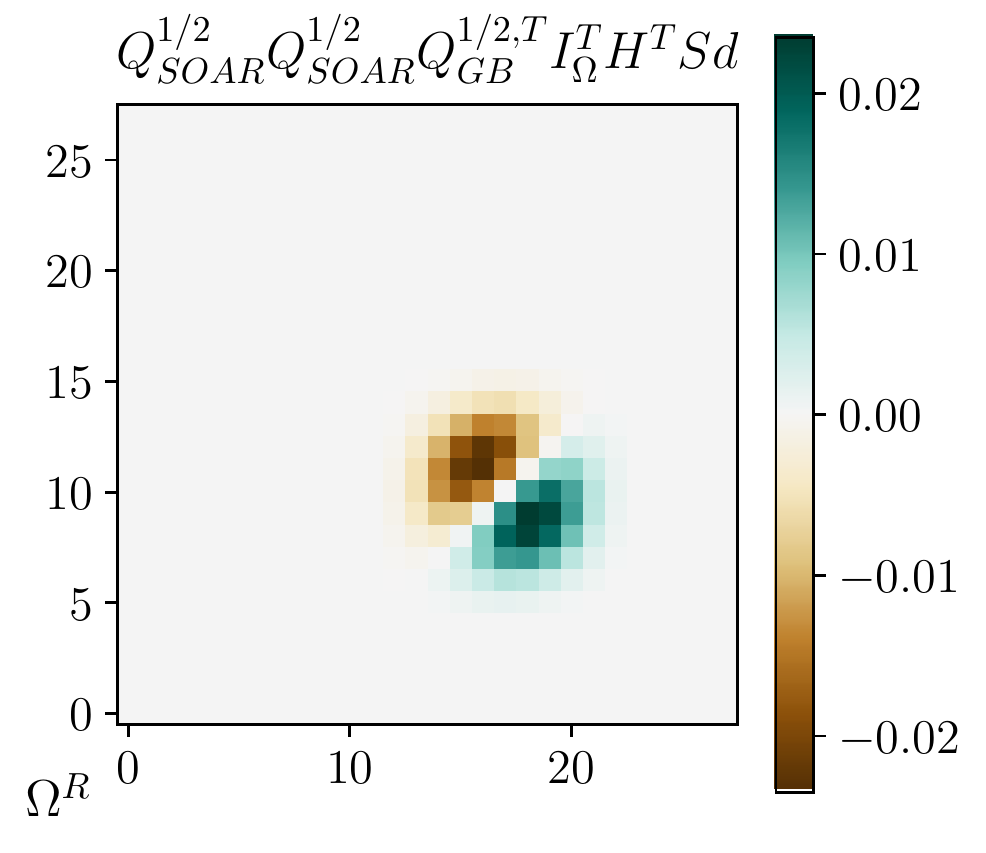}
        \caption{Correlation applied through $Q^{1/2}_\SOAR$ again.}
        \label{fig:optproppull:5}
    \end{subfigure}
    
    \begin{subfigure}[t]{0.7\textwidth}
        \centering
        \includegraphics[width=\textwidth]{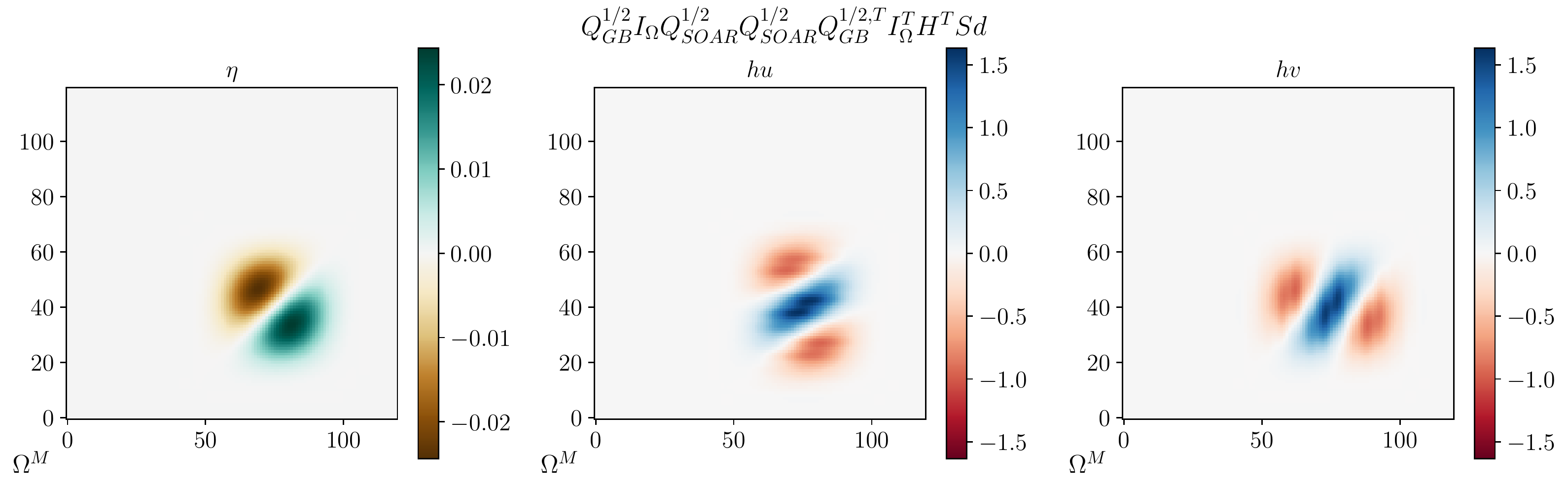}
        \caption{The resulting pull in $\eta$ is obtained by interpolating the previous result with $I_{\Omega}$, and applying $Q^{1/2}_{GB}$ to get the pull for $hu$ and $hv$.}
        \label{fig:optproppull:6}
    \end{subfigure}    
    \quad
    \begin{subfigure}[t]{0.2\textwidth}
        \centering
        \includegraphics[width=\textwidth]{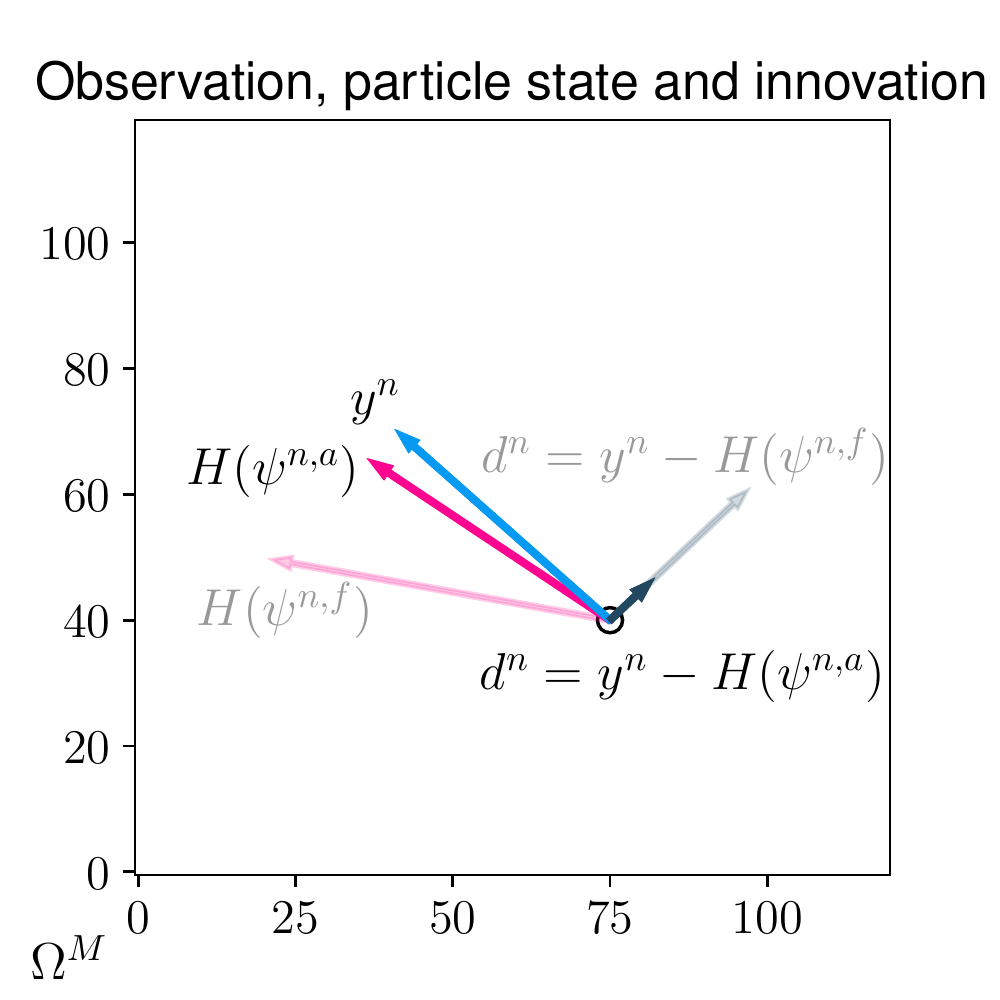}
        \caption{By looking at the optimal proposal state in observation space, we see that the new particle state is much more similar to the observation from before.}
        \label{fig:optproppull:7}
    \end{subfigure}    
    
    \caption{The process of constructing the pull used to obtain the optimal proposal particle state, required in the second step of the IEWPF algorithm. Note that the shown example is exaggerated for illustrative purposes.}
    \label{fig:optimalproposalpull}
\end{figure*}

We start by expanding the expression for the mean of the optimal proposal density in \eref{opd_update} by using $Q = Q^{1/2}Q^{1/2,T}$:
\begin{equation}
    \begin{split}
        \state_i^{n,a} &= M(\state_i^{n-1}) + QH^T(HQH + R)^{-1}\innovation_i^n \\
                       &= \state_i^{n,f} + Q^{1/2}_{GB} \interpolationoperator Q^{1/2}_\SOAR Q_\SOAR^{1/2} Q_{GB}^{1/2,T} \coarseningoperator H^T S \innovation_i^n.
    \end{split}
    \label{eq:detailedOPDUpdate}
\end{equation}
To see how the state of particle $i$ is modified, we go through this expression step-by-step starting from the right.
This process is also illustrated in \reffig{optimalproposalpull}.
\begin{description}
    \item[$S \innovation_i^n$:] The innovation is the difference between observed and modelled current at the location of the drifter (\reffig{optproppull:1}). This measure is scaled by the combined uncertainty from the observation and the model, represented by $S$.
    \item[$\coarseningoperator H^T \gray{S \innovation_i^n}$:] The adjoint observation operator $H^T$ acts on the two-dimensional vector $S\innovation_i^n$ by mapping its two values into state space at the indices representing $hu_{j,k}$ and $hv_{j,k}$. The coarse grid $\Omega^R$ is then positioned with an offset so that the center of the cell $\Omega_{j,k}^M$ containing the observation is aligned with a point value in the coarse grid (\reffig{optproppull:2}). 
    \item[$Q_{GB}^{1/2,T} \gray{\coarseningoperator H^T S \innovation_i^n}$:] The adjoint of the geostrophic balance operator spreads the information given by the fields representing the coarse $hu$ and $hv$ onto a single field representing coarse $\eta$  (\reffig{optproppull:3}), as described by \eref{adjointGeoBalance}.
    \item[$Q_\SOAR^{1/2} \gray{Q_{GB}^{1/2,T} \coarseningoperator H^T S \innovation_i^n}$:] The correlation in the surface elevation given by the SOAR function in \eref{local_soar_on_eta} is applied (\reffig{optproppull:4}), as the final part of the adjoint covariance operator $Q^{1/2,T}$.
    \item[$Q^{1/2}_\SOAR \gray{Q_\SOAR^{1/2} Q_{GB}^{1/2,T} \coarseningoperator H^T S \innovation_i^n}$:] The SOAR function is applied again (\reffig{optproppull:5}), as part of $Q^{1/2}$.
    \item[$\interpolationoperator \gray{Q^{1/2}_\SOAR Q_\SOAR^{1/2} Q_{GB}^{1/2,T} \coarseningoperator H^T S \innovation_i^n}$:] We interpolate the result from $\Omega^R$ to $\Omega^M$, which gives us the final modification applied to $\eta$ ($\eta$ in  (\reffig{optproppull:6})).
    \item[$Q^{1/2}_{GB} \gray{\interpolationoperator Q^{1/2}_\SOAR Q_\SOAR^{1/2} Q_{GB}^{1/2,T} \coarseningoperator H^T S \innovation_i^n}$:] The modifications for $hu$ and $hv$ are found according to the geostrophic balance (\reffig{optproppull:6}) described by \eref{geostrophic_balance_disc}.
\end{description}
Due to the correlation pattern we have chosen for the model error, all pulls that are added to the particle states are constructed as dipoles to generate a local geostrophically balanced current in a given direction at a given point, without making any other assumptions.

\implementation{$S \innovation_i^n$ is calculated on the host before it is passed on to a GPU kernel for calculating the adjoint model error operations $Q^{1/2, T}$, and this temporary result (\reffig{optproppull:4}) is written into the buffer originally allocated for normal distributed random numbers, $\xi$.
The remaining operations (applying $Q^{1/2}$ and adding the result to the current particle state) are now identical to imposing the covariance structure of the model error for perturbing the particle state, and this functionality is therefore re-used.}

After stepping through the expression in \eref{detailedOPDUpdate}, it is also easier to describe how the matrix $S$ is constructed, by first expanding its definition 
\begin{equation}
    S = \left(H Q^{1/2}_{GB} \interpolationoperator Q^{1/2}_\SOAR Q_\SOAR^{1/2} Q_{GB}^{1/2,T} \coarseningoperator H^T + R \right)^{-1}.
    \label{eq:S_extended}
\end{equation}
The observation operator can be considered in matrix form as a $N_{\state} \times 2$ matrix consisting of the value 1 in the positions corresponding to the state values $hu_{j,k}$ and $hv_{j,k}$ in the first and second column, respectively, and zeros elsewhere.
The process just described in list form and depicted in \reffig{optimalproposalpull} can therefore be followed by replacing $S\innovation_i^n$ by $[0, 1]^T$ and $[1, 0]^T$, and applying the observation operator to the final result. 
This process gives us the two columns of $S$.

When the observation consists of $N_D > 1$ drifters, we assume that the observations of the drifters are independent of each other, making $R$ diagonal.
By also assuming that the drifters are sufficiently far from each other, the resulting matrix from $HQH^T$ becomes block diagonal, which also means that $\left(HQH^T + R \right)^{-1}$ is block diagonal with $N_D$ blocks of the $2 \times 2$ matrix $S$ from before.
The optimal proposal pull is essentially a local manipulation of the particle state, as seen in \reffig{optimalproposalpull}, and the influence area of this operation is approximately $5 \coarsedx$ since we have applied the SOAR function twice.
In the experiments, however, we do not validate the proximity assumptions for the drifters, and contributions from drifters close to each other are both added to the particle state.
This way, the process just described can therefore be applied at each drifter location independently, as seen in step two of \reffig{iewpfAlgorithm}.


Note that the expression for $c_i$ defined in \eref{iewpf_ci} and \eref{iewpfPhi} can be calculated almost for free during this step.
As each drifter is handled independently, the contributions from all the drifters are summed as
\begin{equation}
    c_i = -\log \left(w_i^{n-1}\right) + \sum_{d=1}^{N_D} \innovation_{i,d}^{n,T} S \innovation_{i,d}^n.
    \label{eq:ciMultiDrifter}
\end{equation}

\implementation{The most computationally efficient way of calculating the optimal proposal pull would be to add the contributions for all drifters to the coarse grid before applying the second SOAR function and the interpolation, so that $Q^{1/2}$ only would have to be applied once.
However, it is essential that the optimal proposal pull is applied at the drifter position with the precision of the computational grid $\Omega^M$, which means that the offset to align the drifter cell to the coarse grid point might be different for each drifter.
A coloring scheme could be constructed to maximize parallel processing of the drifters, but this performance optimization has not been realized in our implementation.}

\subsection{Sampling perpendicular random vectors}
\label{sec:iewpf_perpendicular_vectors}

The next step is to sample $\xi_i, \nu_i \sim N(0,I)$ in such a way that they become perpendicular. 
This is achieved by first sampling $\xi_i, \tilde{\nu}_i \sim N(0,I)$ independently.
We then decompose $\tilde{\nu}_i = \tilde{\nu}_{i, \parallel} + \tilde{\nu}_{i, \perp}$, so that $\tilde{\nu}_{i,\parallel}$ and $\tilde{\nu}_{i,\perp}$ become parallel and perpendicular to $\xi_i$, respectively, meaning that
\begin{equation}
	\tilde{\nu}_{i,\perp} = \tilde{\nu}_i - \tilde{\nu}_{i,\parallel} = \tilde{\nu}_i - \frac{\tilde{\nu}_i^T \xi_i}{\xi_i^T \xi_i} \xi_i.
    \label{eq:nuTildePerpendicular}
\end{equation}
We then scale $\tilde{\nu}_{i,\perp}$ to have the same length as $\tilde{\nu}_i$, and get
\begin{equation}
	\nu_i = \sqrt{ \frac{ \tilde{\nu}_i^T \tilde{\nu}_i }{ \tilde{\nu}_{i,\perp}^T\tilde{\nu}_{i,\perp} } } \tilde{\nu}_{i,\perp} .
	\label{eq:tempPerpendicular}
\end{equation}
By using \eref{nuTildePerpendicular} for $\tilde{\nu}_{i,\perp}$ in \eref{tempPerpendicular}, $\nu_i$ can be expressed as
\begin{equation}
    	\nu_i = \sqrt{ \frac{ \tilde{\nu}_i^T \tilde{\nu}_i }{ \tilde{\nu}_i^T \tilde{\nu}_i - a_i  \tilde{\nu}_i^T \xi_i}}
	        \left( \tilde{\nu}_i - a_i \xi_i \right), \qquad a_i = \frac{\tilde{\nu}_i^T \xi_i}{\xi_i^T \xi_i}.
	 \label{eq:nuPerpendicular}
\end{equation}
\implementation{This shows that we need to compute the three dot products $\xi_i^T \xi_i$, $\tilde{\nu}_i^T \tilde{\nu}_i$ and $\tilde{\nu}_i^T \xi_i$.
Since these dot products have overlapping data dependencies, they can be efficiently found within a single kernel using a common tree-based reduction approach~\cite{harris2007optimizing}.
Finally, $\tilde{\nu}_i$ can be transformed to $\nu_i$ element-wise and in-place.}
Note that this process resembles the Gram-Schmidt orthogonalization process, with preserved vector sizes.
\nomenclature{$a_i$}{To simplify the expression in \eref{nuPerpendicular}, we use $a_i = \frac{\tilde{\nu}_i^T \xi_i}{\xi_i^T \xi_i}$.}

During these computations, we store the values for $\xi_i^T \xi_i$ and $\nu_i^T \nu_i = \tilde{\nu}_i^T \tilde{\nu}_i$, as they are needed for the parameters $\gamma_i$ and $\zeta_i$, respectively, for solving the implicit equation in step 5.
However, as discussed in \refsec{modelerrors}, the normal distributed random numbers in $\xi_i$ and $\nu_i$ do not represent the entire state vector. 
The derivation of the IEWPF algorithm from \refsec{iewpf} and \appref{iewpfDetails} assume that $\xi_i^T \xi, \nu_i^T \nu_i \approx N_{\state} \pm \sqrt{2N_{\state}}$.
Since our $\nu_i, \xi_i \in \mathbb{R}^{N_R}$, this assumption is not satisfied directly.
To remedy this, we apply an appropriate scaling to the two dot products, and use
\begin{equation}
    \gamma_i = \xi_i^T \xi_i \frac{N_{\state}}{N_R}, \qquad \zeta_i = \nu_i^T \nu_i \frac{N_{\state}}{N_R}.
    \label{eq:scalingGammaZeta}
\end{equation}

\subsection{Target weight and $\beta$}
\label{sec:iewpf_target_weight}
To calculate the target weight $w_{target}$ and $\beta$, we need to obtain $c_i$, $\gamma_i$ and $\zeta_i$ for all particles $i = 1,2, ..., N_e$ in the ensemble.
This step represents a global synchronization point in the algorithm.
Once all three parameters are provided by all particles, we can calculate $w_{target}$ and $\beta$ from \eref{iewpf_targetWeight} and \eref{betaTwoStage}, respectively.

\subsection{Solving the implicit equation}
\label{sec:iewpf_implicit_equation}
The final two stages of the algorithm are again independent for all particles.
First, $c_i^{\star}$ is found according to \eref{cistarTwoStage} and constitutes the final piece for the implicit equation for $\alpha_i$, given by \eref{iewpfAlphaSol}.
As described in \appref{iewpfDetails}, the solution for $\alpha_i$ is obtained by using the Lambert W function, which is a scalar operation for each particle.

\subsection{Posterior particle states}
\label{sec:iewpf_posterior_states}


The final step of IEWPF is to perturb the particles so that they all obtain the target weight, giving the ensemble the correct posterior variance.
We need to apply the covariance structure of $P$ to the random fields $\nu_i$ and $\xi_i$, meaning that we seek an expression for $P^{1/2}$ in terms of (preferably) local operations.
Instead of using $P$ on the form given in \eref{opd_covariance}, it can be written as
\begin{equation}
    \begin{split}
        P =& Q - Q H^T \left( HQH^T + R \right)^{-1} H Q \\
          =& Q^{1/2} \left(I - Q^{1/2, T} H^T \left( HQH^T + R \right)^{-1} H Q^{1/2} \right) Q^{1/2,T}.
    \end{split}
    \label{eq:PcovExpression}
\end{equation}
Since there is no easy way to express the operator square root of the expression in the parenthesis in \eref{PcovExpression}, we consider the operations as matrices and seek its singular value decomposition (SVD) by constructing matrices $U$ and $V$ and a diagonal matrix $\Sigma$ so that 
\begin{equation}
    \begin{split}
    U\Sigma V^H &= I - Q^{1/2, T} H^T \left(HQH^T + R \right)^{-1} H Q^{1/2} \\
                &= I - Q_\SOAR^{1/2} Q_{GB}^{1/2,T} \coarseningoperator H^T S H Q^{1/2}_{GB} \interpolationoperator Q^{1/2}_\SOAR.
    \end{split}
    \label{eq:svdOfParenthesis}
\end{equation}
This allows us to apply the covariance structure $P$ to a sample $\xi_i \sim N(0, I)$ by
\begin{equation}
    P^{1/2} \xi_i =  Q^{1/2} U \Sigma^{1/2} \xi_i.
    \label{eq:Psquareroot}
\end{equation}
\nomenclature{$U$}{Part of the SVD result $U\Sigma V^H$}
\nomenclature{$V$}{Part of the SVD result $U\Sigma V^H$}
\nomenclature{$\Sigma$}{Part of the SVD result $U\Sigma V^H$}

\begin{figure*}[t!]
    \centering
    
    \begin{subfigure}[t]{0.23\textwidth}
        \centering
        \includegraphics[width=\textwidth, trim=0 0 0 0.75cm, clip]{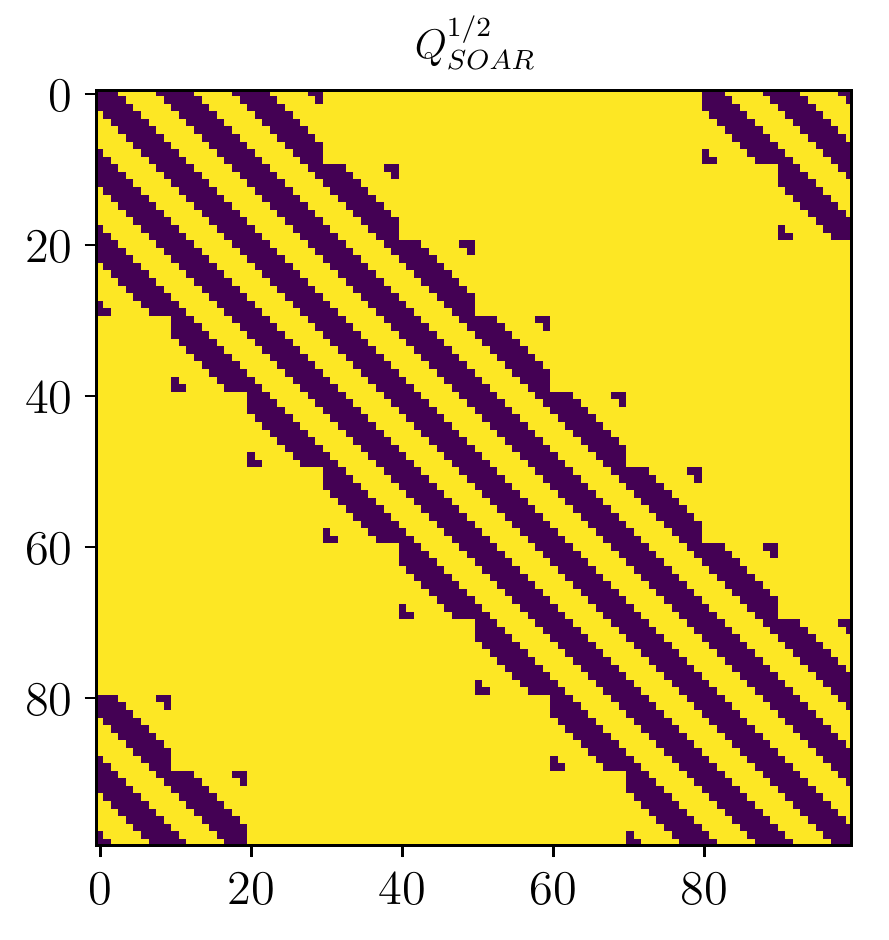}
        \caption{$Q^{1/2}_\SOAR$}
        \label{fig:localsvd:1}
    \end{subfigure}%
    \quad
    \begin{subfigure}[t]{0.21\textwidth}
        \centering
        \includegraphics[width=\textwidth, trim=-2cm 0 -2cm 0.62cm, clip] {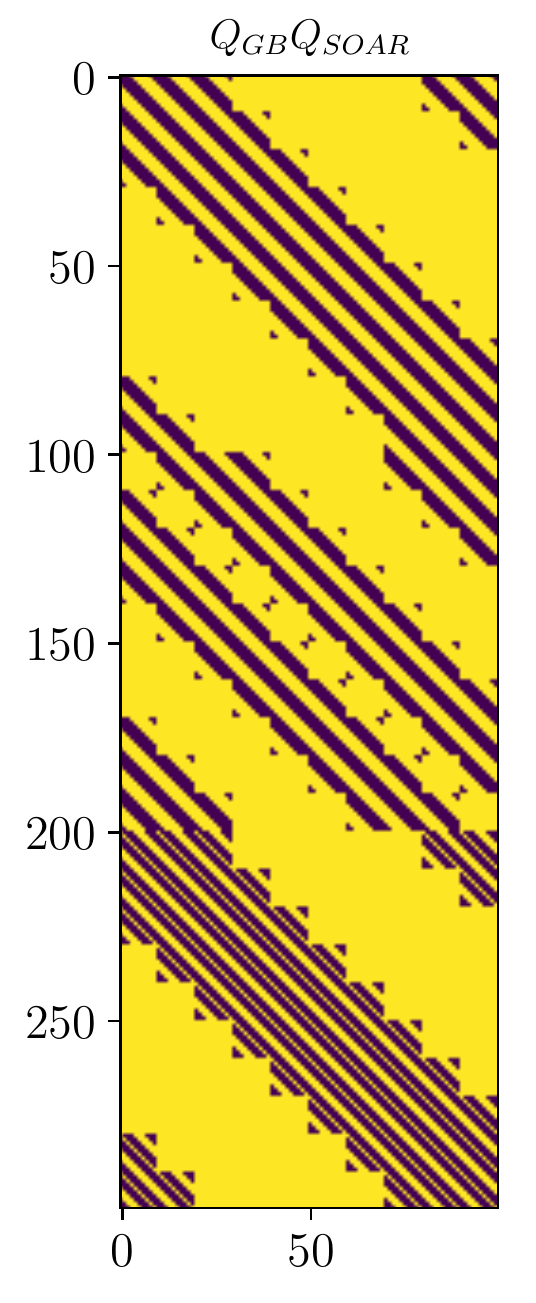}
        \caption{$Q^{1/2}_{GB} Q^{1/2}_\SOAR$}
        \label{fig:localsvd:2}
    \end{subfigure}    
    
    \vspace{0.6em}
    \begin{subfigure}[t]{0.45\textwidth}
        \centering
        \includegraphics[width=\textwidth, trim=0.5cm 0 0 0.75cm, clip]{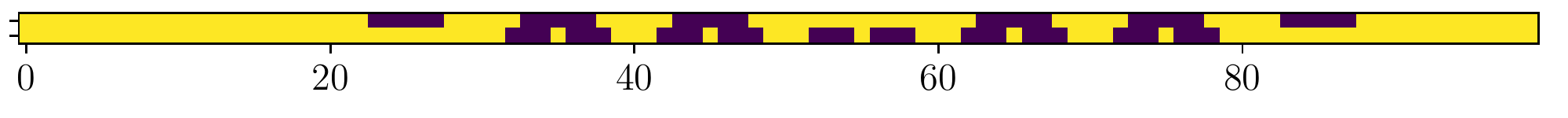}
        \caption{$S H Q^{1/2}_{GB} Q^{1/2}_\SOAR$}
        \label{fig:localsvd:3}
    \end{subfigure}
    
    \begin{subfigure}[t]{0.21\textwidth}
        \centering
        \includegraphics[width=\textwidth, trim=-2cm 0 -2cm 0.75cm, clip] {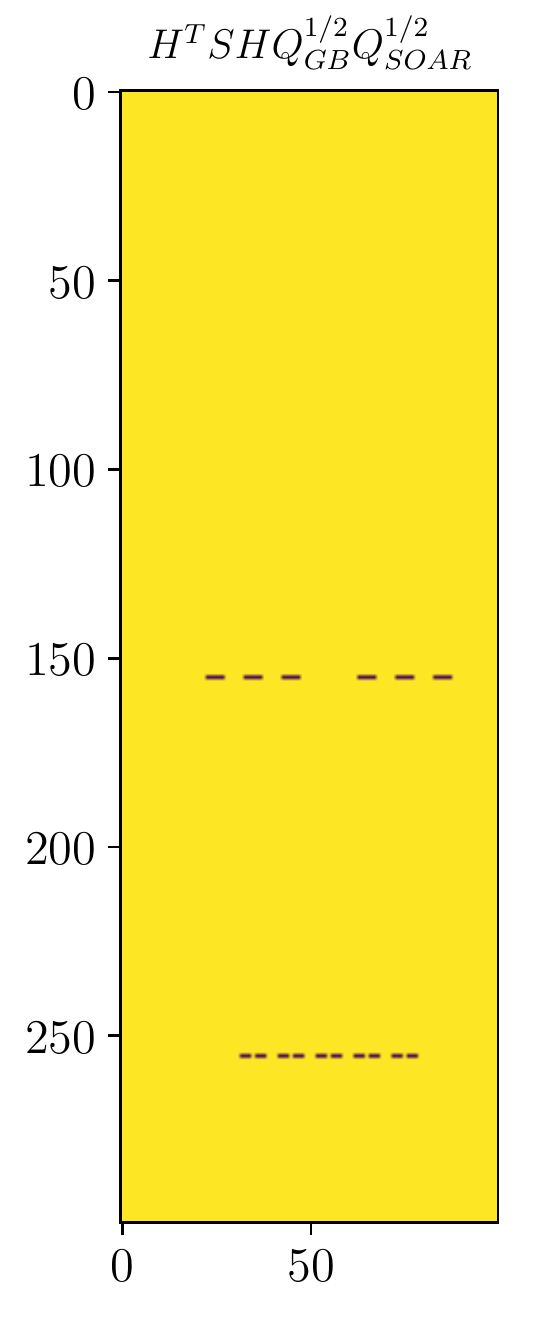}
        \caption{$H^T S H Q^{1/2}_{GB} Q^{1/2}_\SOAR$}
        \label{fig:localsvd:4}
    \end{subfigure}
    \quad
    \begin{subfigure}[t]{0.28\textwidth}
        \centering
        \includegraphics[width=\textwidth, trim=0 0 0 0.75cm, clip] {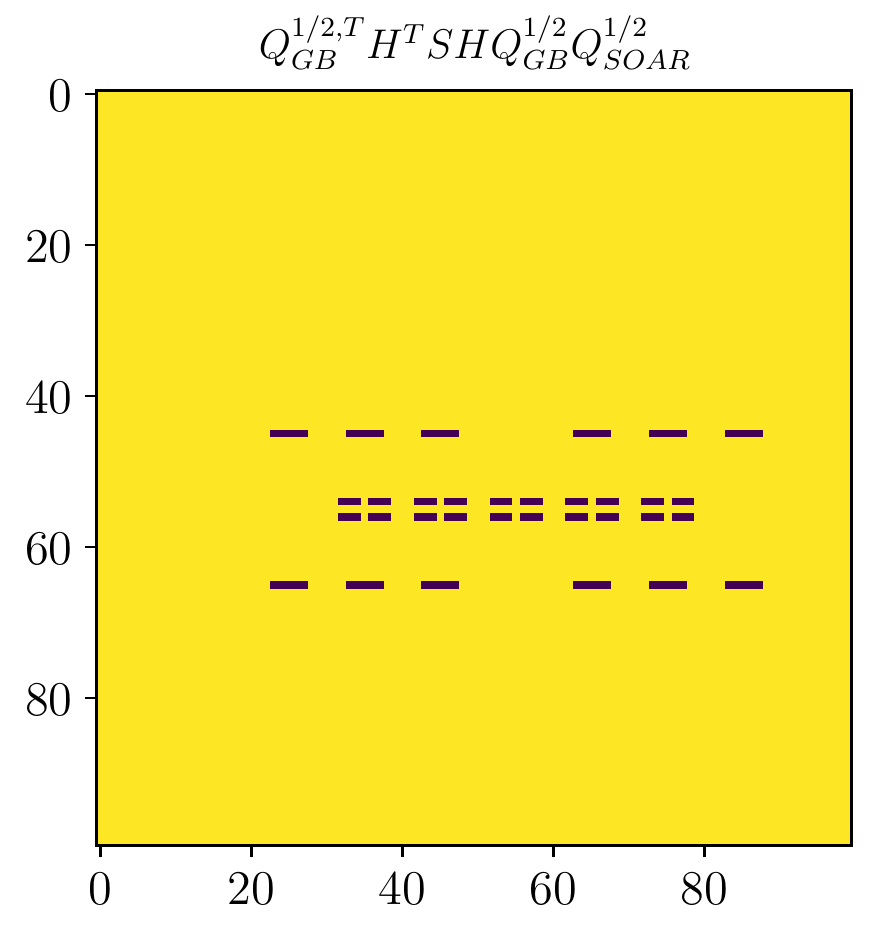}
        \caption{$Q^{1/2,T}_{GB} H^T S H Q^{1/2}_{GB} Q^{1/2}_\SOAR$}
        \label{fig:localsvd:5}
    \end{subfigure}
    \quad
    \begin{subfigure}[t]{0.33\textwidth}
        \centering
        \includegraphics[width=\textwidth, trim=-0.9cm 0 -1cm 0.75cm, clip] {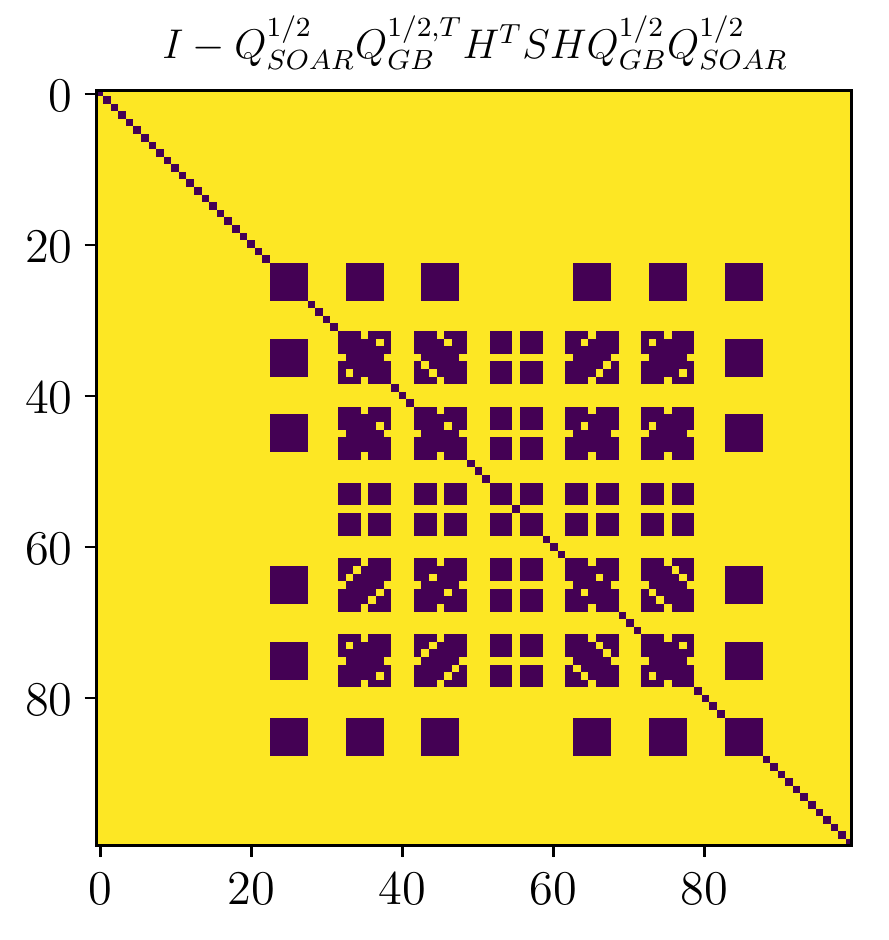}
        \caption{$I - Q^{1/2}_\SOAR Q^{1/2,T}_{GB} H^T S H Q^{1/2}_{GB} Q^{1/2}_\SOAR$}
        \label{fig:localsvd:6}
    \end{subfigure}    

    \vspace{0.6em}
    \begin{subfigure}[t]{0.42\textwidth}
        \centering
        \includegraphics[width=\textwidth, trim=-3cm 0 -3cm 0.75cm, clip ]{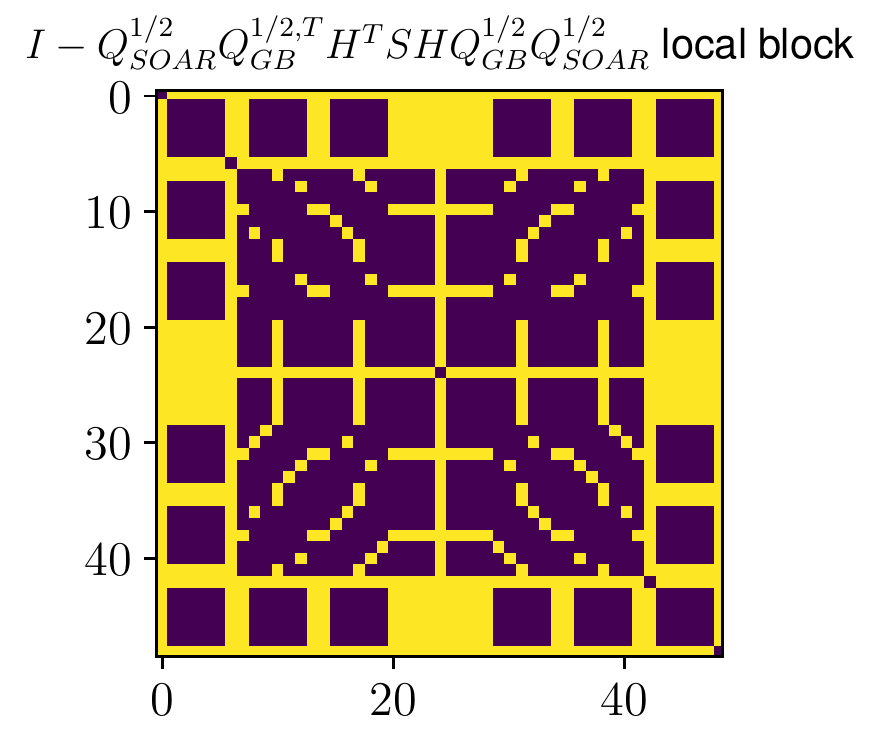}
        \caption{The local $7^2 \times 7^2$ block required to represent \\$I - Q^{1/2}_\SOAR Q^{1/2,T}_{GB} H^T S H Q^{1/2}_{GB} Q^{1/2}_\SOAR$}
        \label{fig:localsvd:7}
    \end{subfigure}    

    \caption{The non-zero patterns that emerge when computing the parenthesis expression for the covariance $P$ in \eref{PcovExpression} on a small domain consisting of $10 \times 10$ cells. (a) The covariance operators $Q^{1/2}_\SOAR$ and $Q_{GB}^{1/2}$ are interpreted as matrices, meaning that $Q^{1/2}_\SOAR$ becomes a $100 \times 100$ matrix. (b) After applying $Q^{1/2}_{GB}$ we get an extra 200 rows, representing $hu$ and $hv$ in addition to $\eta$ in every cell. (c) We extract the rows corresponding to the observation and scale them by $S$, before (d) the values are mapped back to state space. (e) We then apply $Q^{1/2, T}_{GB}$, before (f) applying $Q^{1/2}_\SOAR$ and subtracting the result from the identity. (g) We ignore the part that is equal to the identity and are left with a covariance matrix describing the $7\times 7$ cell block centered on the observation. Note that by increasing the domain to $100 \times 100$ cells, the matrix representing $Q^{1/2}_\SOAR$ will become $10~000 \times 10~000$, but the dense local block (g) will still remain the same.}
    \label{fig:localsvd}
\end{figure*}

For the computation of the SVD, we consider the case in which $\Omega^M = \Omega^R$, meaning that the interpolation and coarsening operators are simplified to the identity. 
Starting from the right in the parenthesis expression in \eref{PcovExpression}, we step through the operations interpreted as matrices and investigate the structure of non-zero values. 
This process is illustrated for a small domain consisting of $10 \times 10$ cells in \reffig{localsvd}.
\begin{description}
    \item[$Q^{1/2}_\SOAR$:] 
    Symmetric matrix of size $N_M \times N_M$, describing the covariance structure defined by the SOAR function in \eref{soar_function} and \eref{local_soar_on_eta}. By using $c_\SOAR = 2$, the value in each grid cell is given a correlation with a grid cell block of size $5 \times 5$ centered on itself. This means that each row of $Q^{1/2}_\SOAR$ has 25 non-zero values (\reffig{localsvd:1}).
    \item[$Q^{1/2}_{GB} \gray{Q^{1/2}_\SOAR}$:] 
    A $3N_M \times N_M$ matrix, in which the first $N_M$ rows are equal to $Q^{1/2}_\SOAR$. The middle and lower $N_M$ rows are the results from applying a central difference formula on values of $Q^{1/2}_\SOAR$ in the $y$- and $x$-direction, respectively. These rows have 35 non-zero values on column indices representing $7 \times 5$ and $5 \times 7$ grid blocks for the middle and lower matrix block, respectively (\reffig{localsvd:2}).
    \item[$H \gray{Q^{1/2}_{GB} Q^{1/2}_\SOAR}$:] 
    The observation operation extracts values of the rows representing $hu_{j,k}$ and $hv_{j,k}$ only, giving us a $2 \times N_M$ matrix with 35 non-zero values for each row (\reffig{localsvd:3}).
    \item[$S \gray{H Q^{1/2}_{GB} Q^{1/2}_\SOAR}$:] All values are scaled by the matrix $S$ representing model and observation uncertainty. The non-zero pattern is not affect by this operation (\reffig{localsvd:3}).
    \item[$H^T \gray{S H Q^{1/2}_{GB} Q^{1/2}_\SOAR}$:]
    The two rows are mapped back into state space, and inserted into an otherwise zero matrix of size $3N_M \times N_M$ at the rows with indices representing $hu_{j,k}$ and $hv_{j,k}$ (\reffig{localsvd:4}).
    \item[$Q_{GB}^{1/2,T} \gray{H^T S H Q^{1/2}_{GB} Q^{1/2}_\SOAR}$:] 
    The adjoint of the geostrophic balance operator maps the rows representing volume transport to the $\eta$-field based on adjoint central differences, resulting in an $N_M \times N_M$ matrix. This means that the row representing $hu_{j,k}$ has non-zero values in rows representing cells $\Omega_{j,k-1}$ and $\Omega_{j,k+1}$, and similarly the row representing $hv_{j,k}$ has non-zero data in the row representing $\Omega_{j-1,k}$ and $\Omega_{j+1,k}$. There are now four rows with 35 non-zero values each (\reffig{localsvd:5}).
    \item[$Q_\SOAR^{1/2} \gray{Q_{GB}^{1/2,T} H^T S H Q^{1/2}_{GB} Q^{1/2}_\SOAR}$:] 
    Finally, we apply the SOAR function and each of the existing four non-zero rows are mapped to 25 rows representing a $5 \times 5$ grid cell block in the resulting matrix. Considering the overlap between these blocks, we get an $N_M \times N_M$ matrix with 45 non-zero rows, each containing 45 non-zero values. The rows represent a $7 \times 7$ grid cell block centered in cell $\Omega_{j,k}$ with a single cell missing in each of the four corners.
    \item[$I - \gray{Q_\SOAR^{1/2} Q_{GB}^{1/2,T} H^T S H Q^{1/2}_{GB} Q^{1/2}_\SOAR}$:] 
    The final matrix is a $N_M \times N_M$ matrix equal to the identity except for the 45 rows representing the $7 \times 7$ grid cell block centered in the cell in which the observation was made. 
\end{description}
Due to the structure of the problem, the computations for finding the SVD can be greatly simplified by ignoring all rows that are equal to the identity.
To further simplify the structure of the code, we include the corners and consider the complete $7 \times 7$ grid cell block. 
This results in a  $49 \times 49$ matrix described by the above process (\reffig{localsvd:7}), and we can obtain the SVD from this much smaller matrix instead of from the full covariance matrix.
When applying $P^{1/2}$ to $\xi_i$ we can then apply the obtained $U\Sigma^{1/2}$ locally according to the observed location of the drifter, before applying $Q^{1/2}$ to values defined in the entire domain as before.
In fact, by assuming constant equilibrium depth $H$, constant Coriolis force $f$, and double periodic boundary conditions, the structure of the $49 \times 49$ non-identity block is always the same for any drifter position.
This allows us to pre-compute the local SVD matrix ahead of the data-assimilation loop. 

Now, if we consider a case with $\Omega^M \neq \Omega^R$, we would need to handle two significant issues.
First, the interpolation would result in a much larger local non-zero structure, and thus a larger local matrix $U\Sigma^{1/2}$, requiring more storage, and becoming more expensive to apply.
Second, since $\xi_i$ and $\nu_i$ are defined for all coarse grid points, we need to use the same offset for co-locating points across the entire domain.
Since the drifters are most likely to be located in cells that require different offsets, we will not be able to apply the SVD structure accurately on top of all drifters.
Because of these two reasons, we have chosen to define $U \Sigma^{1/2}$ on the coarse grid only, also when $\Omega^M \neq \Omega^R$, which enables us to apply the $49 \times 49$ pre-computed matrix $U \Sigma^{1/2}$ to the random field.
The $Q^{1/2}$ operator then spreads this information to all three conserved variables on the computational grid.
Since this term structure is applied to values that are sampled randomly ($\xi_i$ and $\nu_i$), the simplification does not introduce significant errors.

Finally, we note that $Q^{1/2}$ and $U \Sigma^{1/2}$ are linear operations.
Instead of applying the covariance structure first to $\xi_i$ and then to $\nu_i$, we add the scaled random fields before applying $P^{1/2}$.
The final posterior particle states in \eref{updateEquationTwoStage} are then obtained by
\begin{equation}
	\state_i^n = \state_i^{n,a} + P^{1/2} \left( \beta^{1/2} \nu_i + \alpha_i^{1/2} \xi_i \right).
    \label{eq:optUpdateEquationTwoStage}
\end{equation}

%
%
%


\section{Experimental results}
\label{sec:results}

Here, we describe an experimental setup for experimenting with the data-assimilation method discussed in the previous sections.
First, we produce rank histograms to show that the IEWPF method produces statistically sound forecasts and thereby is a valid method for data assimilation.
We continue with drift trajectory forecasts through a series of experiments using different numbers of observations from drifters and moorings.
This is followed by an illustration of how the standard particle filter collapses for the same case, even when starting from an ensemble that is centered around the true state with low variance. 
Finally, we measure the computational performance to determine the workload in the IEWPF compared to simply advancing the model.

To get a synthetic but near-realistic ocean model state to represent $\state_{true}$, we take inspiration from a test case for validating shallow-water models on a rotating sphere suggested by Galewsky et al.~\cite{galewsky_2004_test_case}.
This test case describes a steady-state solution in which an eastward atmospheric jet is balanced by a smoothed step function in the thickness of the fluid layer due to the sphere's rotation.
When a small perturbation is introduced in $\eta$, the jet develops instabilities after some time and produces a state that is dominated by complex currents and eddies.
By running a simulation from the steady-state but adding random model errors, the case has a chaotic behaviour in which instabilities develop in different places and in different ways for independent simulation runs.
This enables us to produce a challenging test case with realistic features for data-assimilation experiments.

\begin{figure}
	\begin{center}
	\includegraphics[width=0.95\linewidth, trim= 0 0 0 0, clip]{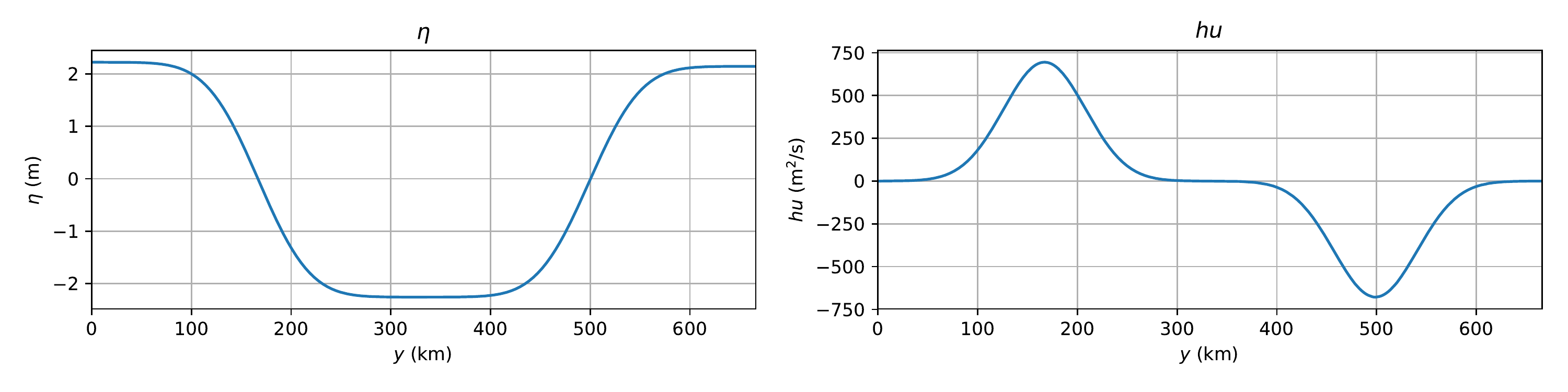}
    \caption{The cross-section along the $y$-axis of the steady-state initial conditions for the unstable double jet case.}
    \label{fig:doubleJet_initialconditions}
    \end{center}
\end{figure}

\begin{figure}
	\begin{center}
	\includegraphics[width=\linewidth, trim=0.8cm 0 0 1.6cm, clip]{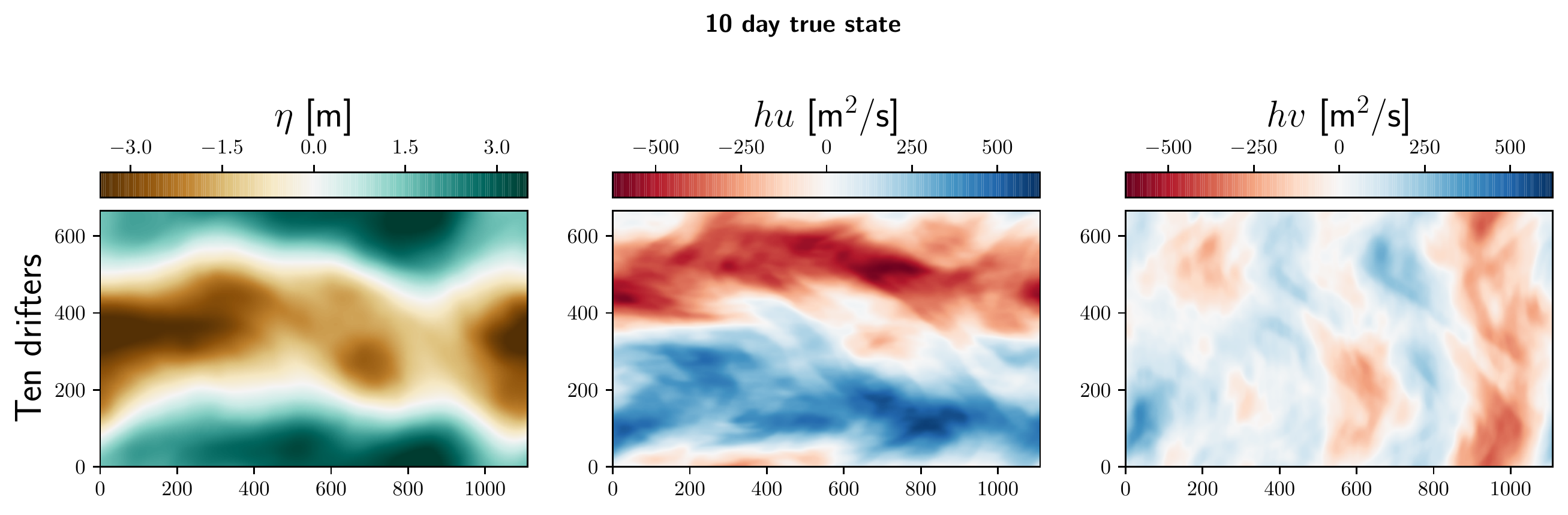}	
    \caption{A possible model state after 10 days, resulting from running the shallow-water simulation with additive model errors from the steady-state shown in \reffig{doubleJet_initialconditions}. From left to right, the figures show the surface elevation $\eta$, and the volume transport $hu$ and $hv$ in $x$- and $y$-direction, respectively. All $x$- and $y$-axes are given in km. The realized model state displayed here is also used as the true state for the drift forecast experiments in \refsec{driftTrajectoryForecasting}.}
    \label{fig:doubleJet_tenDays}
    \end{center}
\end{figure}

We transform the Galewsky test case to a flat two-dimensional rectangular domain with a constant Coriolis parameter.
To make the case even more interesting, a second jet is introduced in the opposite direction south of the original jet, so that the balancing ocean surface ends up at an equivalent level at the northern and southern boundary, allowing us to use periodic boundary conditions at all four boundaries.
We introduce parameters so that the case represents oceanic flow, rather than atmospheric, and model the domain after the Barents Sea, using a rectangular domain that covers $1110 \mathrm{~km} \times 666 \mathrm{~km}$, divided into $500 \times 300$ cells with $\dx = \dy = 2220$~m.
Further, $g = 9.806~\mathrm{m/s}$, $f = 1.405 \cdot 10^{-4}~\mathrm{s}^{-1}$ (corresponding to 75 degrees north), and a constant equilibrium depth $H_{eq} = 230~\mathrm{m}$.
Cross sections of the initial steady state for $\eta$ and $hu$ are shown in \reffig{doubleJet_initialconditions}, and the initial condition for $hv$ is zero.
The model time step is chosen as $\dt = 60~\mathrm{s}$, and the time step in the numerical scheme $\dt_{scheme}$ is dynamically adjusted according to the CFL-condition in \eref{swecfl} with a Courant number of 0.8.
We use a model error amplitude $q_0 = 2.5\cdot 10^{-4}$, coarsening factor $c_{\Omega} = 5$, and model error length scale $L_0 = \tfrac{3}{4}\coarsedx$.
\reffig{doubleJet_tenDays} shows an example of a model state produced by these parameters after ten simulation days. 
This is also the true state that is used in the drift trajectory forecasting experiments in \refsec{driftTrajectoryForecasting}.

In all following experiments we consider observations from drifters and moorings, according to the description in \refsec{truthAndObservations}.
The observation error consists of a measurement error and a representation error and is rarely trivial to quantify in geophysical systems.
The measurement error is related to the precision of the instruments that are used to make the observation, e.g., the precision of the GPS used for drifter experiments, and is typically given by the instruction handbook for the relevant instrument. 
The representation error, on the other hand, should reflect how well the observed measure represents the simulated variables, e.g., how well the mean current along the drifter trajectory represents the cell-averaged depth-integrated water transport.
As we here use an identical twin experiment, we can consider the representation error to be small, and we assume that the instruments involved have good accuracy.	
All experiments hence use $R = I$. 

\begin{figure}
    \begin{center}
    \resizebox{0.9\textwidth}{!}{%
      \makeatletter
\DeclareRobustCommand{\rvdots}{%
  \vbox{
    \baselineskip4\p@\lineskiplimit\z@
    \kern-\p@
    \hbox{.}\hbox{.}\hbox{.}
  }}
\makeatother

\newcommand{\epsfont}{\large}

\begin{tikzpicture}
[header_style/.style={
        text width=2.5cm,
        font=\scriptsize,
        align=center
    },
    fancytitle/.style={
        fill = white,
        text = black,
        draw=black!50, 
        rounded corners=0.1cm,
        font=\scriptsize
    }
]


\tikzstyle{particle}=[auto=left,circle,draw=black,fill=white,minimum size=10pt,inner sep=0pt, font=\tiny]

\tikzstyle{observation}=[auto=left,circle,draw=black!50,fill=black!20,minimum size=4pt,inner sep=0pt, font=\tiny]


\coordinate (truth-ic) at (0,2.8); 
\node (truth-warm) at (1.5,2.8)    [particle] {};
\node (truth-fc-end) at (6.5,2.8)   [particle] {};

\begin{pgfonlayer}{foreground}
    \coordinate (obs0coord) at (2.0, 2.8);
    \node (obs0) at (obs0coord) [observation] {};
    \node[above=9pt of obs0coord, anchor=center, font=\small] (obs0label) {$\observation^0$};
    
    \coordinate (obs1coord) at (2.5, 2.8);
    \node (obs1) at (obs1coord) [observation] {};
    \node[above=9pt of obs1coord, anchor=center, font=\small] (obs1label) {$\observation^1$};

    \coordinate (obsmcoord) at (5.0, 2.8);
    \node (obsn) at (obsmcoord) [observation] {};
    \node[above=9pt of obsmcoord, anchor=center, font=\small] (obsmlabel) {$\observation^m$};

    \coordinate (obshidden) at (3.65, 2.8);
    \node[above=9pt of obshidden, anchor=center, font=\small] (obsmlabel) {$...$};
\end{pgfonlayer}

\coordinate (m0-ic) at (0,2); 
\node (m0-warm) at (1.5,2)    [particle] {};
\node (m0-fc-start) at (5,2)   [particle] {};
\node (m0-fc-end) at (6.5,2)   [particle] {};

\coordinate (m1-ic) at (0,1.5); 
\coordinate (m1-spinup) at (0.75, 1.5);
\node (m1-warm) at (1.5,1.5)    [particle] {};
\coordinate (m1-da) at (3.25, 1.5);
\node (m1-fc-start) at (5,1.5)   [particle] {};
\coordinate (m1-fc) at (5.75, 1.5);
\node (m1-fc-end) at (6.5,1.5)   [particle] {};

\coordinate (mn-ic) at (0,0.5); 
\coordinate (mn-spinup) at (0.75, 0.5);
\node (mn-warm) at (1.5,0.5)    [particle] {};
\coordinate (mn-da) at (3.25, 0.5);
\node (mn-fc-start) at (5,0.5)   [particle] {};
\coordinate (mn-fc) at (5.75, 0.5);
\node (mn-fc-end) at (6.5,0.5)   [particle] {};

\begin{pgfonlayer}{foreground}
    \draw[rounded corners, fill=white] (-0.05, 0.4) rectangle (0.15, 2.9) {};
    \node[rotate=90, font=\scriptsize] at (0.05, 1.6) {Initial conditions}; 
\end{pgfonlayer}

\node[left=1em of truth-ic, anchor=east, font=\small] (truth-label) {$\state_{true}$};
\node[left=1em of m0-ic, anchor=east, font=\small] (m0-label) {$\state_0$};
\node[left=1em of m1-ic, anchor=east, font=\small] (m1-label) {$\state_1$};
\node[left=1em of mn-ic, anchor=east, font=\small] (mn-label) {$\state_{N_e}$};

\path (m1-label) -- node[auto=false]{\rvdots} (mn-label);
\path (m1-spinup) -- node[auto=false]{\rvdots} (mn-spinup);
\path (m1-da) -- node[auto=false]{\rvdots} (mn-da);
\path (m1-fc) -- node[auto=false]{\rvdots} (mn-fc);

\begin{pgfonlayer}{background}
    \coordinate (truthlowerleft) at (1.65, 2.55);
    \coordinate (truthupperright) at (6.45, 3.25);
    
    \node[rounded corners=8pt, draw=black!50, fill=yellow!20, fit=(truthlowerleft) (truthupperright)] (truth) {};
\end{pgfonlayer}

\begin{pgfonlayer}{background}
    \coordinate (spinuplowerleft) at (0.1, 0.3);
    \coordinate (spinupupperright) at (1.35, 3.25);
    
    \node[rounded corners=8pt, draw=black!50, fill=yellow!20, fit=(spinuplowerleft) (spinupupperright)] (spin-up) {};
\end{pgfonlayer}

\begin{pgfonlayer}{background}
    \coordinate (dalowerleft) at (1.65, 0.3);
    \coordinate (daupperright) at (4.85, 2.2);
    
    \node[rounded corners=8pt, draw=black!50, fill=red!20, fit=(dalowerleft) (daupperright)] (data-assimilation) {};
\end{pgfonlayer}

\begin{pgfonlayer}{background}
    \coordinate (fclowerleft) at (5.15, 0.3);
    \coordinate (fcupperright) at (6.45, 2.2);
    
    \node[rounded corners=8pt, draw=black!50, fill=green!20, fit=(fclowerleft) (fcupperright)] (forecast) {};
\end{pgfonlayer}

\node[fancytitle] at (spin-up.south) (spin-up-title) {Spin-up};
\node[fancytitle] at (data-assimilation.south) (spin-up-title) {Data assimilation};
\node[fancytitle] at (forecast.south) (spin-up-title) {Forecast};
\node[fancytitle] at (truth.north) (truth-title) {Generate truth};

\draw[thick,shorten >=2pt, ->]
    (truth-ic) edge (truth-warm)
    (truth-warm) edge (truth-fc-end)
    
    (m0-ic) edge (m0-warm)
    (m0-warm) edge (m0-fc-start)
    (m0-fc-start) edge (m0-fc-end)

    (mn-ic) edge (mn-warm)
    (mn-warm) edge (mn-fc-start)
    (mn-fc-start) edge (mn-fc-end)
    
    (m1-ic) edge (m1-warm)
    (m1-warm) edge (m1-fc-start)
    (m1-fc-start) edge (m1-fc-end);
  
\draw[->] (obs0coord) -- (obs0coord |- data-assimilation.north);
\draw[->] (obs1coord) -- (obs1coord |- data-assimilation.north);
\draw[->] (obsmcoord) -- (obsmcoord |- data-assimilation.north);

\draw[->] (0, -0.2) -- (7, -0.2);
\node [anchor=west, font=\small] (truth-label) at (7.1, -0.2) {$t$ [days]};
\draw (0.0, -0.3) -- (0.0, -0.1);
\draw (1.5, -0.3) -- (1.5, -0.1);
\draw (5.0, -0.3) -- (5.0, -0.1);
\draw (6.5, -0.3) -- (6.5, -0.1);
\node[header_style]  (ic) at (0.0, -0.45) {0};
\node[header_style]  (ic) at (1.5, -0.45) {3};
\node[header_style]  (ic) at (5.0, -0.45) {10};
\node[header_style]  (ic) at (6.5, -0.45) {13};

%

\end{tikzpicture}
    }
    \caption{Overview of the drift trajectory ensemble forecast experiments. Before the experiments start, an ensemble of size $N_e$ is spun up from a common initial state, and regular observations from the synthetic truth is generated (yellow). Experiments start at day three, with a seven day period of data assimilation, during which the observations are used to guide the ensemble towards the true state (red). At day 10, there are no more observations, and the ensemble runs a drift trajectory forecast with the latest observed drifter positions as starting point (green).}
    \label{fig:EPSInit}
    \end{center}
\end{figure}
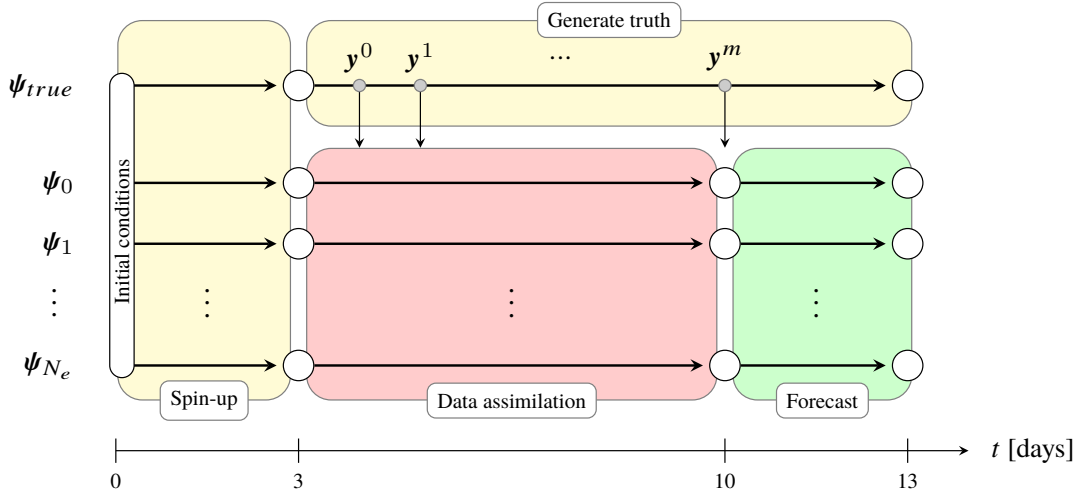

The true state for our experiments is pre-generated by a single thirteen day simulation, during which drifters and moorings are added to the model at the start of day three, and imperfect observations from them are written to file every five minutes.
We have used 64 drifters and 240 moorings which are initiated throughout the domain in an $8 \times 8$ and a $12 \times 20$ pattern, respectively.
The experiments use different subsets of these observations for data assimilation.
The experimental setup is then divided into three phases (see \reffig{EPSInit}):
\begin{description}
    \item[Day 0--2:] Spin-up period to let the ensemble members start to develop independent instabilities in the two jets. 
    We only perform the spin-up of the ensemble once, so that all experiments start from the same initial ensemble at the beginning of day three. 
    \item[Day 3--9:] Observations from the pre-generated truth are assimilated into the ensemble. 
    \item[Day 10--12:] This is the forecasting period. Drifters are added to all ensemble members at the observed drifter positions at the start of day ten. Each ensemble member runs independently to generate three-day  drift trajectory forecasts for all drifters.
\end{description}
We use an ensemble size of $N_e = 100$ where not stated otherwise, as this size will typically fit on a single desktop with a commodity-level GPU.



\subsection{Rank histograms for IEWPF}
\label{sec:rankHistogram}
Prior to using the IEWPF scheme for forecasting experiments, we aim to evaluate the quality of the method.
One way to do so is by creating a so-called rank histogram by running a large number of independent data-assimilation experiments, and for each of them find the \emph{rank} of a chosen observed variable within the ensemble.
For instance, choosing an observed variable $hu_{j,k}$, the ensemble members are sorted based on the value of $(hu_i)_{j,k} + \epsilon_i$, in which $\epsilon_i \sim N(0,R)$ represents the observation error, from lowest to highest.
The rank is then the position that the observed value for $hu_{j,k}$ takes when inserted into this sorted list.
A histogram is then generated from the number of appearances for each of the $N_e + 1$ possible ranks over all the forecasting experiments.
Since a rank can be considered to be a sample from the inverse cumulative distribution of the state density function, the rank histogram is expected to be flat when the data-assimilation method works as intended.


For simplicity of discussion, assume for a moment that there is no observation error and that the posterior distribution of $(hu_i)_{j,k}$ is Gaussian.
This corresponds to a high concentration of ensemble members with values close to some mean value, and gradually fewer ensemble members further away from this mean.
The mean value is our best estimate for $(hu_{true})_{j,k}$, whereas the spread in the ensemble represents the uncertainty of this estimate.
Most likely, the truth should be close to the mean, but since the density of $(hu_i)_{j,k}$ is high around the mean, the resulting rank is sensitive to small variations of $(hu_{true})_{j,k}$.
On the other hand, there is a chance of finding the truth in the tail of the ensemble as well.
The probability for a tail value to be exactly equal to the truth is very low, but since the density of values in the tail also is low, the truth can take a larger range of values and still obtain the same rank.
Ideally, the probability should be the same for obtaining any rank for any forecast experiment, meaning that the rank histogram created from a large number of such experiments should resemble a uniform distribution.
If the ensemble constantly fails to represent the uncertainty of the observed state, the histogram takes other forms.
For instance, if all ensemble members are too close to the mean, the ranks corresponding to ensemble values at the tail of the ensemble will be over-represented, creating a U-shaped rank histogram. This indicates that the ensemble is under-dispersed.
On the other hand, a hill-shaped rank histogram means that the spread in the ensemble forecast is too large to represent the true values, meaning that the ensemble is over-dispersed.
This discussion holds for any posterior distribution, not only Gaussian.
For a more detailed discussion on the interpretation of rank histograms, see Hamill~\cite{hamill2001_rankhistograms}.
\nomenclature{$\mean{hu_{j,k}}$}{The ensemble mean for a given variable, in this case $hu_{j,k}$.}

\begin{figure}
	\begin{center}
	\includegraphics[width=0.95\linewidth, clip]{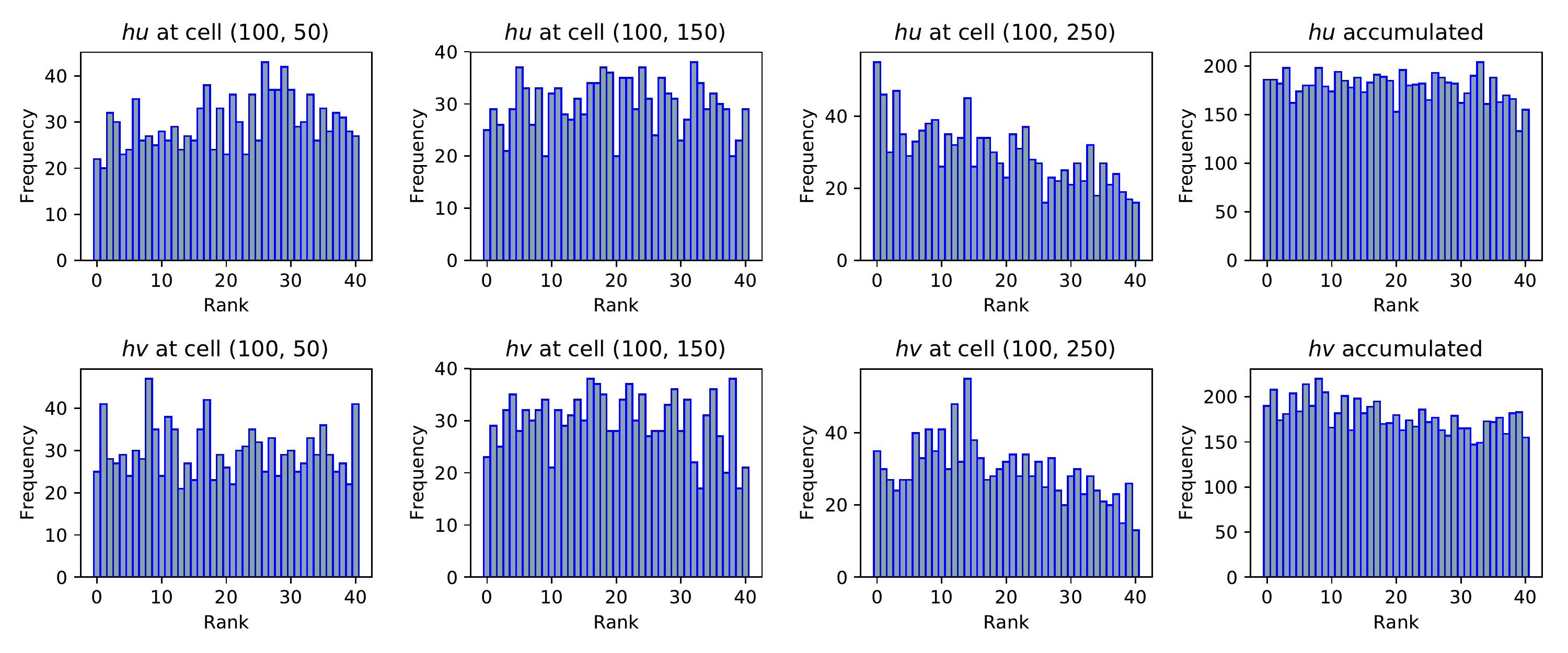}
    \caption{Rank histograms based on 1222 experiments of one hour ensemble forecasts. The rank histograms are generated for $hu$ and $hv$ at cells $(100, y)$ for $y = \{0, 50, 100, ..., 250\}$. The accumulated rank histograms are the sum of the rank at all these cells, and they are considered independent of each other. Most of the generated rank histograms resembles uniform distributions, such as for cells $(100, 50)$ and $(100, 150)$, whereas some are more irregular, such as for cell $(100, 250)$. The accumulated rank histograms are flat.}
    \label{fig:rankhistograms}
    \end{center}
\end{figure}

We generate rank histograms from 1222 data-assimilation experiments using $N_e = 40$ ensemble members and observations from independently generated truths every five minutes from the 120 moorings in the western half of the domain.
The ensemble is initialized by a random sample from the hundred spun up initial conditions generated for the forecast experiments and is run with data assimilation for six hours. 
We find the ranks from a one hour ensemble forecast from this state at simulation time three days and seven hours.
In this time range, we assume that variables located 50 cells apart from each other can be considered as independent and have therefore generated rank histograms for $hu$ and $hv$ at cells $(100, y)$ for $y = \{0, 50, 100,..., 250\}$.
Additionally, since these values are assumed to be independent, an accumulated rank histogram consisting of the sum of all these ranks is created as well.
\reffig{rankhistograms} displays a selection of the generated rank histograms.
Most of them resemble uniform distributions, such as those shown for cells $(100, 50)$ and $(100, 150)$, but there are also some that display a more irregular trend, such as the one for cell $(100, 250)$. 
The accumulated rank histograms shown in the rightmost column of the figure also resemble a uniform distribution.
In total, these results indicate that our implementation of the IEWPF gives ensemble forecasts with good statistical quality.

\subsection{Drift trajectory forecasting experiments}
\label{sec:driftTrajectoryForecasting}
\begin{figure*}[t!]
    \centering
    
    \begin{subfigure}[t]{0.85\textwidth}
        \centering
        \includegraphics[width=\linewidth, trim=0 0.2cm 0 1.5cm, clip]{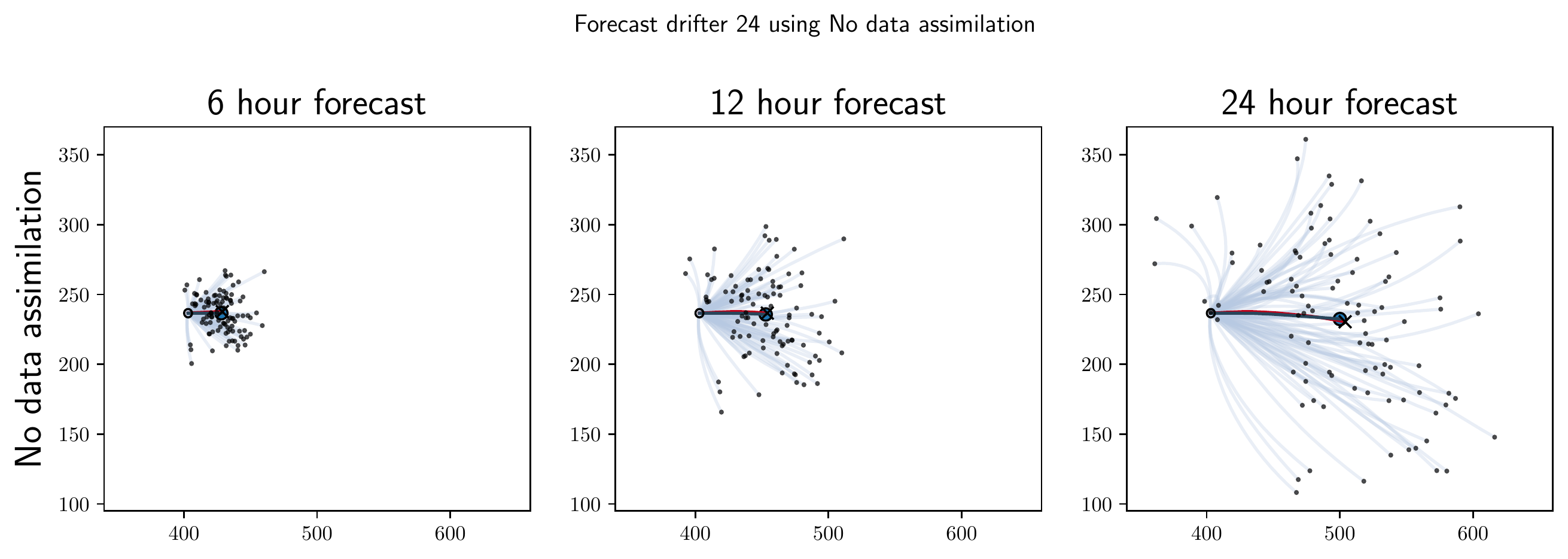}
    \end{subfigure}%
	\\
	\begin{subfigure}[t]{0.85\textwidth}
        \centering
    	\includegraphics[width=\linewidth, trim=0 0.2cm 0 2.0cm, clip]{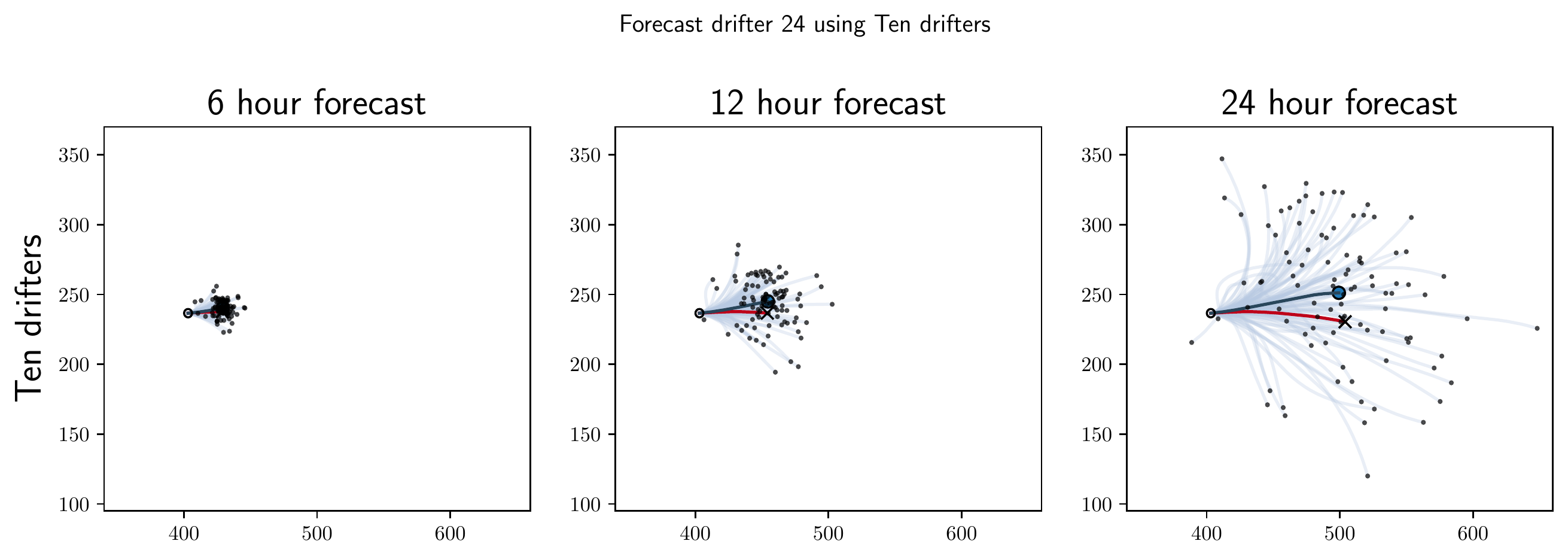}
    \end{subfigure}    
    \\
	\begin{subfigure}[t]{0.85\textwidth}
        \centering
    	\includegraphics[width=\linewidth, trim=0 0.2cm 0 2.0cm, clip]{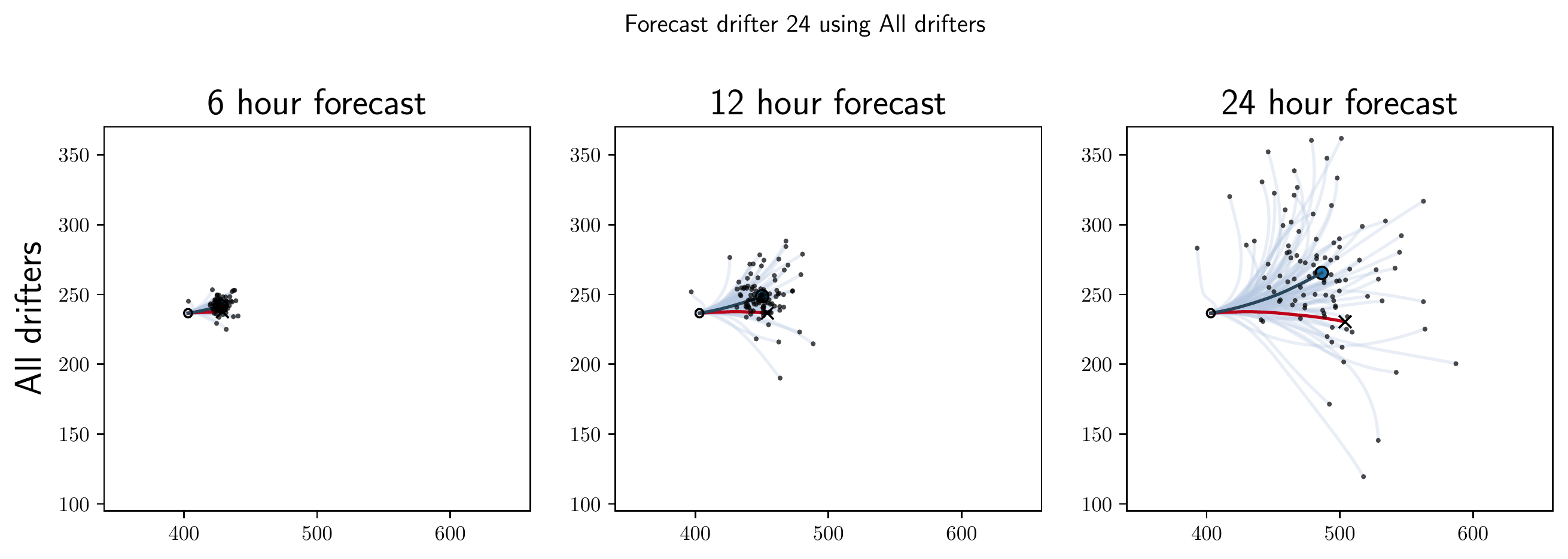}
    \end{subfigure}  
    \caption{Short-range ensemble drift trajectory forecasts after six, twelve, and 24 hours for drifter 24, using no data assimilation (top row), observations from ten drifters (middle row), and observations from all 64 drifters (bottom row). Trajectories from each ensemble member is shown as a light blue line ending in a small black circle, whereas the dark blue lines represent the ensemble mean. The red line ending in an x is the true drift trajectory. The values along the $x$- and $y$-axes are given in km, and only the relevant part of the domain is considered. The forecast at six hours is greatly improved by using observations from ten drifters, but the advantage is almost lost after 24 hours. Further improvements are made using observations from all 64 drifters.
    Even though the ensemble mean is perfectly on top of the true trajectory in the experiment without data assimilation, the spread is very large.}
    \label{fig:driftTrajectoryForecastDrifter24Short}
\end{figure*}

We now turn to drift trajectory forecasting experiments and compare how well the ensemble manages to represent the truth using different sets of observations.
 We start by investigating how well the ensemble is able to capture the true state at day ten (see \reffig{doubleJet_tenDays}), by looking at the ensemble means in \reffig{ensembleMeans}, and ensemble variances in \reffig{ensembleVars}.
We then proceed to look at the drift trajectory forecasts for two selected drifters, shown in Figures \ref{fig:driftTrajectoryForecastDrifter24Short}--\ref{fig:driftTrajectoryForecastDrifter02Long}, in which the dark lines illustrate the drift trajectory obtained in the generated truth.
All Figures \ref{fig:ensembleMeans}--\ref{fig:driftTrajectoryForecastDrifter02Long} are organized so that each row corresponds to a single experiment.
In \reffig{ensembleMeans} and \ref{fig:ensembleVars}, the columns represent the ensemble mean/variance of state variables $\eta$, $hu$ and $hv$, from left to right, whereas in Figures \ref{fig:driftTrajectoryForecastDrifter24Short} - \ref{fig:driftTrajectoryForecastDrifter02Long} each column shows forecasts for a given time range.

With 64 drifters available, it is infeasible to show and discuss forecast results for all of them, and we have therefore chosen to illustrate the results using two different drifters.
Drifter number 24 represents a drifter that is located in an area still dominated by one of the jets, and its true drift trajectory follows a smooth path with high velocity.
For this drifter, we look at both the long-term and short-term ensemble forecasts in Figures \ref{fig:driftTrajectoryForecastDrifter24Short} and \ref{fig:driftTrajectoryForecastDrifter24Long}, respectively. 
In the end of this experiment section, we show that this drifter is representative for the majority of drifters.
Drifter number 2, on the other hand, represents a particularly difficult drifter to forecast, as it was at rest at the start of the forecast, before changing direction.
Its long-term forecasts are shown in \reffig{driftTrajectoryForecastDrifter02Long}.

\newcommand{\meanwidth}{0.83}
\begin{figure*}[t!]
    \centering
    
    \begin{subfigure}[t]{\meanwidth\textwidth}
        \centering
        \includegraphics[width=\linewidth, trim=0 0.2cm 0 1.45cm, clip]{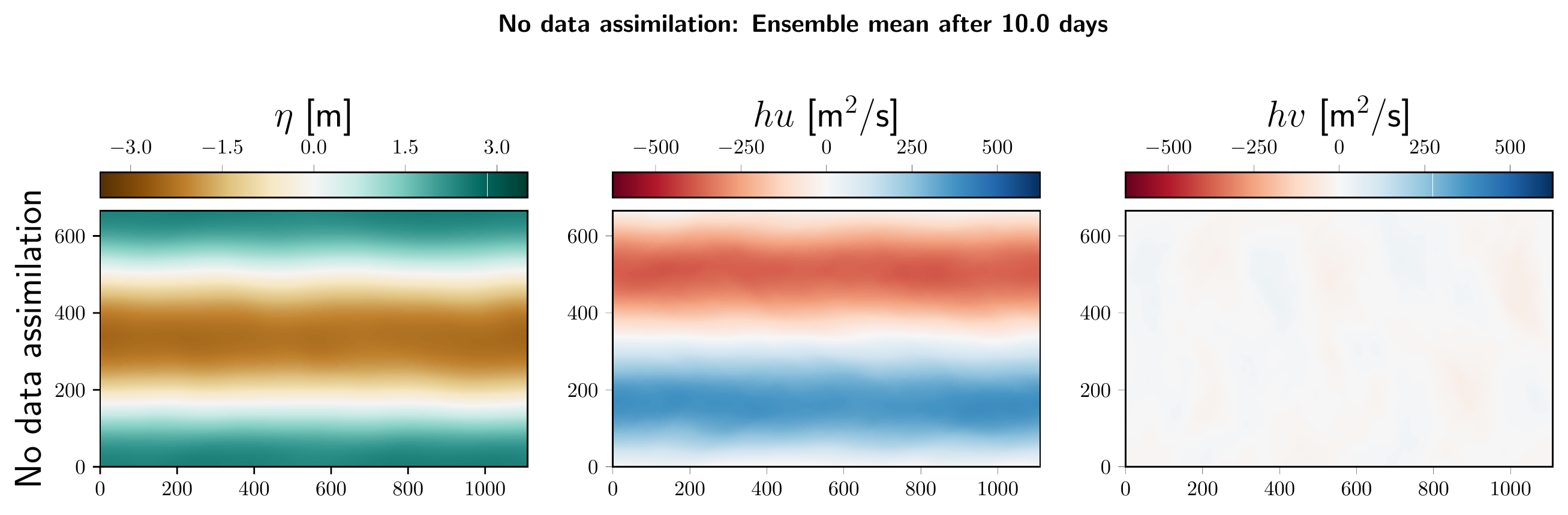}
    \end{subfigure}%
	\\
	\begin{subfigure}[t]{\meanwidth\textwidth}
        \centering
    	\includegraphics[width=\linewidth, trim=0 0.2cm 0 3.4cm, clip]{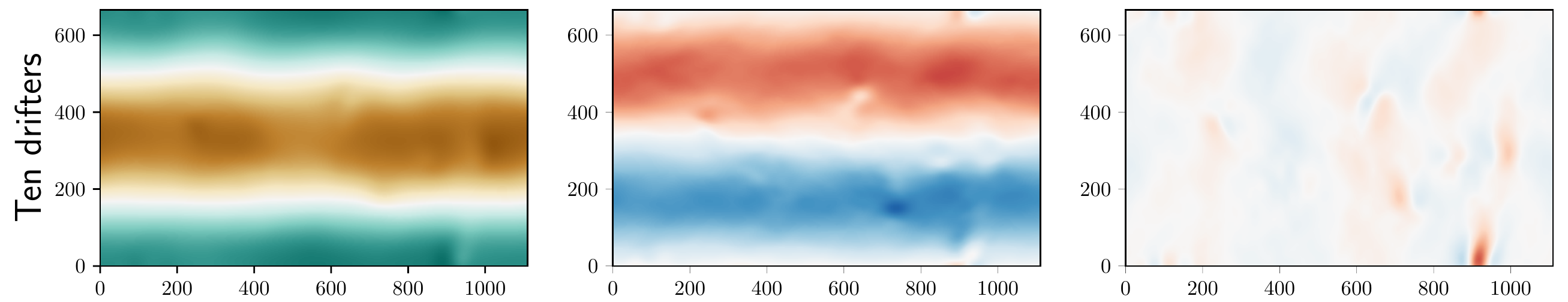}
    \end{subfigure}    
    \\
	\begin{subfigure}[t]{\meanwidth\textwidth}
        \centering
    	\includegraphics[width=\linewidth, trim=0 0.2cm 0 3.4cm, clip]{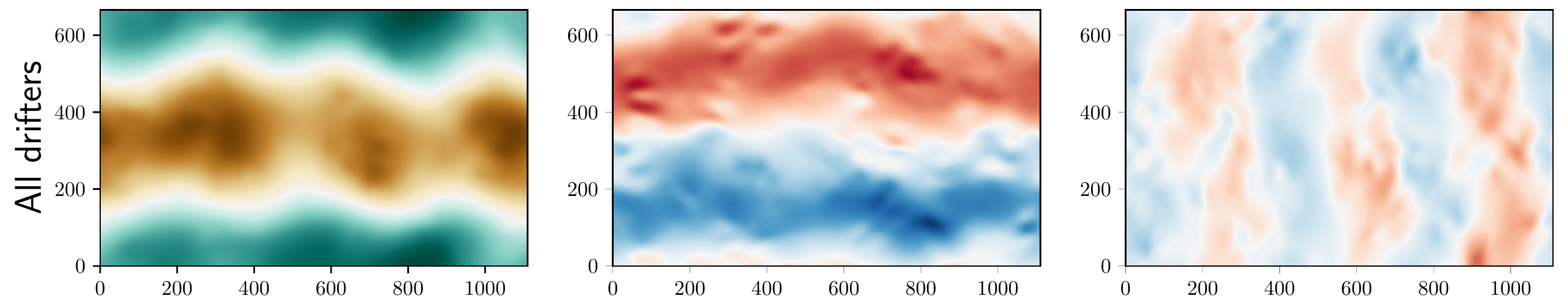}
    \end{subfigure}  
    \\
	\begin{subfigure}[t]{\meanwidth\textwidth}
        \centering
    	\includegraphics[width=\linewidth, trim=0 0.2cm 0 3.4cm, clip]{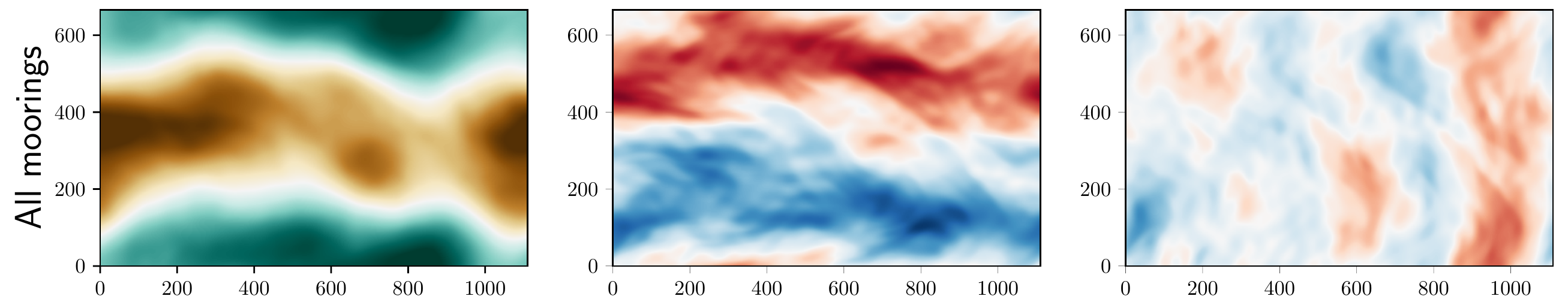}
    \end{subfigure}  
    \\
	\begin{subfigure}[t]{\meanwidth\textwidth}
        \centering
    	\includegraphics[width=\linewidth, trim=0 0.2cm 0 3.4cm, clip]{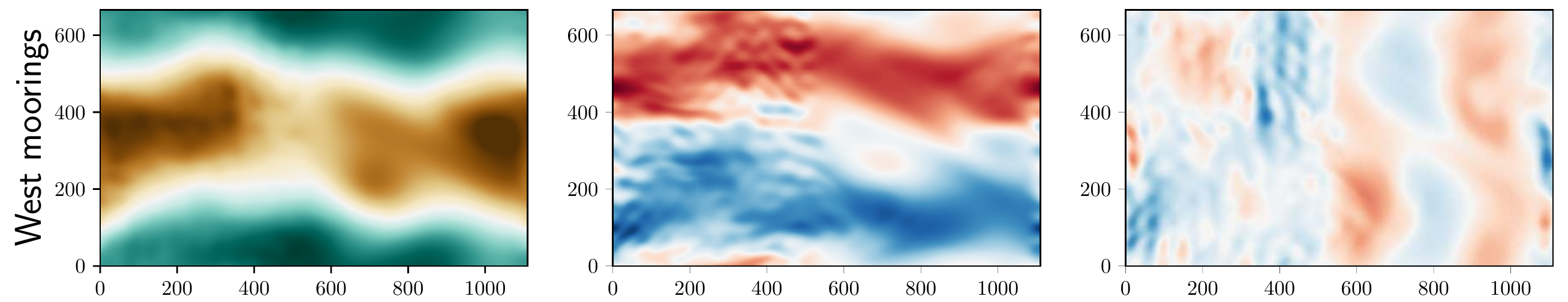}
    \end{subfigure}  
    \\
	\begin{subfigure}[t]{\meanwidth\textwidth}
        \centering
    	\includegraphics[width=\linewidth, trim=0 0.2cm 0 3.4cm, clip]{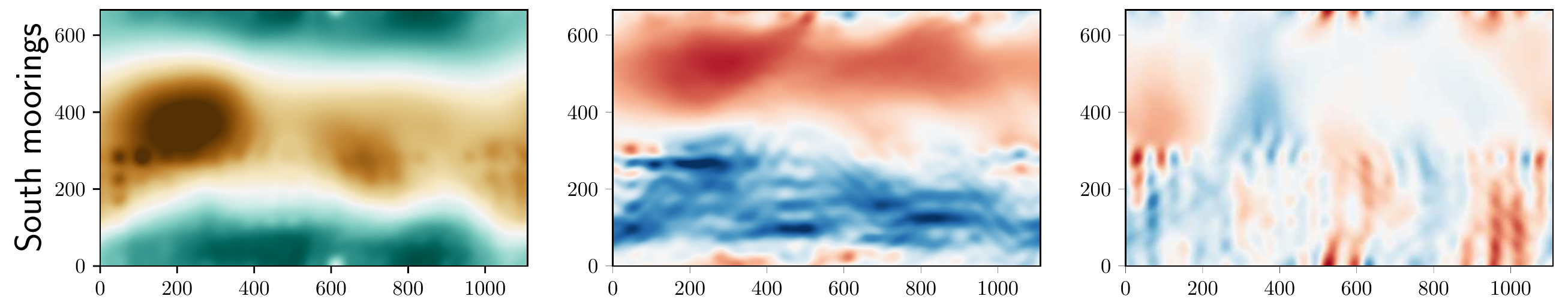}
    \end{subfigure}  
    \caption{Ensemble means for the state variables sea-surface level ($\eta$), jet flow ($hu$), and cross-jet flow ($hv$) at simulation day ten, when the data-assimilation period ends and the forecasting starts. The rows represent the six forecast experiments.
    The $x$- and $y$-axes are in km, and all figures cover the entire computational domain. The top row illustrates the chaotic nature of the test case, as the experiment without using data-assimilation results in a steady-state ensemble mean. Through observations from drifters, some localized details are captured, as seen in rows two and three. The ensemble mean from the experiment using all mooring observations in row four gives a very good representation of the true state. The final two rows show the impact of flow-dependent information transport, as only half of the domain is observed in these experiments.}
    \label{fig:ensembleMeans}
\end{figure*}

\begin{figure*}[t!]
    \centering
    
    \begin{subfigure}[t]{\meanwidth\textwidth}
        \centering
        \includegraphics[width=\linewidth, trim=0 0.2cm 0 1.45cm, clip]{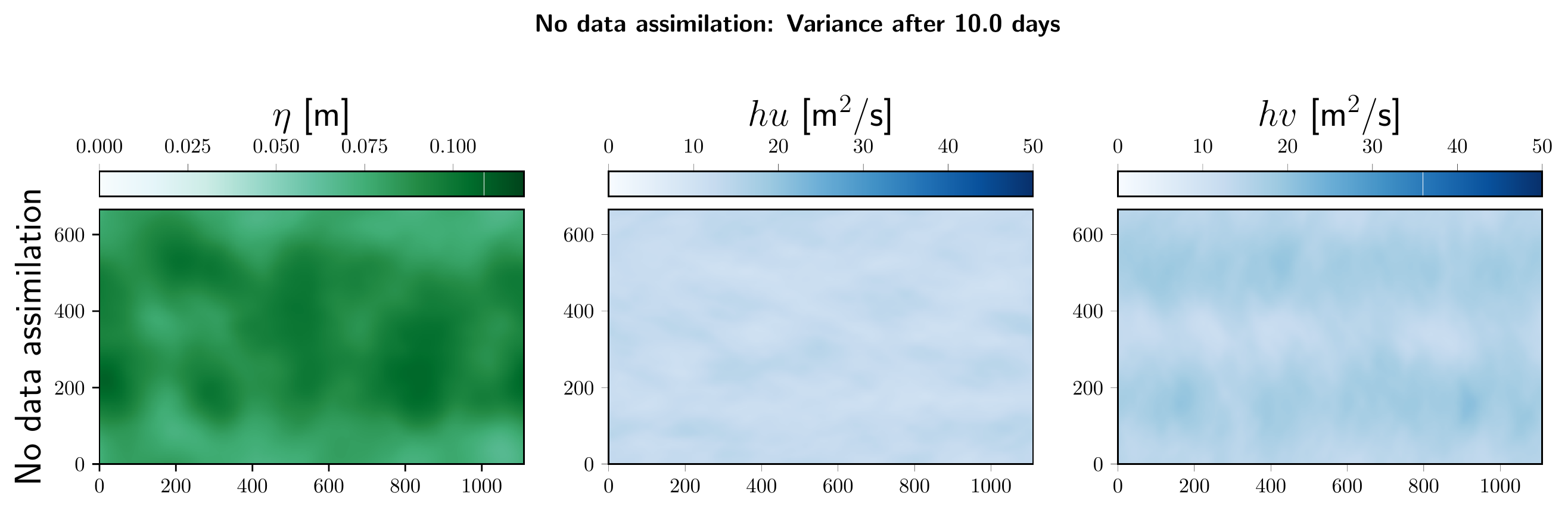}
    \end{subfigure}%
	\\
	\begin{subfigure}[t]{\meanwidth\textwidth}
        \centering
    	\includegraphics[width=\linewidth, trim=0 0.2cm 0 3.4cm, clip]{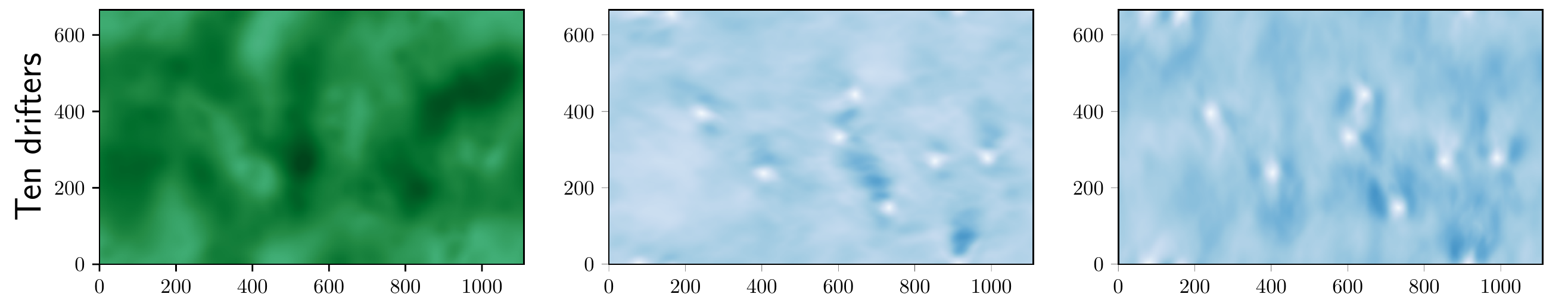}
    \end{subfigure}    
    \\
	\begin{subfigure}[t]{\meanwidth\textwidth}
        \centering
    	\includegraphics[width=\linewidth, trim=0 0.2cm 0 3.4cm, clip]{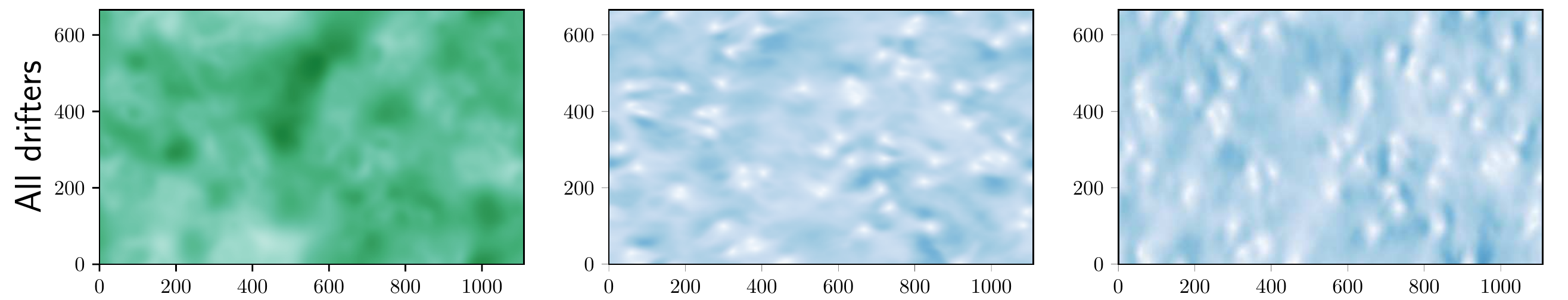}
    \end{subfigure}  
    \\
	\begin{subfigure}[t]{\meanwidth\textwidth}
        \centering
    	\includegraphics[width=\linewidth, trim=0 0.2cm 0 3.4cm, clip]{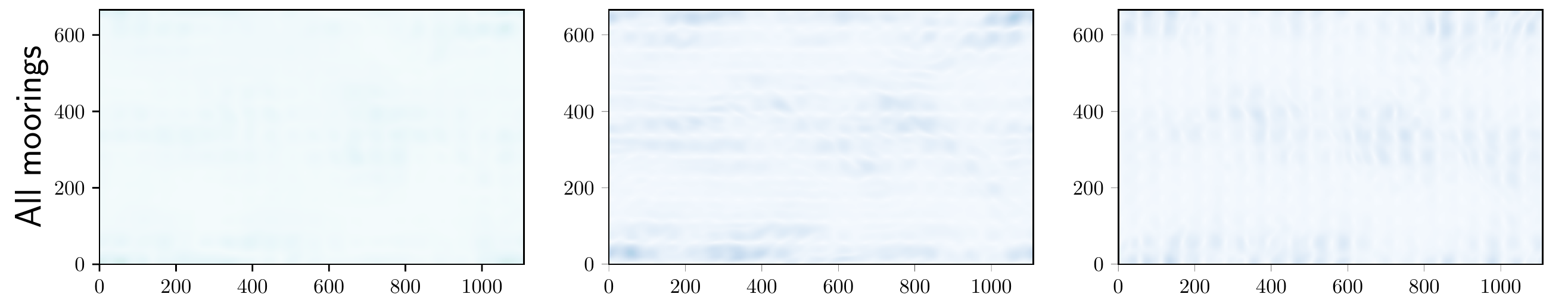}
    \end{subfigure}  
    \\
	\begin{subfigure}[t]{\meanwidth\textwidth}
        \centering
    	\includegraphics[width=\linewidth, trim=0 0.2cm 0 3.4cm, clip]{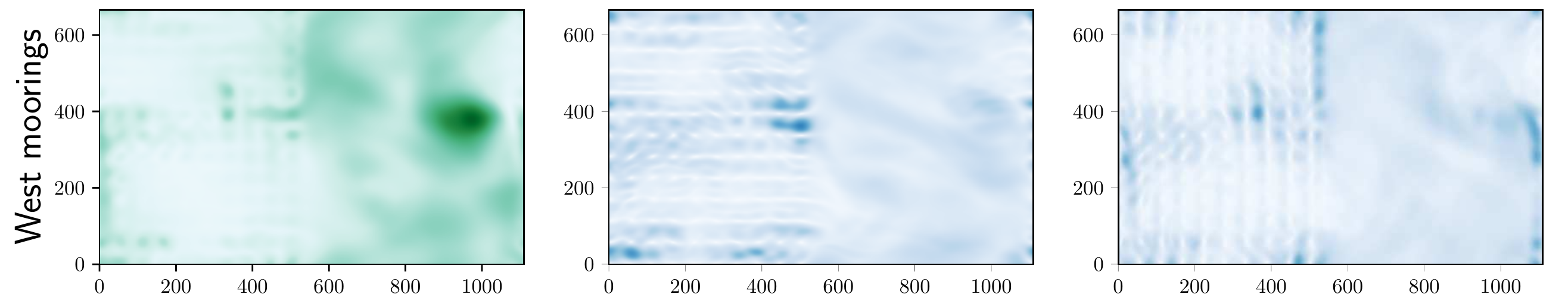}
    \end{subfigure}  
    \\
	\begin{subfigure}[t]{\meanwidth\textwidth}
        \centering
    	\includegraphics[width=\linewidth, trim=0 0.2cm 0 3.4cm, clip]{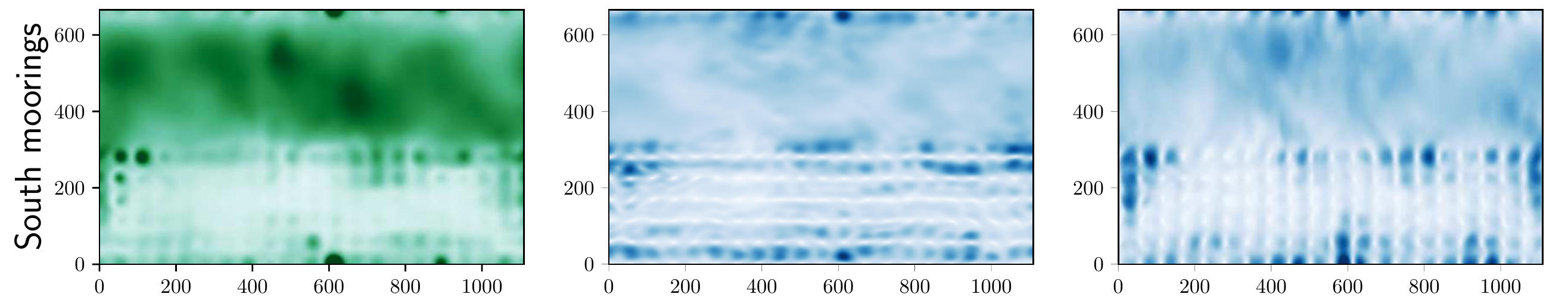}
    \end{subfigure}  
    \caption{Ensemble variance for state variables sea-surface level ($\eta$), jet flow ($hu$), and cross-jet flow ($hv$) at simulation day ten, when the data-assimilation period ends and the forecasting starts. The rows represent the six forecast experiments.
    All $x$- and $y$-axes are in km, and cover the entire computational domain. The top row shows almost equal variance throughout the domain when no data assimilation is applied.
    In rows two and three, the assimilated drifter observations can be clearly seen as local areas with low variance.
    Row four uses observations from all moorings, resulting in very low variance throughout the domain.
    The variance increases between the moorings when only half of the them are observed, as seen in rows five and six. 
    We also see that there is a large benefit in observing parts of both jets, contrary to observing one jet fully, as the variance is lower in the fifth row than in the sixth.}
    \label{fig:ensembleVars}
\end{figure*}

\newcommand{\forecastwidth}{0.88}
\begin{figure*}[t!]
    \centering
    
    \begin{subfigure}[t]{\forecastwidth\textwidth}
        \centering
        \includegraphics[width=\linewidth, trim=0 0.2cm 0 2.5cm, clip]{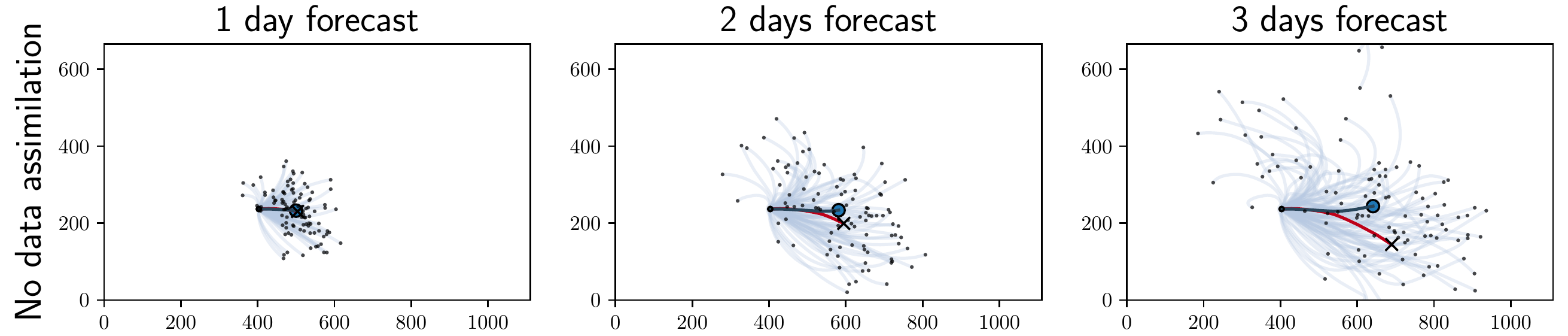}
    \end{subfigure}%
	\\
	\begin{subfigure}[t]{\forecastwidth\textwidth}
        \centering
    	\includegraphics[width=\linewidth, trim=0 0.2cm 0 3.2cm, clip]{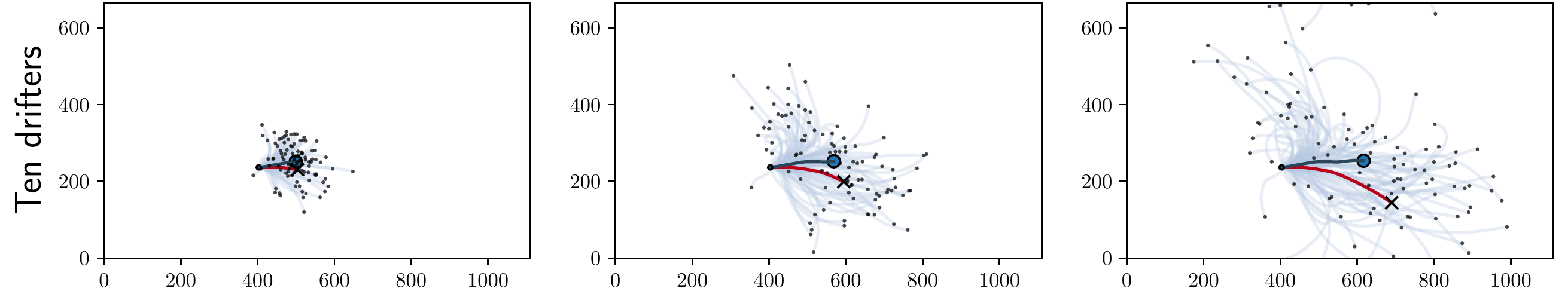}
    \end{subfigure}    
    \\
	\begin{subfigure}[t]{\forecastwidth\textwidth}
        \centering
    	\includegraphics[width=\linewidth, trim=0 0.2cm 0 3.2cm, clip]{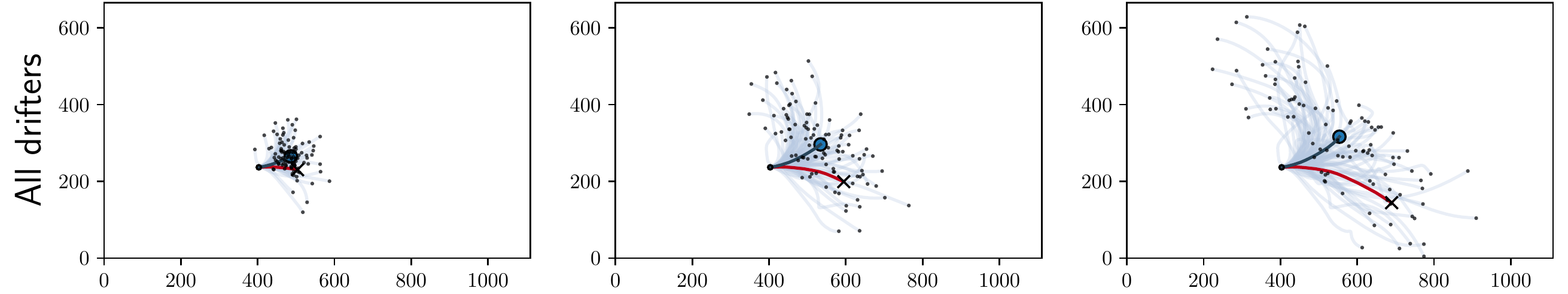}
    \end{subfigure}  
    \\
	\begin{subfigure}[t]{\forecastwidth\textwidth}
        \centering
    	\includegraphics[width=\linewidth, trim=0 0.2cm 0 3.2cm, clip]{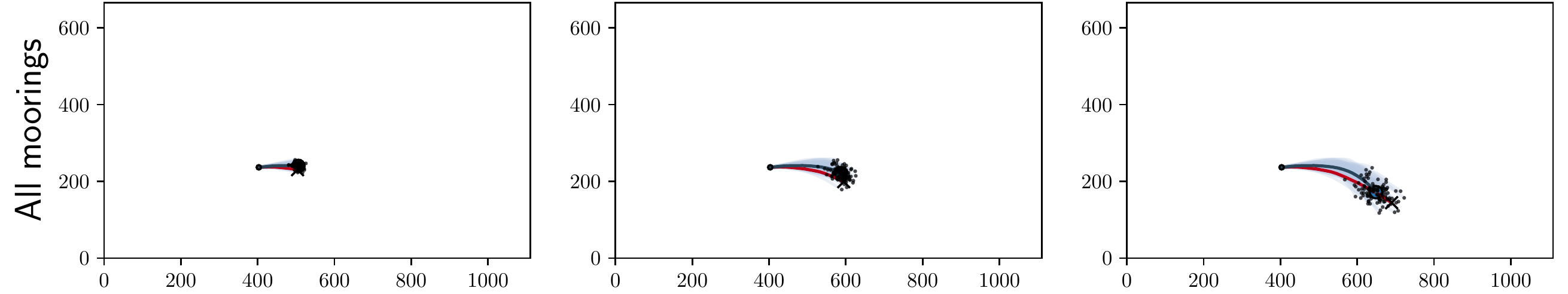}
    \end{subfigure}  
    \\
	\begin{subfigure}[t]{\forecastwidth\textwidth}
        \centering
    	\includegraphics[width=\linewidth, trim=0 0.2cm 0 3.2cm, clip]{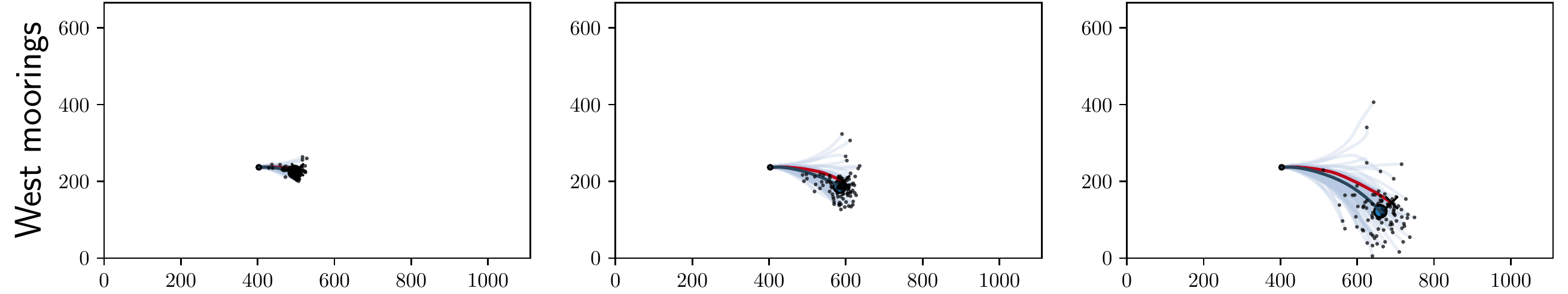}
    \end{subfigure}  
    \\
	\begin{subfigure}[t]{\forecastwidth\textwidth}
        \centering
    	\includegraphics[width=\linewidth, trim=0 0.2cm 0 3.2cm, clip]{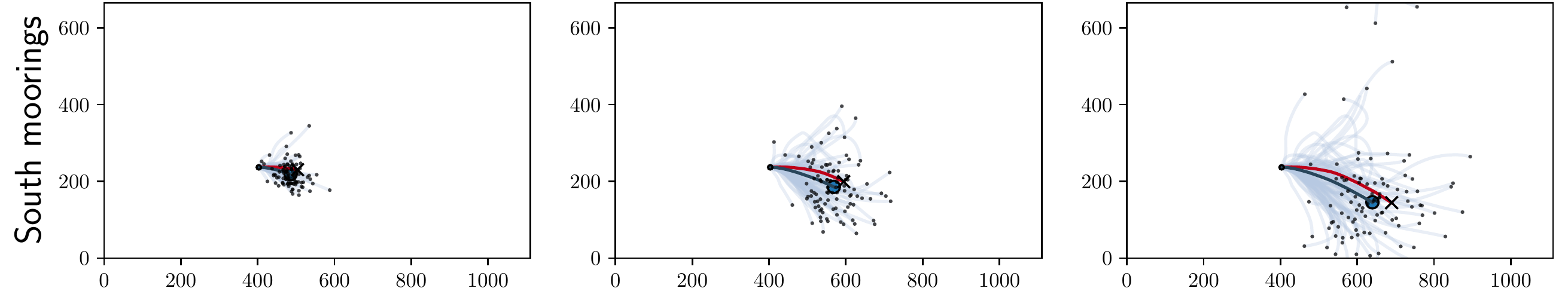}
    \end{subfigure}  
    \caption{Long-range drift trajectory forecasts after one, two and three days for drifter 24. Each row corresponds to a different set of observations.
    Each figure shows the entire computational domain, with values in km on both axes.
    Forecast trajectories from each ensemble member is shown as a light blue line ending with a small black circle, the dark blue lines represent the ensemble means, and the red lines ending in x are the true drift trajectory.
    In this time range, observations obtained from the drifters are of limited value, as the top three rows are qualitatively similar.
    The use of mooring observations in the fourth row, however, makes the forecast very accurate even for as long as three days.
    The two last rows show experiments using mooring observations from half the domain, and illustrates the benefit by partly observing both jets (west moorings), compared to fully observing one jet (south moorings).}
    \label{fig:driftTrajectoryForecastDrifter24Long}
\end{figure*}

\newcommand{\forecastwidthtwo}{0.85}
\begin{figure*}[t!]
    \centering
    
    \begin{subfigure}[t]{\forecastwidthtwo\textwidth}
        \centering
        \includegraphics[width=\linewidth, trim=0 0.2cm 0 2.5cm, clip]{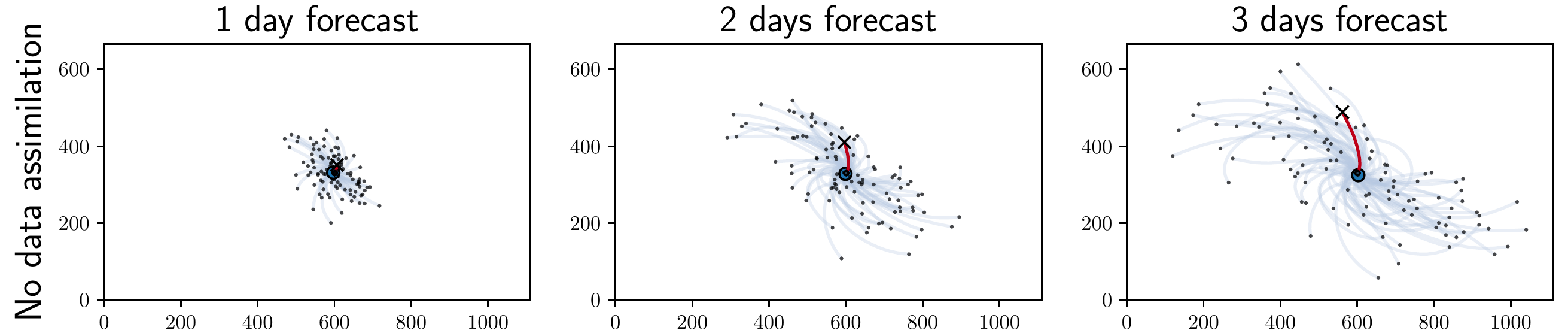}
    \end{subfigure}%
	\\
	\begin{subfigure}[t]{\forecastwidthtwo\textwidth}
        \centering
    	\includegraphics[width=\linewidth, trim=0 0.2cm 0 3.2cm, clip]{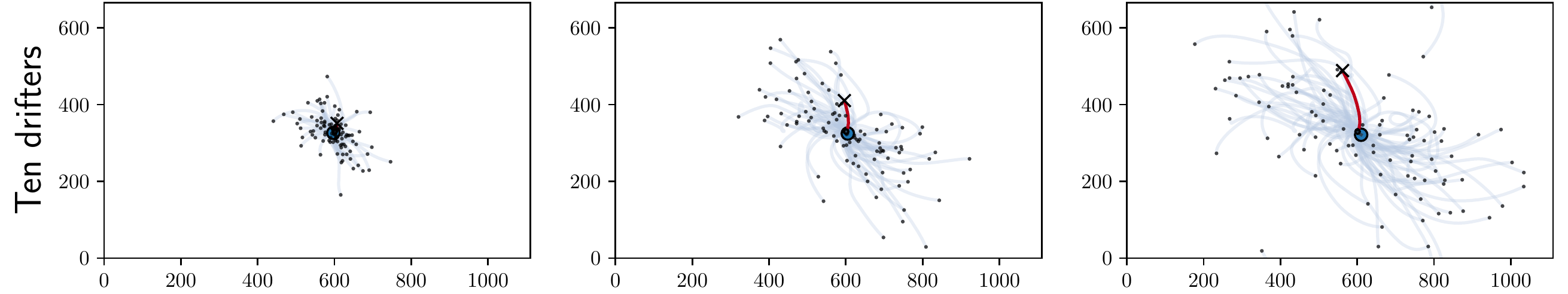}
    \end{subfigure}    
    \\
	\begin{subfigure}[t]{\forecastwidthtwo\textwidth}
        \centering
    	\includegraphics[width=\linewidth, trim=0 0.2cm 0 3.2cm, clip]{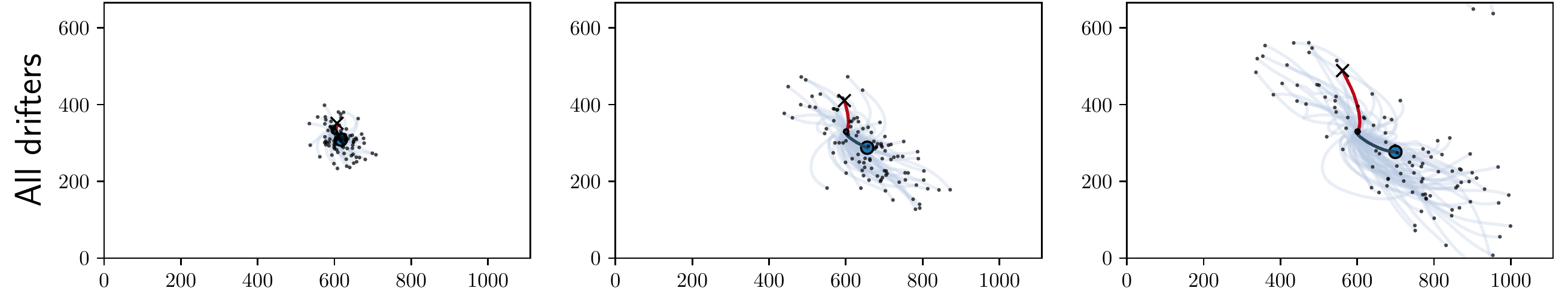}
    \end{subfigure}  
    \\
	\begin{subfigure}[t]{\forecastwidthtwo\textwidth}
        \centering
    	\includegraphics[width=\linewidth, trim=0 0.2cm 0 3.2cm, clip]{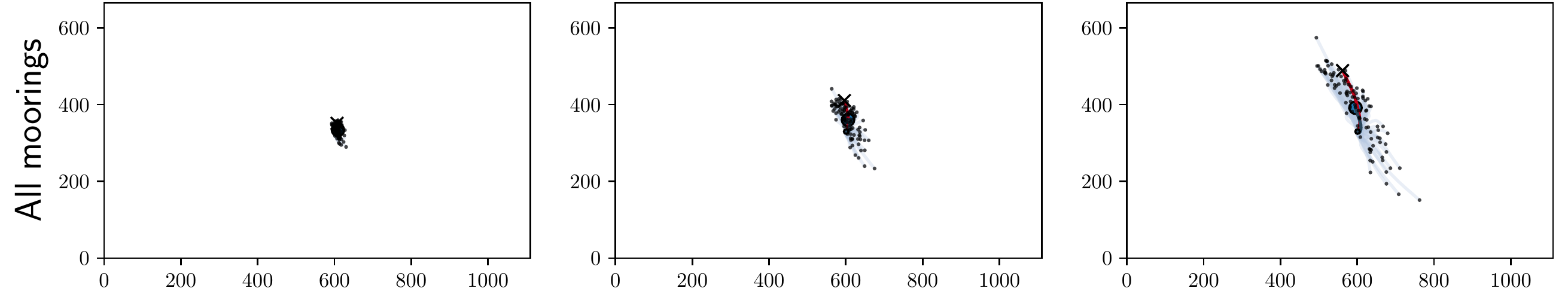}
    \end{subfigure}  
    \\
	\begin{subfigure}[t]{\forecastwidthtwo\textwidth}
        \centering
    	\includegraphics[width=\linewidth, trim=0 0.2cm 0 3.2cm, clip]{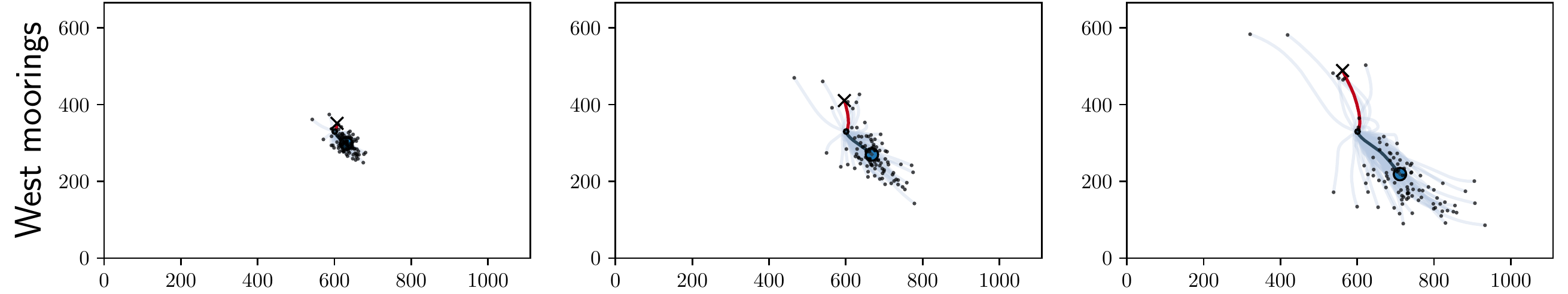}
    \end{subfigure}  
    \\
	\begin{subfigure}[t]{\forecastwidthtwo\textwidth}
        \centering
    	\includegraphics[width=\linewidth, trim=0 0.2cm 0 3.2cm, clip]{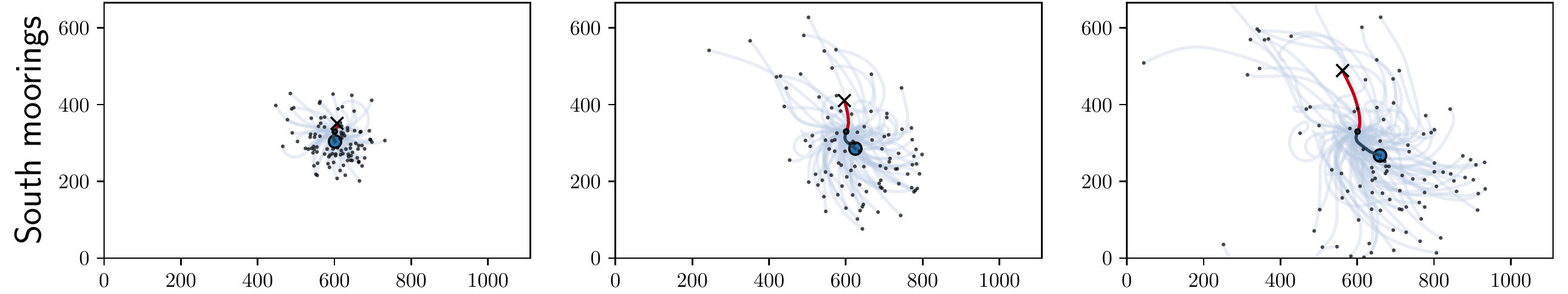}
    \end{subfigure}  
    \caption{Long-term drift trajectory forecast after one, two and three days for drifter 2. 
    This drifter is particularly hard to forecast, as it is at a complete stop while changing direction at the start of the forecast. 
    Each figure shows the entire computational domain, with values in km on both axes.
    Forecasted trajectories from all ensemble members are shown in light blue lines ending with a small black circle, the dark blue lines represent the ensemble means, and the red lines ending in x are the true drift trajectory.
    The use of drifter observations give a limited improvement in the forecast on these time ranges, and the ensemble means in the first two experiment are static at the drifter's initial positions.
    Observations of all moorings give a large impact on the forecast quality, as seen in the fourth row, and the forecast shows a 75\% chance of the drifter drifting northwards.
    The experiments using observations from half the domain only, give forecasts that show higher probability for southward drift, but there are still a few ensemble members allowing for northwards drift in both cases.}
    \label{fig:driftTrajectoryForecastDrifter02Long}
\end{figure*}

\subsubsection{Experiment A: No data assimilation}
The first experiment is a pure Monte Carlo forecasting experiment, in which all ensemble members run independently without any knowledge of the truth.
Without any observations to guide the ensemble, the entire space of possible model states that can be reached from the instable initial conditions are explored, and the ensemble mean at day ten shown in the top row in \reffig{ensembleMeans} illustrates the chaotic nature of the chosen test case.
The ensemble mean is almost completely smooth, dominated by two opposing jets in the $x$-direction and almost no action for $hv$, and resembles a smoothed version of the initial conditions.
The ensemble variance, shown in the top row of \reffig{ensembleVars}, is more or less the same all over the domain.

In the top row of \reffig{driftTrajectoryForecastDrifter24Short}, we see that the ensemble forecast suggests that drifter 24 is heading eastward. The ensemble mean trajectory accurately describes the true trajectory, but the variance in the ensemble is very large, indicating high uncertainty in the contribution of north/south currents. 
As time goes on, the forecast diverges to cover a large portion of the entire domain, as shown in the three day forecast in \reffig{driftTrajectoryForecastDrifter24Long}.
For drifter 2 the one day forecast in \reffig{driftTrajectoryForecastDrifter02Long} suggests that there are possible drift trajectories in directions, resulting in an ensemble mean that is statically located at the drifter's initial position.
As the forecasted trajectories hit the dominant jets, the long-term forecast covers the diagonal from the northwest to the southeast corner of the domain.

\subsubsection{Experiment B: Assimilating data from ten drifters}
We now observe the locations of ten drifters, and assimilate the underlying currents based on their movements.
The ten drifters are hand-picked to cover as large portion of the domain as possible, and both drifter 2 and drifter 24 are included.
The second row of \reffig{ensembleMeans} shows the ensemble mean at day ten, which is more similar to the mean obtained using no data assimilation than the true state itself (see \reffig{doubleJet_tenDays}).
There are however some localized patches of features in the mean for $hu$, corresponding to the last observed drifter positions.
By looking at the model state variance in the second row of \reffig{ensembleVars}, the latest drifter positions can be clearly identified by the areas of very low variance in $hu$ and $hv$.
Note, however, that the highest values of ensemble variance are also found close to the drifters.
As we assimilate an observed current by adding a correcting local dipole at the drifter locations, we might introduce a larger error close to the drifter as a side effect, where opposite currents are needed for maintaining the geostrophic balance.
It is also possible to spot traces of the drifters' movements from the patterns in the variance plots, as some of them have tails of low or high variance.

In the short-term forecast in the second row of \reffig{driftTrajectoryForecastDrifter24Short}, we see great improvement in the six hour forecast over the assimilation-free forecast, as almost the entire ensemble of drift trajectories starts moving straight eastwards with lower spread.
The forecast after 12 hours is improved  as well, but when turning to the long-term forecast in \reffig{driftTrajectoryForecastDrifter24Long}, it becomes harder to see any significant differences in the forecast quality. 
The same applies to the forecast for drifter 2 in \reffig{driftTrajectoryForecastDrifter02Long}, again with a non-moving ensemble mean trajectory.

\subsubsection{Experiment C: Assimilating data from all 64 drifters}
By using observations from all 64 drifters, we see a large change in the ensemble mean at day ten, presented in the third row of \reffig{ensembleMeans}.
The border between the mean eastward and westward jets in $hu$ is no longer a straight line, and there are more features seen for $hv$.
Even though some of the features resemble the truth, such as the shape of the main parts of the eastward current and the location of the north and south bands in $hv$, there are other features that are less correct, e.g., the continuity of the north and south bands in $hv$.
As in experiment B, the drifter locations can be seen from the variance of $hu$ and $hv$, in the third row of \reffig{ensembleVars}, and we also note that there are larger areas between the drifters with lower variance than before.
This indicates that with an increased number of drifters we are able to improve the model state in a larger portion of the domain. 

The short-term forecasts for drifter 24 in the third row of \reffig{driftTrajectoryForecastDrifter24Short} have slightly lower spread compared to using ten drifters only, as could be expected.
At six hours, the forecast is quite confident in the location of the drifter, and there is less uncertainty in the twelve and 24 hour forecasts as well.
By looking at the long-term forecast in \reffig{driftTrajectoryForecastDrifter24Long}, however, the quality drops and is again comparable to the previous two experiments.
However, we see that the ensemble mean trajectory is no longer static, showing that the majority of ensemble members move south-east.

\subsubsection{Experiment D: Assimilating data from all 240 moorings}
In this experiment, we assimilate observations from all the 240 moorings that are placed equidistantly throughout the domain.
The distances between the moorings are 55 km, corresponding to 25 grid cells.
Even though we observe only approximately 0.1\% of the state variable, the observations are dense enough for the covariance structures from two neighbouring observations to be overlapping, meaning that the observational coverage is quite good.
This is also seen in the ensemble mean after ten days in the fourth row of \reffig{ensembleMeans}, which is almost indistinguishable from the true state shown in \reffig{doubleJet_tenDays}.
Even the unobserved variable $\eta$ seems to be correctly captured by the ensemble.
The variance is very low throughout the domain, as seen in the forth row of \reffig{ensembleVars}.

The trajectory for drifter 24 shown in \reffig{driftTrajectoryForecastDrifter24Long} is very good, with a very confident forecast and accurate ensemble mean trajectory even at day three.
The trajectory of all ensemble members show the same general characteristics by a steady eastern flow with a southward bend, disagreeing only slightly on the strength of these currents.
The forecast for the challenging drifter 2 in \reffig{driftTrajectoryForecastDrifter02Long} is also much more accurate than the previous three experiments, but the forecast has a higher spread than for drifter 24 at day three.
Most ensemble members stay close to the initial position during the first day. 
The forecast is then divided, with approximately 75\% of the drifters moving northwest, and the last quarter moving southeast.
We see that the true trajectory is found among the most likely outcome, to the north.
Note that we do not show the short-term forecast trajectories for this and the following two experiments, since the long-term forecasts in  Figures \ref{fig:driftTrajectoryForecastDrifter24Long} and  \ref{fig:driftTrajectoryForecastDrifter02Long} show sufficient information to discuss their results.

\subsubsection{Experiment E: Assimilating data from moorings in only the western half of the domain}
The last two experiments explore how well the ensemble mean is able to represent the true state if observations come from only half of the domain.
We start by using the 120 moorings in the western half only, resulting in the ensemble mean and variance shown in the fifth rows of Figures \ref{fig:ensembleMeans} and \ref{fig:ensembleVars}, respectively.
The first thing to notice is that the ensemble mean appears to be less smooth than the true state in the observed area.
These slightly noisy features are most dominant in the southwest and northeast corners of the observed area, corresponding to where unobserved water enters the observed part of the domain.
The reason for this can be that the signal flowing into the observed area likely need a stronger correction by the data-assimilation system, compared to the signal that have been observed and corrected for some time already.
Note especially that the variance in $hv$ is higher at the jets' entry points to the observed area, compared to the rest of the observed area.
Finally, we point out that the ensemble means for both $hu$ and $hv$ seem to capture the main features of the truth in the eastern part of the domain as well, even though this area is never observed.
This is due to the transport of information that is assimilated into the system, along with the currents.

The drift trajectory forecast in \reffig{driftTrajectoryForecastDrifter24Long} is another indication of how well features are kept in the system even after the assimilation is ended.
Drifter 24 starts close to the outflow of the southernmost jet in the western half of the domain, meaning that its underlying current has been influenced by the assimilation system for some time before the start of the forecast.
This is reflected in the one-day forecast, which is almost as good as the forecast using all moorings, but the spread in the ensemble increases as we reach the two- and three-day forecasts.
Most of the ensemble members still show the correct characteristics and therefore maintain the flow characteristics even without using further observations. 
The forecast also opens up for the possibility that the true drifter can turn north instead of south, but only with a very small probability, and we see that the ensemble mean trajectory is slightly south of the truth.
For drifter number 2, however, row five of \reffig{driftTrajectoryForecastDrifter02Long} shows that the ensemble is quite confident that the drifter will move southwards.
This drifter starts just on the outside of the observed area, and between the two dominating jets.
Still, the forecast has a significantly lower spread than the experiments using all drifter observations.
The forecast for this drifter turns out to be wrong, however, as a large majority of the ensemble trajectories are towards the southeast.
The forecast does still leave a small probability for northwards drift, as we know to be the true trajectory.

\subsubsection{Experiment F: Assimilating data from moorings in only the southern half of the domain}
This time we use observations from the southern half of the domain, capturing only one of the initial jets.
These observations are not sufficient to capture the true state, as can be seen from the obtained ensemble mean in the lower row of \reffig{ensembleMeans}.
In the observed area, the mean is dominated by small-scale eddies that are not found in the truth, whereas the unobserved part of the domain hardly have any features at all. 
From the variance plots in \reffig{ensembleVars}, we see larger values than in any of the other experiments along the boundary of the observed area.
This experiment illustrates better than the previous one how the ensemble needs to make larger adjustment on the ensemble members in the outskirts of the observed parts of the domain. 
Since there is limited information transport between the observed and unobserved areas, this experiment leads to much weaker results than in Experiment E.

At the start of the drift trajectory forecast, drifter number 24 is just within the observed area, whereas drifter number 2 is just outside.
Our choice of drifters should therefore not favour one of the half-domain experiments more than the other.
Still, we see from Figures~\ref{fig:driftTrajectoryForecastDrifter24Long} and \ref{fig:driftTrajectoryForecastDrifter02Long} that the forecasts produced by observations in the southern half of the domain have a much larger spread than the forecasts made after using observations in the western half.
Particularly, the long-term trajectory characteristics of drifter 2 are very different from all previous experiments.

\subsubsection{Comparison of forecast errors}
To show that the forecast results for drifter 24 just discussed are reasonably representative for the majority of drifters, we investigate general forecasting statistics by defining a forecast error norm.
First, let the error in the ensemble forecast for drifter $d$ at time $t^n$ be defined as 
\begin{equation}
    E_{d}(t^n) = \frac{1}{N_e} \sum_{i=1}^{N_e} \left[ \left( x_{i,d}^n - x_{true, d}^n \right)^2 + \left(y_{i,d}^n - y_{true, d}^n \right)^2 \right].
    \label{eq:errorPerEnsembleDrifter}
\end{equation}
Furthermore, let the forecast error be the square root of the $E_d^n$ mean over all drifters,
\begin{equation}
    E(t^n) = \sqrt{ \frac{1}{N_D} \sum_{d=1}^{N_D} E_d(t^n)}.
    \label{eq:errorNorm}
\end{equation}
Similarly, we define the root-mean-square error \textit{RMSE}$(t^n)$ in the same fashion as \eref{errorPerEnsembleDrifter} and \eref{errorNorm}, but use the ensemble mean instead of the true drifter position. 
A low RMSE indicates low spread in the ensemble forecast, whereas a low error confirms that the true drifter location is within the low-spread forecast.
On the contrary, if the RMSE is low, but the error is large, the ensemble gives a confident but wrong forecast.  
\nomenclature{$E_{d}(t^n)$}{Ensemble forecast error for drifter $d$.}
\nomenclature{$E_{d}(t^n)$}{Mean ensemble forecast error for all drifters.}
\nomenclature{\textit{RMSE}$_{d}(t^n)$}{Ensemble mean squared forecast error for drifter $d$.}
\nomenclature{\textit{RMSE}$_{d}(t^n)$}{Ensemble root-mean-squared forecast error over all drifters.}

\begin{figure*}[t!]
    \centering
    
    \begin{subfigure}[t]{0.5\textwidth}
        \centering
        \includegraphics[width=0.95\linewidth, clip]{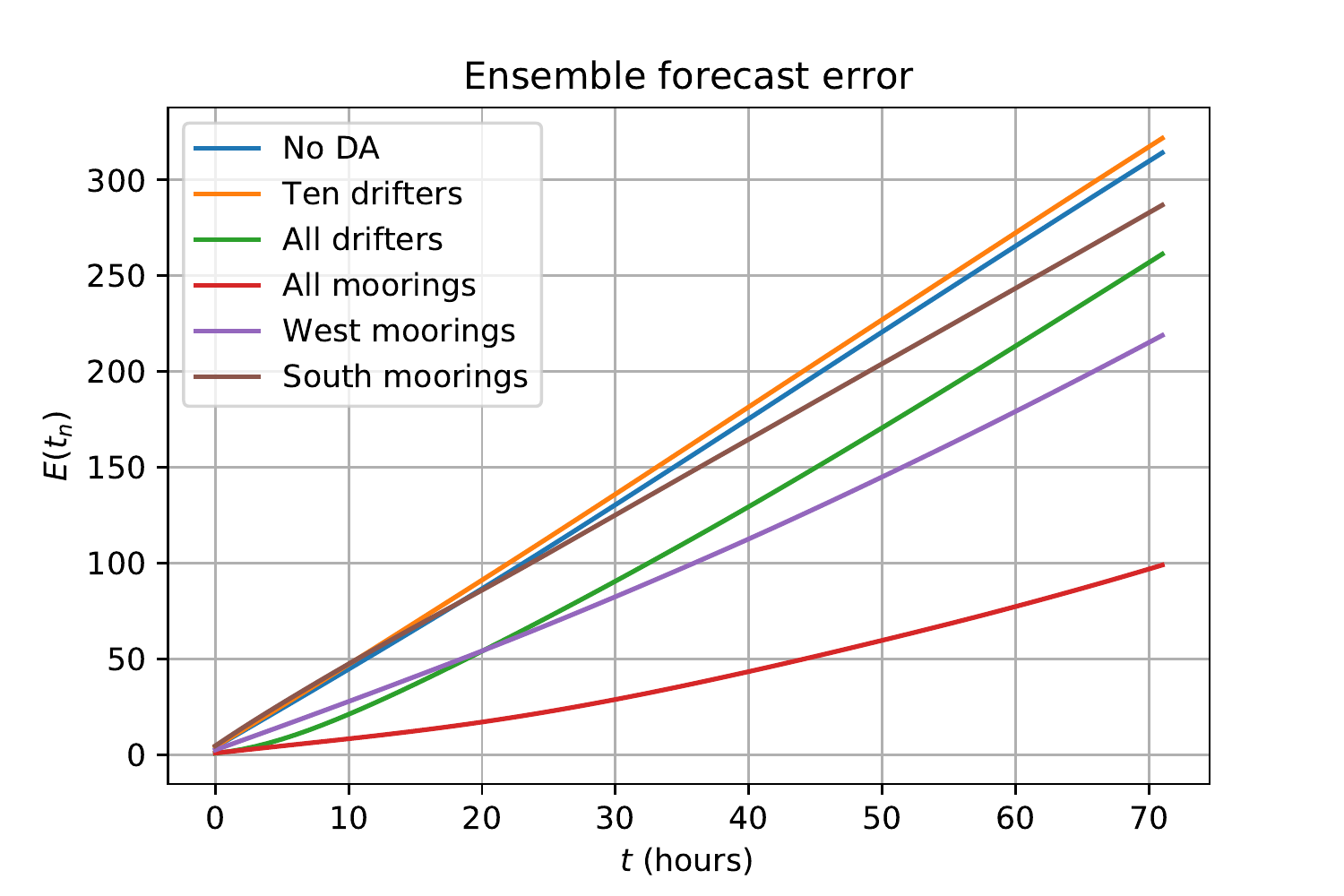}
        \caption{}
        \label{fig:meanForecastErrorAll}
    \end{subfigure}%
	~
	\begin{subfigure}[t]{0.5\textwidth}
        \centering
    	\includegraphics[width=0.95\linewidth, clip]{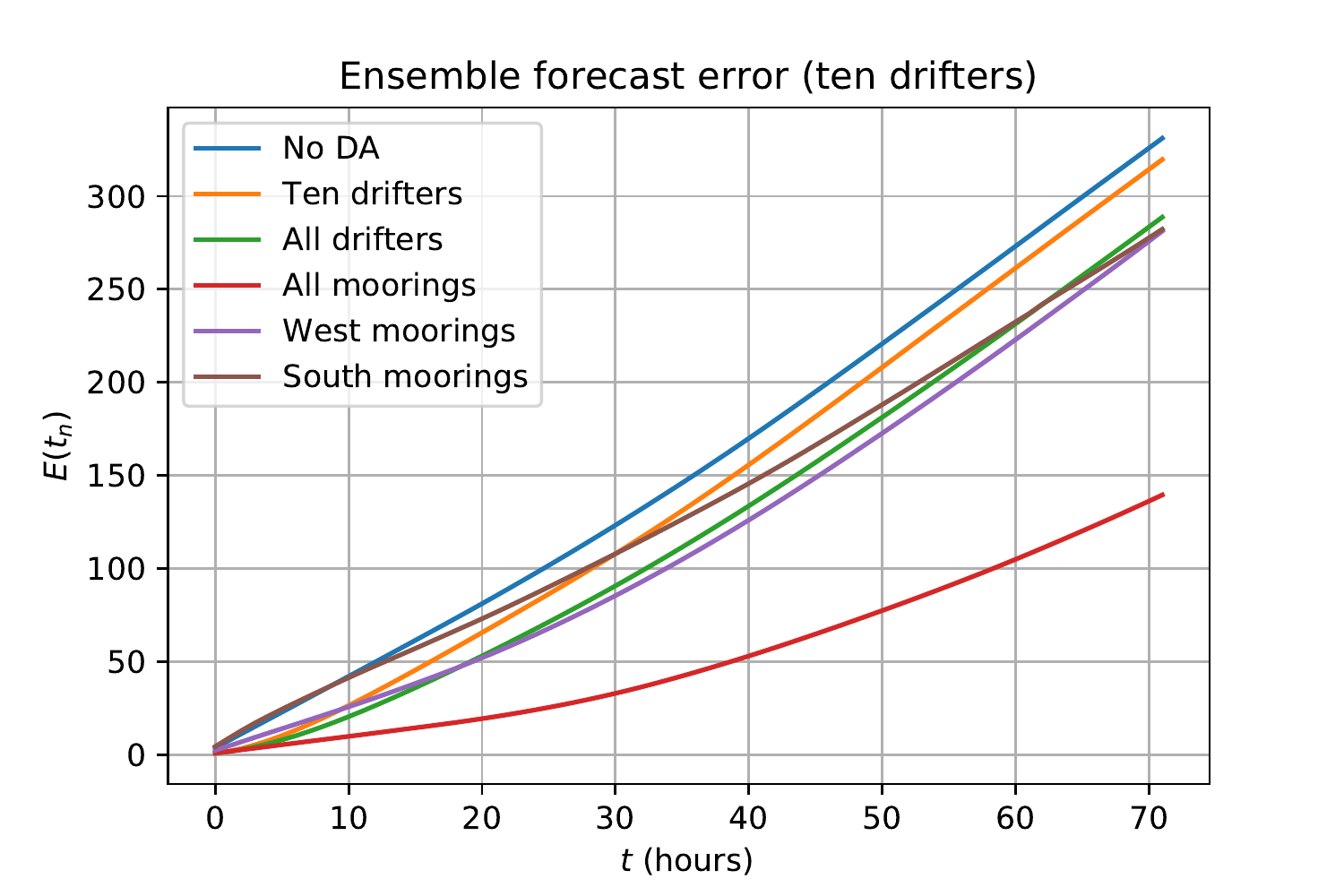}
        \caption{}
        \label{fig:meanForecastErrorDrifterSet}
    \end{subfigure}    
    \\
    \begin{subfigure}[t]{0.5\textwidth}
        \centering
        \includegraphics[width=0.95\linewidth, clip]{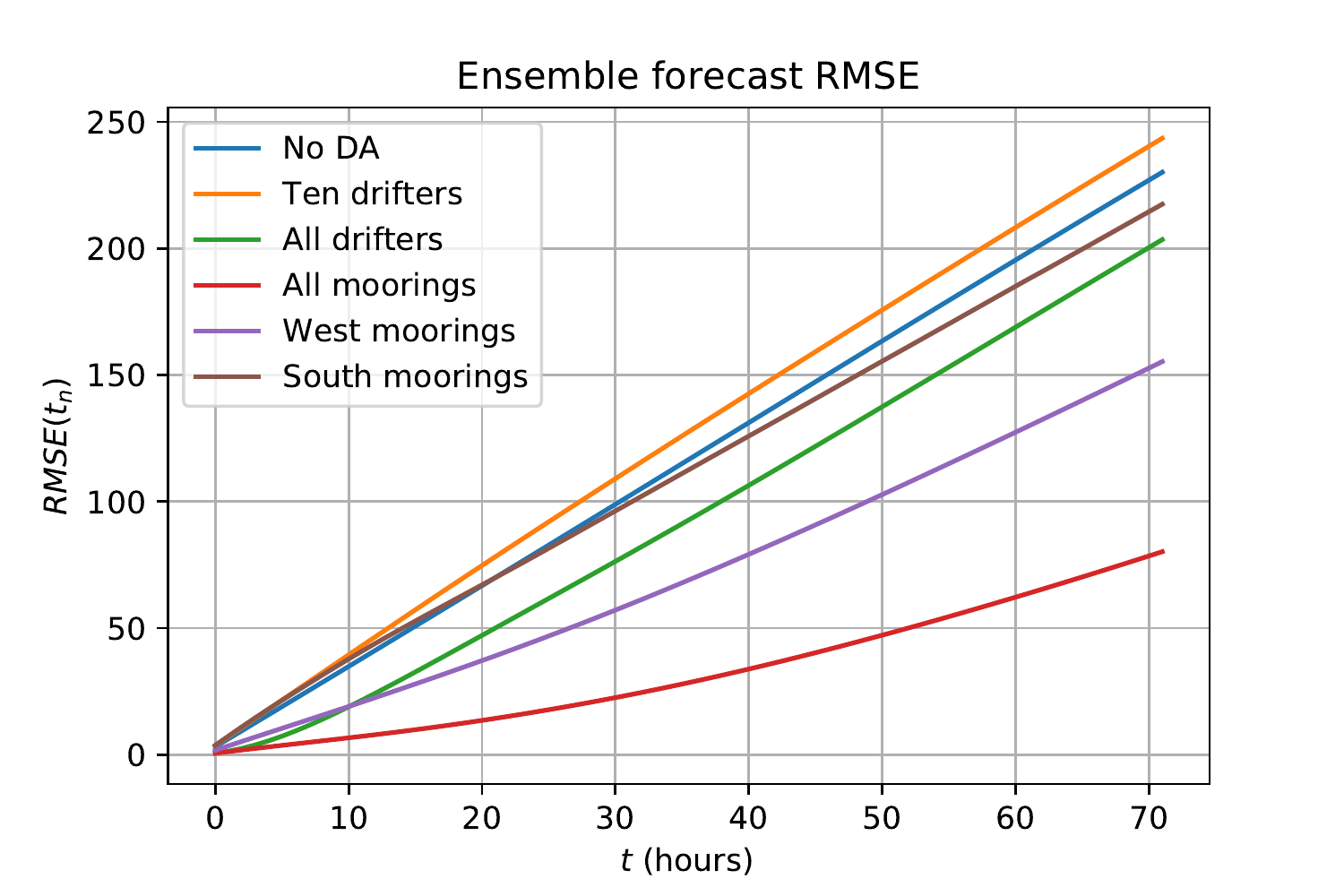}
        \caption{}
        \label{fig:meanForecastRMSEAll}
    \end{subfigure}%
	~
	\begin{subfigure}[t]{0.5\textwidth}
        \centering
    	\includegraphics[width=0.95\linewidth, clip]{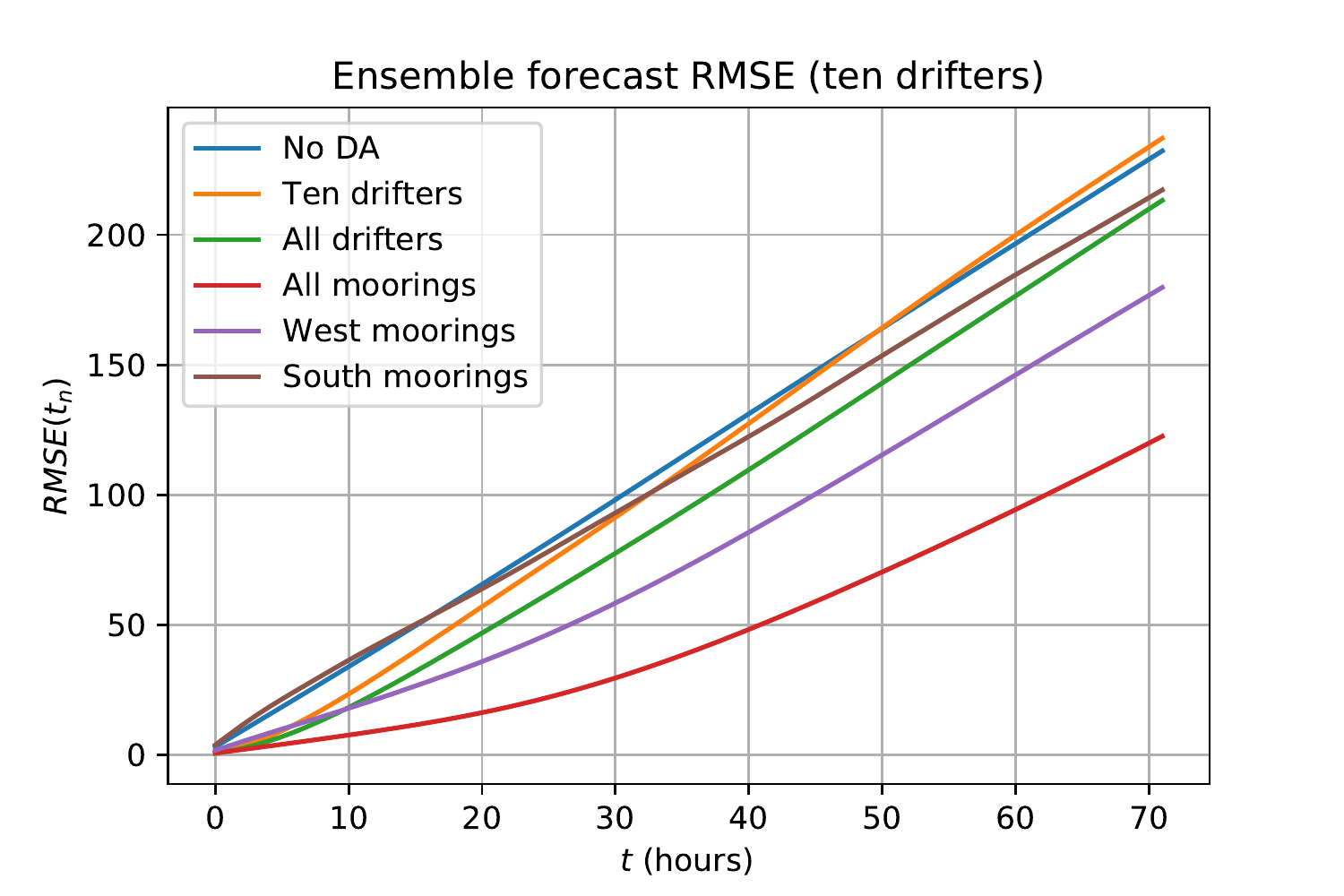}
        \caption{}
        \label{fig:meanForecastRMSEDrifterSet}
    \end{subfigure}    
    \caption{
    Mean forecast error for all six forecast experiments, considering the forecast for (a) all 64 drifters and (b) only the ten handpicked drifters used in experiment B. The second row shows RMSE as the comparable measure in terms of the forecast mean instead of the true drifter trajectory, again for (c) all 64 drifters and (d) the ten handpicked drifters only.  
    The forecast errors are lowest for the experiment using observations from all moorings, whereas observations of some drifters do not improve the forecast much compared to the forecast without data assimilation. }
    \label{fig:meanForecastErrors}
\end{figure*}

\reffig{meanForecastErrors} shows how the forecast error $E(t^n)$ and \textit{RMSE}$(t^n)$ develop over time for the six different forecast experiments.
The figures to the left (\ref{fig:meanForecastErrorAll} and \ref{fig:meanForecastRMSEAll}) shows the error when we consider all drifters, and the figures to the right (\ref{fig:meanForecastErrorDrifterSet} and \ref{fig:meanForecastRMSEDrifterSet}) consider only the ten drifters that are used in experiment B.
The figures show that good results are consistently obtained by using observations from all the moorings, as the forecast error for this experiment is significantly lower than for the others.
Also, we see that the error and the RMSE are consistent for all experiments, which means that the ensemble mean often is a good representation of the true drift trajectory.
The exception is the west moorings experiment, which has a higher error than RMSE relative to the other experiments when considering the forecast for ten drifters only.
The reason for this behavior is that the true trajectory often is forecasted as an ensemble outlier, such as seen for the west moorings experiment for drifter 2 in \reffig{driftTrajectoryForecastDrifter02Long}.
When comparing Figures \ref{fig:meanForecastRMSEAll} and \ref{fig:meanForecastRMSEDrifterSet}, we see that this behaviour is slightly over-represented among the ten selected drifters.



In general, the forecasts are best when all moorings are observed, followed by observation of moorings in the west of the domain (runner-up for long-term forecast), and observations of all drifters (runner-up for short-term forecast).
It is worth noting the differences between the two experiments that use observations from moorings in only half of the domain.
Even though both experiments use the same kind and same number of observations, the west moorings enable a much better forecast as they observe a larger portion of the information flow over time. 

Perhaps more interesting is the relationship between the forecast errors from using observations from ten drifters, and the forecast errors when using no observations. 
In \reffig{meanForecastErrorDrifterSet}, we see that using the drifters give a significantly better short-term forecast, whereas there is only a slight improvement on the long-term forecast.
When the forecast for all 64 drifters are taken into account, however, we see that the long-term forecast becomes slightly worse by using these ten observations, compared to using no data assimilation at all.
This could be due to our choice of local covariance structures, which is enforced on the ensemble through the IEWPF method.
An observation is assimilated in the ensemble members through adding a dipole that gives the correct current at the drifter position.
A side effect may be that the dipole induces a wrong current a small distance away from the drifter, causing the forecast for unobserved drifters at that location to be worse than if no data assimilation had been performed.

\subsection{Collapse of the standard particle filter}

Collapse of the standard particle filter for high-dimensional observations is well known in the literature (see~\cite{snyder2008_obstacles_highdimPF, pjvl_2009_pf_review, snyder2015_perfboundsPF_optprop}). 
To illustrate its inefficiency, we look at the weight distribution using observations from a varying number of drifters.

Normalized weights are calculated using \eref{standardParticleFilterWeight}, which depends only on the size of the innovation $\innovation^n_i$ and the observation covariance matrix $R$, and not the size of the model error covariance matrix $Q$.
Since our experiments start after a three day spin-up period, the ensemble has the highest variance during the initial data-assimilation cycles.
In the IEWPF, this corresponds to a low target weight during the first iterations, while the ensemble is gradually adjusted according to the observations.
With the standard particle filter, however, the large spread in the spin-up ensemble makes it very prone to collapse already in the first assimilation cycle. 
This is indeed what we observe, even with observations from only a single drifter.

\begin{figure}
    \begin{center}
    \includegraphics[width=0.7\linewidth]{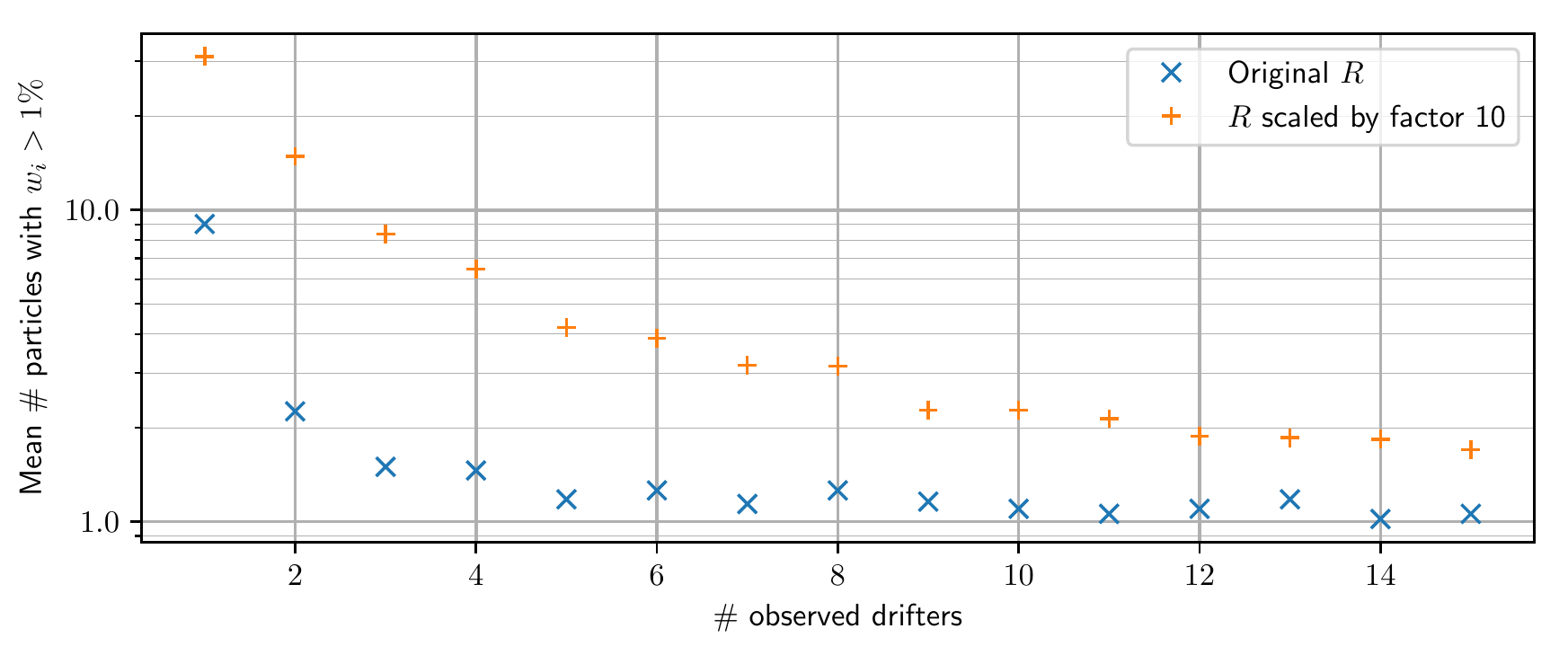}
    \caption{The number of particles guaranteed to be resampled when using a 100-member standard particle filter with residual resampling, for observations from different numbers of drifters. Blue crosses use the original weights based on the same uncertainty as the forecast experiments, and the weight is distributed on very few particles even for very low-dimensional observations. The yellow plus signs show the weight distributions assuming a tenfold increase in the observation covariance matrix $R$, but even with less reliable observations, the ensemble collapses for any more than six observed drifters. }
    \label{fig:sirCollapse}
    \end{center}
\end{figure}

To give the standard particle filter a fair chance, we run an experiment for three simulation days using the IEWPF method on observations from all 64 drifters, and thus obtain a well-distributed ensemble with a low spread and good representations of the underlying ocean current at the drifter locations. 
Under the restriction of fitting on a single commodity-level GPU, the ensemble size is kept as $N_e = 100$.
The ensemble then runs to the next observation time, and we calculate the innovation vector using all drifters.
We use subsets of the innovation vector to calculate normalized weights for different numbers of observed drifters.
For each observation size, 50 drifter subsets are chosen at random, and for each subset we find the number of particles that have a normalized weight larger than $1/N_e$ (in this case, $1\%$), which guarantees that the given particle is kept in the ensemble when using residual sampling~\cite{Liu98sequentialmonte}.
\reffig{sirCollapse} shows the mean number of particles which are guaranteed to be resampled for different observation sizes.
The figure shows that if we observe one drifter only, we can expect about nine drifters to obtain a weight larger than $1/N_e$, but already when observing two drifters we see that this weight level reached by two particles only, which results in an ensemble collapse.
Since the weight distribution depends largely on the size of the observation error, we make the same weight calculations assuming that the uncertainty in the observations are ten times as larger.
The result is that the weight is distributed on more particles for all the observation sizes, but when observing six or more drifters, most of the weight is still on only three particles.
This experiment clearly confirms how the standard particle filter cannot be used for the application at hand.

\subsection{Computational performance}
\label{sec:performance}
The baseline for evaluating the computational performance of the data-assimilation system is the efficiency of running just the model, consisting of the numerical scheme for solving the shallow-water equations.
The implementation is based on the same approach as a GPU implementation of a very similar scheme~\cite{sw12}, and has been profiled and optimized to maximize its performance and the occupancy of the GPU. 
The scheme has also been tuned to use the optimal block-size configuration applicable to the specific GPU used in this work, an Nvidia GeForce GTX 780.
The efficiency of all other kernels will be evaluated through a comparison to the deterministic model step, to search for limitations and bottlenecks for relevant applications, such as the experiments in \refsec{driftTrajectoryForecasting}.

We start by evaluating the computational performance of the stochastic model errors by comparing the run-time required for generating $\modelerror$ and deterministically evolving the model one time step.
We analyse a benchmark application using $500 \times 300$ grid cells, with model errors added every $\dt_{scheme}$ and a coarsening factor similar to the above experiments, $c_{\Omega} = 5$.
Profiling reveals that 74\% of the GPU compute time is spent evaluating the numerical scheme, 22\% is spent on interpolation, 1.4\% on the SOAR function, and 1.4\% on generating random numbers.
This indicates that the implementation of the model error is sufficiently efficient compared to the model step and does not represent a major performance bottleneck.
It should also be noted that whereas the relationship between the deterministic model step and the interpolation does not change when the number of grid cells is increased, the relative amount of compute time spent in the other two kernels becomes negligible.

\begin{figure*}[t!]
    \centering
    
    \begin{subfigure}[t]{0.25\textwidth}
        \centering
        \includegraphics[width=\textwidth, trim=0 0 6.2cm 0, clip]{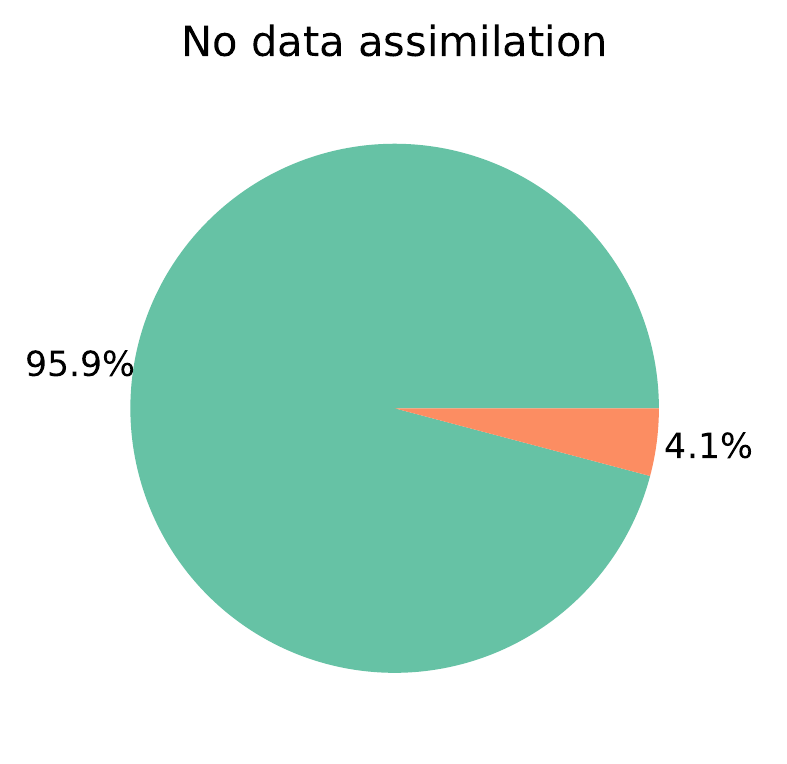}
    \end{subfigure}%
	~
	\begin{subfigure}[t]{0.24\textwidth}
        \centering
        \includegraphics[width=\textwidth, trim=0 0 6.2cm 0, clip]{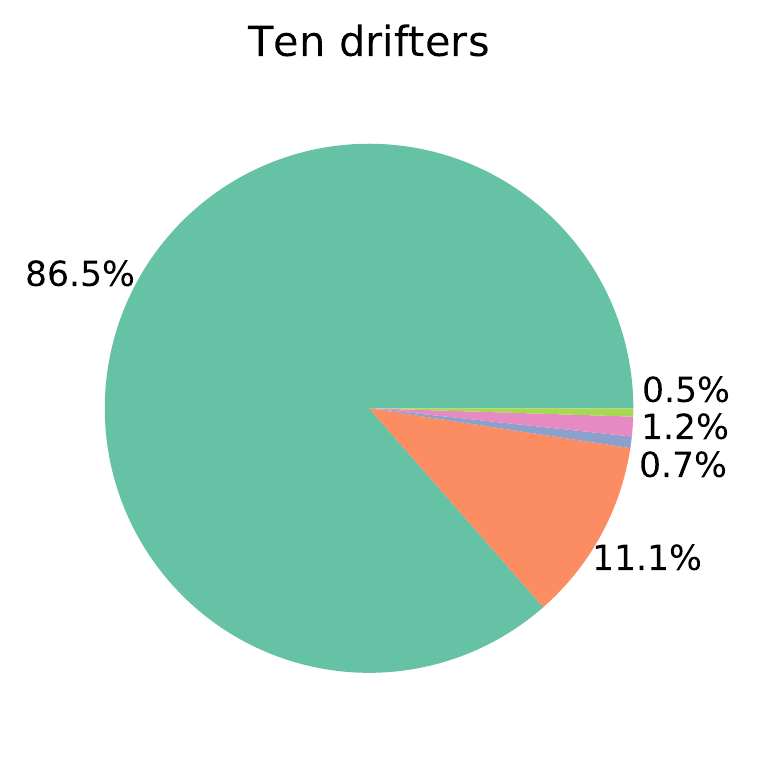}
    \end{subfigure}    
    ~
    \begin{subfigure}[t]{0.43\textwidth}
        \centering
        \includegraphics[width=\textwidth]{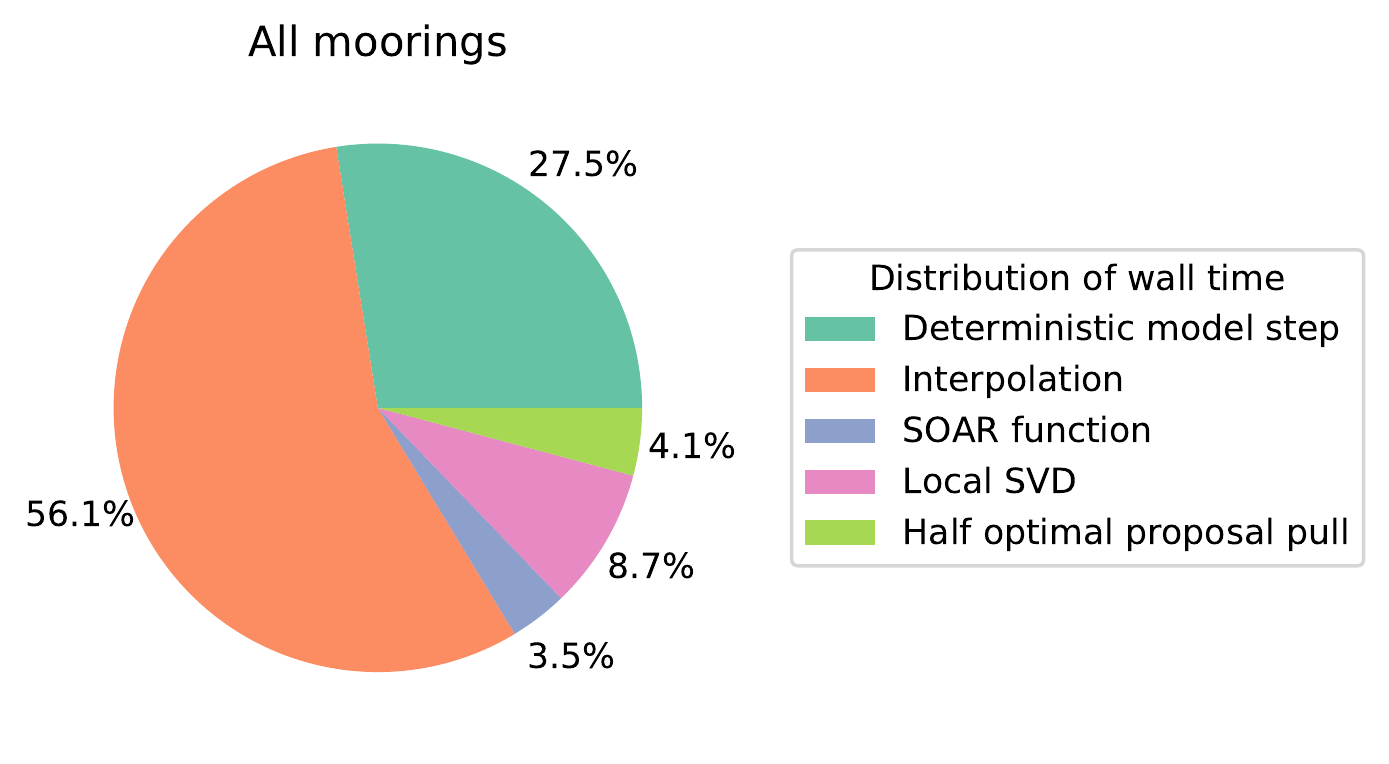}
    \end{subfigure}
    \caption{The pie charts show the distribution of total GPU compute time spent in the different CUDA kernels during the data-assimilation part of three chosen experiments. To the left, we see that only a small part of the compute time is spent on generating the model error, and that the time spent in the SOAR function is negligible. The center chart shows that the overhead from data assimilation on a small drifter set triples the amount of time spent in kernels that do not contribute to solving the deterministic model. With a large number of moorings, however, the majority of the time is spent in the interpolation kernel, and other kernels related to the data assimilation also play a significant part of the compute time, as seen to the right.
    }
    \label{fig:kernelDistribution}
\end{figure*}

\reffig{kernelDistribution} shows the distribution of GPU compute time during some data-assimilation cycles for three of the experiments from \refsec{driftTrajectoryForecasting}, restricted to $N_e = 10$ to make it feasible to run short experiments through the profiler.
During these experiments, the model error is added every model time step, which typically consists of eight steps of the numerical scheme.
In the experiment with no data assimilation, the model error amounts to only 4.1\% of the GPU compute time.
With assimilation of ten drifters, shown in the center pie chart, the fraction of time spent on interpolation increases to 11.1\%, whereas an additional 2.4\% is spent on other assimilation-related kernels.
The majority of the time is nevertheless still spent on the deterministic model step, meaning that there is limited value in optimizing the particle filter kernels for this problem size.
When assimilating all 240 moorings, we reach a situation in which the interpolation represents 56.1\% of the GPU compute time, and a further effort in optimizing the IEWPF implementation should be considered.
Some ideas for this are discussed at the end of this section.

\begin{figure}
    \begin{center}
    \includegraphics[width=0.95\linewidth]{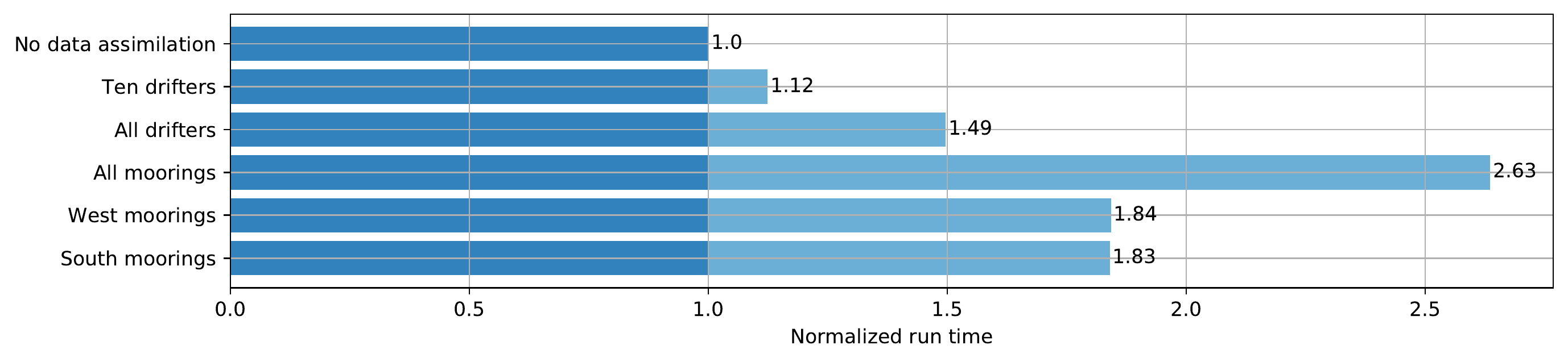}
    \caption{Wall clock run-time measured for the data-assimilation part for each of the six forecast experiments, normalized with respect to the experiment without data assimilation. The lighter color indicates time used on the data assimilation. The assimilation of observations from ten drifters gives a 12\% overhead, whereas using all 240 moorings adds 160\% to the total wall clock time.}
    \label{fig:relativeWallClockRuntime}
    \end{center}
\end{figure}

\reffig{relativeWallClockRuntime} shows the wall clock time for each of the six experiments with $N_e = 100$, normalized with respect to the experiment without data assimilation.
Note that the additional time spent on the data assimilation per drifter observed is constant for the mooring experiments.
For the drifters, there is a small overhead with 64 drifters, but for ten drifters the data assimilation takes almost twice as long per drifter as for the mooring experiments.
These observations are well in accordance with the algorithmic complexity outlined in \reffig{iewpfAlgorithm}.

The wall clock run-time for simulating one hour of data assimilation, consisting of twelve data-assimilation cycles for 100 particles, is 41 seconds on the Nvidia GeForce GTX 780.
This GPU represents a commodity-level graphics card, which has been used for five years at the time of writing, and thereby represents a class of GPUs that is widely available.
By upgrading to a modern high-end GPU, such as the Nvidia Tesla P100, we have observed a 3 times speed up without adjusting any implementation configurations. 

The profiler shows that the occupancy (the concurrent utilization of the available resources on the GPU) is 97.3\% for sufficiently large domains, without exposing any clear strategy for further optimization.
At this point, the kernel has already been tuned by balancing occupancy and register spilling to achieve optimal performance.
To increase the performance for the experiments with a large number of observations, the main focus should therefore be on optimizing the use of the bicubic interpolation kernel.
A high-level performance optimization would be to introduce parallel processing of drifters during the optimal proposal pull, as the interpolation is currently done once per drifter during this step.
This would require that drifters with the same offset configuration are identified, and that those drifters are color coded according to their location within the domain to avoid overlapping memory access.
For this strategy to be fruitful, the number of drifters must be sufficiently large compared to the number of possible offset configurations, $c_{\Omega}^2$, so that the extra computational work required to color code the drifters is compensated by the expected amount of increased parallelization.
This trade-off is less of an issue with mooring observations, as the constant location of the moorings would mean that the color coding can be pre-computed, rather than updated for every observation time step.

\section{Summary and conclusions}
\label{sec:conclusion}

We have presented a GPU implementation of the state-of-the-art implicit equal-weights particle filter applied to an ensemble of simplified ocean models and used it to forecast drift trajectories.
The observations are obtained from the positions of passive drifters and direct ocean current measurements from moored buoys in a synthetic true state.
Forecasts of drift trajectories have been generated for a near-realistic unstable jet experiment, for which the instabilities develop chaotically due to random model error realizations.
All parts of the data-assimilation system (model, model errors, and particle filter) have been designed to take advantage of fine-grained data parallelism, and we have shown that the most computationally expensive components are able to efficiently utilize the resources on a GPU. 

We have shown how the forecast quality is improved as more drifter and mooring observations are assimilated through the forecast experiments.
The best results are achieved when information is assimilated from all 240 available moorings equally distributed throughout the domain.
Even though the observations cover only approximately 0.1\% of the state space, the ensemble mean at the start of the forecast is a very good representation of the true state.
Since the ensemble contains a very accurate description of the true ocean currents, the forecast is shown to be both accurate and confident, even in the long-term up to three days.
Two of the experiments assimilated mooring observations from only the southern half or only the western half of the domain, respectively.
As the dominating currents are in the east-west direction, these experiments illustrate the importance of considering information transport in the system.
The ensemble mean after the data-assimilation period and the general quality of the drift trajectory forecasts are significantly better when both the jets were partially observed (west moorings) compared to observations of one full jet (south moorings) only.

With fewer drifter observations, we have seen that the ensemble is not able to capture the model state with the same accuracy compared to using lots of moorings.
However, the short-term forecasts are significantly improved for the first 12 hours, which is an important time scale for search and rescue operations.
The drifter experiments are also more realistic in terms of equipment than the mooring experiments.
In an operational setting, drifters could be released in the area of interest by, e.g., a search and rescue vessel, to sample relevant observations.
With our approach consisting of an efficient data-assimilation system applied to simplified models, these observations can be used to perform in-situ drift trajectory forecasts, using the most recent traditional ocean forecasts as starting points.


Although the results from the particle filter are good, some issues remain. 
Since we assimilate single-point mass transport, the update takes a dipolar structure in sea-surface height around the observation location. 
The size of these dipoles is limited to the length scales in the model error covariances and can be smaller than the length scale of actual eddies, potentially leading to unrealistic updates some distance away from the observation localtions. 
This is indeed what we see when only 10 drifters are present. 
Different structures for the model errors should improve this issue.

All experiments are conducted with a barotropic ocean model. 
The resulting currents would not be representative of realistic situations with strong bottom topography, and hence a reduced gravity set up would be more appropriate. 
This is not conceptually different from our current approach and, since this also would allow us to use a larger time step, it could further contribute to accelerate the model forecasts. 
A more extensive alternative to a reduced gravity model would be to extend our method to multilayered systems.
But again, no major obstacles are expected for such an extension. 
In fact, it might result in a better balance between data assimilation effort and forecast effort.



\section*{Acknowledgments}
HHH and MLS thanks the Research Council of Norway for funding the GPU Ocean project, with grant number 250935.
PJvL thanks the European Research Council for funding the CUNDA grant 694509 under the European Union's Horizon 2020 research and innovation programme.
Some of the computations were performed on resources provided by UNINETT Sigma2 -- the National Infrastructure for High Performance Computing and Data Storage in Norway under project number nn9550k.
Furthermore, the authors would like to thank Andr\'{e} Rigland Brodtkorb for valuable discussions, and Kai H\aa{}kon Christensen and Knut-Andreas Lie for feedback on the manuscript.

\section*{Supplementary material}
The source code for the methods and experiments described in this paper is available under an GNU open source license under the DOI 10.5281/zenodo.3458291.
The complete datasets representing the ensemble results presented in this paper are available under a GNU free and open source license under the DOI 10.5281/zenodo.3457538.



\appendix
\numberwithin{equation}{section}

\section{A modified implicit equal-weights particle filter}
\label{app:iewpfDetails}
As mentioned in the main text, the update equation for each particle $\state_i$ in the original implicit equal-weights particle filter (IEWPF)~\cite{pjvl_2016_iepfw} is
\begin{equation}
	\state_i^n = \state_i^{n,a} + \alpha_i^{1/2} P^{1/2} \xi_i.
	\label{eq:iewpfUpdateAgain}
\end{equation}
Because $\state_i^{n,a}$ is a deterministic move of the particles according to \eref{opd_update}, this is a transformation of coordinates from $\state$ to $\xi$, so we can write
\begin{equation}
	q(\state^n | \state_{1:N_e}^{n-1}, \observation^n) = \frac{q(\xi)}{\left\Vert \frac{\mathrm{d} \state}{\mathrm{d} \xi} \right\Vert}.
    \label{eq:implicitProposalDensity}
\end{equation}
The denominator represents the absolute value of the determinant of the Jacobian, and can be found through the mapping between $\xi_i$ and $\state_i^n$. This mapping is complicated because $\alpha_i$ also depends on $\xi_i$, but in an up-to-now unknown way. Using \eref{implicitProposalDensity}, the expression for the weights from \eref{sampledProposalDensity} becomes
\begin{equation}
	w_i^{n} = \frac{ p(\observation^n | \state_i^n) p(\state_i^n | \state_i^{n-1})}{N_e p(\observation^n) q(\xi)} \left\Vert \frac{\mathrm{d} \state_i^n}{\mathrm{d} \xi_i} \right\Vert.
    \label{eq:iewpfWeights}
\end{equation}

By assuming that $\alpha_i$ only depends on $\xi_i$ through its magnitude $\xi_i^T \xi_i = \gamma_i$, \eref{iewpfWeights} can be written as the scalar implicit equation
\begin{equation} 
	-\log \left(w_i^n\right) = (\alpha_i - 1)\gamma_i - 2 \log \left[ \alpha_i^{N_{\state}/2} \left| 1 + \frac{\gamma_i}{\alpha^{1/2}_i} \frac{\partial \alpha^{1/2}_i}{\partial \gamma_i}  \right|   \right] + c_i,
	\label{eq:unfinishedImplicitEquation}
\end{equation}
in which 
\begin{equation}
	c_i = \phi_i - \log \left(w_i^{n-1}\right)
	\label{eq:iewpf_ci}
\end{equation}
and
\begin{equation}
    \phi_i = (\innovation_i^n)^T \left(HQH^T + R \right)^{-1} \innovation_i^n.
    \label{eq:iewpfPhi}
\end{equation}
The essence of the IEWPF is that in order to ensure a significant weight for all particles, $\alpha_i$ is chosen so that all weights become equal to a target weight, $w_i^n = w_{target}$ for $i = 1,..., N_e$, leading to a nonlinear equation for each $\alpha_i$.
See \cite{pjvl_2016_iepfw} or the appendix of Skauvold et al.~\cite{skauvold2019_2sIEWPF} for further details.
\nomenclature{$\Vert \cdot \Vert$}{Absolute value of the determinant of a matrix}
\nomenclature{$\alpha_i$}{The IEWPF scaling factor to the optimal proposal density covariance $P$ as used for particle $i$.}
\nomenclature{$\xi_i$}{Random $N(0,I)$ vector sampled for the IEWPF scheme for particle $i$.}
\nomenclature{$\gamma_i$}{Defined as $\xi_i^T \xi_i$ and used in the implicit equation in IEWPF}
\nomenclature{$c_i$}{Term used in the implicit equation of IEWPF representing the optimal proposal weight and the weight at the previous time step}
\nomenclature{$\phi_i$}{The optimal proposal weight for particle $i$, used in the IEWPF scheme}

It is important to set the target weight such that all particles can reach it.
Since a smaller $c_i$ leads to a larger weight, and since $c_i$ denotes the best value for the weight that particle $i$ can attain, the target weight has to be related to the maximum of the $c_i$, and it is chosen as
\begin{equation}
	w_{target} = \max_{i=1,...,N_e} \{c_i\}.
    \label{eq:iewpf_targetWeight}
\end{equation}
By setting  $-\log(w_i) = w_{target}$ in \eref{unfinishedImplicitEquation}, the expression for $\alpha_i$ becomes 
\begin{equation}
	\begin{split}
	(\alpha_i - 1)\gamma_i - 2 \log \left[ \alpha^{N_{\state}/2} \left| 1 + \frac{\gamma_i}{\alpha^{1/2}_i} \frac{\partial \alpha^{1/2}_i}{\partial \gamma_i}  \right|   \right] & = w_{target} - c_i.
    \end{split}
	\label{eq:implicitEquation}
\end{equation}
This equation is equivalent to 
\begin{equation}
	\Gamma\left(\frac{N_x}{2}, \frac{\alpha_i \gamma_i}{2} \right) = e^{-c_i^{\star}/2} \Gamma\left(\frac{N_x}{2}, \frac{ \gamma_i}{2} \right),
    \label{eq:numericalEqForAlpha}
\end{equation}
which can be solved numerically for $\alpha_i$ by, e.g., the Newton method, as illustrated by Skauvold et al.~\cite{skauvold2019_2sIEWPF}.
Here, we use that
\begin{equation}
	c_i^{\star} = w_{target} - c_i = \max_{j=1,...,N_e} \{c_j\} - c_i,
    \label{eq:iewpfCStar}
\end{equation}
and $\Gamma(s,x) = \int_{0}^{x} t^{s-1} e^{-t} \mathrm{d}t$ is the incomplete lower gamma function.
Whenever the state space $N_{\state}$ is large, however, \eref{numericalEqForAlpha} becomes harder to solve as the gamma functions become prone to overflow.
In this high-dimensional limit, it is possible to solve \eref{implicitEquation} analytically in terms of the Lambert W function, as showed by Zhu et al.~\cite{pjvl_2016_iepfw}, as
\begin{equation}
	\alpha_i = - \frac{N_{\state}}{\gamma_i} W_{0} \left[ - \frac{\gamma_i}{N_{\state}} e^{-\gamma_i/N_{\state}} e^{-c_i^{\star}/N_{\state}}\right].
	\label{eq:iewpfAlphaSol}
\end{equation}
As pointed out by Skauvold et al~\cite{skauvold2019_2sIEWPF}, only solutions $\alpha_i < 1$ should be accepted, meaning that only the zero branch for the Lambert W function is considered.
\nomenclature{$w_{target}$}{The target weight for the IEWPF}
\nomenclature{$W_{0,1}(\cdot)$}{The Lambert W function, with 0 and 1 denoting different branches.}
\nomenclature{$c_i^{\star}$}{Similar to $c_i$ but related to the target weight.}
\nomenclature{$\Gamma(s,x)$}{The incomplete lower gamma function}

Two weaknesses of the scheme above can be identified. Firstly, for low-dimensional systems it can be shown that the posterior variance is always underestimated. The other weakness occurs for high-dimensional systems. Since all particles have to reach the same target weight, and that target weight has to be chosen as the weight of the weakest particle, the more particles we use the worse the weakest particle will be, so the further away all particles are pushed from the high likelihood values. So, in high dimensions, although the scheme is useful for small ensemble sizes, it degenerates at larger ensemble sizes.

To overcome the above-mentioned challenges, a revised two-stage IEWPF scheme has been proposed~\cite{skauvold2019_2sIEWPF}, which explores the complete proposal density and does not underestimate the posterior variance.
The new update equation is
\begin{equation}
	\state_i^n = \state_i^{n,a} + \beta^{1/2} P^{1/2} \nu_i + \alpha_i^{1/2}P^{1/2}\xi_i,
	\label{eq:updateEquationTwoStageAgain}
\end{equation}
in which $\nu_i$ is a second random vector $\nu_i \sim N(0,I)$, and $\beta$ is a covariance scaling parameter common to all particles.
Using the same assumptions as for the one-stage method, \eref{iewpfWeights} can now be written as
\begin{equation}
    	\log \left(w_i^n\right) = (\alpha_i - 1)\gamma_i + 2\beta^{1/2}\alpha_i^{1/2}\xi_i^T \nu_i + (\beta - 1)\zeta_i - 2 \log \left[ \alpha_i^{N_{\state}/2} \left| 1 + \frac{\gamma_i}{\alpha^{1/2}_i} \frac{\partial \alpha^{1/2}_i}{\partial \gamma_i}  \right|   \right] + c_i,
    \label{eq:unfinishedImplicitEquationTwoStage}
\end{equation}
in which $c_i$ is according to \eref{iewpf_ci}, and $\zeta_i = \nu_i^T \nu_i$.
To solve \eref{unfinishedImplicitEquationTwoStage}, $\nu_i$ is constructed to be perpendicular to $\xi_i$, making the cross term between the two random vectors disappear.
In the case of large $N_{\state}$, and by defining
\begin{equation}
	c_i^{\star} = w_{target} - c_i - (\beta - 1)\zeta_i,
	\label{eq:cistarTwoStage}
\end{equation}
\eref{unfinishedImplicitEquationTwoStage} becomes similar to \eref{implicitEquation}, with solution according to \eref{iewpfAlphaSol}.
We require that $c_i^{\star} \geq 0$, which is equivalent to 
\begin{equation}
	\beta \leq \frac{w_{target} - c_i}{\zeta_i} + 1.
    \label{eq:betaRequirement}
\end{equation}
This equation shows that the introduction of $\beta$ allows us to choose a different target weight. By choosing the target weight to be $w_{target} = \mean{c_i}$, the mean of $c_i$ across the ensemble, $\beta$ can be set to the minimum value of the right-hand-side of \eref{betaRequirement}, 
\begin{equation}
	\beta = \min_{i=1,...,N_e} \left\{ \frac{\mean{c_i} - c_i}{\zeta_i} + 1 \right\}.
	\label{eq:betaTwoStage}
\end{equation}
Since $\mean{c} - c_i \approx N_y \pm \sqrt{2N_y}$ and $\zeta_i \approx N_{\state} \pm \sqrt{2N_{\state}}$, the parameter $\beta^{1/2}$ should remain real as long as $N_{\state} >> N_{y}$, which holds for our high-dimensional application.
\nomenclature{$\beta$}{The second scaling parameter for the optimal proposal covariance used in the two-stage IEWPF}
\nomenclature{$\nu_i$}{Second random vector from $N(0,I)$ used in the two-stage IEWPF scheme}
\nomenclature{$\bar{c}$}{The mean of $c_i$ across all particles, related to the mean of the optimal proposal weights}
\nomenclature{$\zeta_i$}{Defined as $\nu_i^T \nu_i$ and used in the implicit equation in two-stage IEWPF}

Choosing the target weight equal to the mean of $c_i$, is equivalent to choosing it equal to the mean of the optimal proposal weights.
An advantage with this choice is that the target weight will not vary much when $N_e$ increases.
This is contrary to the one-stage scheme, in which the target weight is equal to $\max_{i=1,..,N_e} \{c_i\}$, which becomes larger if $N_e$ increases.
In other words, the one-stage scheme pushes the particles further and further away from the high-probability regions of the posterior.
Because of its choice of $w_{target}$, the two-stage scheme does not have this problem, and is the method of choice in this paper.

\bibliographystyle{ieeetr}
\bibliography{arxiv_references}

\end{document}